\documentclass[longauth]{aa}
\usepackage{float}
\usepackage{placeins}
\usepackage{graphicx}
\usepackage{natbib}
\usepackage{scalerel}
\usepackage{txfonts}
\usepackage{amsmath}
\usepackage{multirow}
\usepackage{siunitx}

\usepackage[table]{xcolor}

\usepackage{txfonts}
\usepackage[pdfencoding=auto,psdextra]{hyperref}
\hypersetup{
    colorlinks=true,
    linkcolor=blue,
    filecolor=magenta,      
    urlcolor=blue,
    citecolor=blue
}
\makeatletter
\renewcommand*\aa@pageof{, page \thepage{} of \pageref*{LastPage}}
\makeatother

\usepackage[utf8]{inputenc}

\usepackage[switch, modulo]{lineno}

\usepackage{euclid}

\titlerunning{\Euclid\/ preparation. XXXIII. Finding strong lenses}
\authorrunning{Euclid Collaboration: Leuzzi et al.}
\begin{document}
%
%
   \title{\Euclid\/ preparation}
\subtitle{XXXIII. Characterization of convolutional neural
networks for the identification of galaxy-galaxy strong-lensing events}

    \newcommand{\orcid}[1]{} 
    \author{\normalsize Euclid Collaboration: L.~Leuzzi$^{1,2}$\thanks{\email{laura.leuzzi3@unibo.it}}, M.~Meneghetti\orcid{0000-0003-1225-7084}$^{2,3}$, G.~Angora\orcid{0000-0002-0316-6562}$^{4,5}$, R.~B.~Metcalf\orcid{0000-0003-3167-2574}$^{1}$, L.~Moscardini\orcid{0000-0002-3473-6716}$^{1,2,3}$, P.~Rosati\orcid{0000-0002-6813-0632}$^{4,2}$, P.~Bergamini\orcid{0000-0003-1383-9414}$^{6,2}$, F.~Calura\orcid{0000-0002-6175-0871}$^{2}$, B.~Clément\orcid{0000-0002-7966-3661}$^{7}$, R.~Gavazzi\orcid{0000-0002-5540-6935}$^{8,9}$
% ...
}
    
    \institute{$^{1}$ Dipartimento di Fisica e Astronomia "Augusto Righi" - Alma Mater Studiorum Universit\'a di Bologna, via Piero Gobetti 93/2, 40129 Bologna, Italy\\
    $^{2}$ INAF-Osservatorio di Astrofisica e Scienza dello Spazio di Bologna, Via Piero Gobetti 93/3, 40129 Bologna, Italy\\
    $^{3}$ INFN-Sezione di Bologna, Viale Berti Pichat 6/2, 40127 Bologna, Italy\\
    $^{4}$ Dipartimento di Fisica e Scienze della Terra, Universit\'a degli Studi di Ferrara, Via Giuseppe Saragat 1, 44122 Ferrara, Italy\\
    $^{5}$ INAF-Osservatorio Astronomico di Capodimonte, Via Moiariello 16, 80131 Napoli, Italy\\
    $^{6}$ Dipartimento di Fisica "Aldo Pontremoli", Universit\'a degli Studi di Milano, Via Celoria 16, 20133 Milano, Italy\\
    $^{7}$ Institute of Physics, Laboratory of Astrophysics, Ecole Polytechnique F\'ed\'erale de Lausanne (EPFL), Observatoire de Sauverny, 1290 Versoix, Switzerland\\
    $^{8}$ Aix-Marseille Universit\'e, CNRS, CNES, LAM, Marseille, France\\
    $^{9}$ Institut d'Astrophysique de Paris, UMR 7095, CNRS, and Sorbonne Universit\'e, 98 bis boulevard Arago, 75014 Paris, France\\
    $^{10}$ Dipartimento di Fisica e Astronomia, Universit\'a di Bologna, Via Gobetti 93/2, 40129 Bologna, Italy\\
    $^{11}$ Department of Physics and Astronomy, University of the Western Cape, Bellville, Cape Town, 7535, South Africa\\
    $^{12}$ South African Radio Astronomy Observatory, 2 Fir Street, Black River Park, Observatory, 7925, South Africa\\
    $^{13}$ INAF-IASF Milano, Via Alfonso Corti 12, 20133 Milano, Italy\\
    $^{14}$ Observatoire de Sauverny, Ecole Polytechnique F\'ed\'erale de Lausanne, 1290 Versoix, Switzerland\\
    $^{15}$ Universit\'e Paris-Saclay, CNRS, Institut d'astrophysique spatiale, 91405, Orsay, France\\
    $^{16}$ Institute of Cosmology and Gravitation, University of Portsmouth, Portsmouth PO1 3FX, UK\\
    $^{17}$ Institut f\"ur Theoretische Physik, University of Heidelberg, Philosophenweg 16, 69120 Heidelberg, Germany\\
    $^{18}$ Max Planck Institute for Extraterrestrial Physics, Giessenbachstr. 1, 85748 Garching, Germany\\
    $^{19}$ INAF-Osservatorio Astrofisico di Torino, Via Osservatorio 20, 10025 Pino Torinese (TO), Italy\\
    $^{20}$ Dipartimento di Fisica, Universit\'a di Genova, Via Dodecaneso 33, 16146, Genova, Italy\\
    $^{21}$ INFN-Sezione di Genova, Via Dodecaneso 33, 16146, Genova, Italy\\
    $^{22}$ Department of Physics "E. Pancini", University Federico II, Via Cinthia 6, 80126, Napoli, Italy\\
    $^{23}$ Instituto de Astrof\'isica e Ci\^encias do Espa\c{c}o, Universidade do Porto, CAUP, Rua das Estrelas, PT4150-762 Porto, Portugal\\
    $^{24}$ Dipartimento di Fisica, Universit\'a degli Studi di Torino, Via P. Giuria 1, 10125 Torino, Italy\\
    $^{25}$ INFN-Sezione di Torino, Via P. Giuria 1, 10125 Torino, Italy\\
    $^{26}$ Institut de F\'{i}sica d'Altes Energies (IFAE), The Barcelona Institute of Science and Technology, Campus UAB, 08193 Bellaterra (Barcelona), Spain\\
    $^{27}$ Port d'Informaci\'{o} Cient\'{i}fica, Campus UAB, C. Albareda s/n, 08193 Bellaterra (Barcelona), Spain\\
    $^{28}$ INAF-Osservatorio Astronomico di Roma, Via Frascati 33, 00078 Monteporzio Catone, Italy\\
    $^{29}$ INFN section of Naples, Via Cinthia 6, 80126, Napoli, Italy\\
    $^{30}$ Dipartimento di Fisica e Astronomia "Augusto Righi" - Alma Mater Studiorum Universit\'a di Bologna, Viale Berti Pichat 6/2, 40127 Bologna, Italy\\
    $^{31}$ Centre National d'Etudes Spatiales -- Centre spatial de Toulouse, 18 avenue Edouard Belin, 31401 Toulouse Cedex 9, France\\
    $^{32}$ Institut national de physique nucl\'eaire et de physique des particules, 3 rue Michel-Ange, 75794 Paris C\'edex 16, France\\
    $^{33}$ Institute for Astronomy, University of Edinburgh, Royal Observatory, Blackford Hill, Edinburgh EH9 3HJ, UK\\
    $^{34}$ Jodrell Bank Centre for Astrophysics, Department of Physics and Astronomy, University of Manchester, Oxford Road, Manchester M13 9PL, UK\\
    $^{35}$ European Space Agency/ESRIN, Largo Galileo Galilei 1, 00044 Frascati, Roma, Italy\\
    $^{36}$ ESAC/ESA, Camino Bajo del Castillo, s/n., Urb. Villafranca del Castillo, 28692 Villanueva de la Ca\~nada, Madrid, Spain\\
    $^{37}$ University of Lyon, Univ Claude Bernard Lyon 1, CNRS/IN2P3, IP2I Lyon, UMR 5822, 69622 Villeurbanne, France\\
    $^{38}$ Mullard Space Science Laboratory, University College London, Holmbury St Mary, Dorking, Surrey RH5 6NT, UK\\
    $^{39}$ Departamento de F\'isica, Faculdade de Ci\^encias, Universidade de Lisboa, Edif\'icio C8, Campo Grande, PT1749-016 Lisboa, Portugal\\
    $^{40}$ Instituto de Astrof\'isica e Ci\^encias do Espa\c{c}o, Faculdade de Ci\^encias, Universidade de Lisboa, Campo Grande, 1749-016 Lisboa, Portugal\\
    $^{41}$ Department of Astronomy, University of Geneva, ch. d'Ecogia 16, 1290 Versoix, Switzerland\\
    $^{42}$ INFN-Padova, Via Marzolo 8, 35131 Padova, Italy\\
    $^{43}$ Universit\'e Paris-Saclay, Universit\'e Paris Cit\'e, CEA, CNRS, AIM, 91191, Gif-sur-Yvette, France\\
    $^{44}$ INAF-Osservatorio Astronomico di Trieste, Via G. B. Tiepolo 11, 34143 Trieste, Italy\\
    $^{45}$ INAF-Osservatorio Astronomico di Padova, Via dell'Osservatorio 5, 35122 Padova, Italy\\
    $^{46}$ University Observatory, Faculty of Physics, Ludwig-Maximilians-Universit{\"a}t, Scheinerstr. 1, 81679 Munich, Germany\\
    $^{47}$ INAF-Osservatorio Astronomico di Brera, Via Brera 28, 20122 Milano, Italy\\
    $^{48}$ INFN-Sezione di Milano, Via Celoria 16, 20133 Milano, Italy\\
    $^{49}$ Institute of Theoretical Astrophysics, University of Oslo, P.O. Box 1029 Blindern, 0315 Oslo, Norway\\
    $^{50}$ Jet Propulsion Laboratory, California Institute of Technology, 4800 Oak Grove Drive, Pasadena, CA, 91109, USA\\
    $^{51}$ von Hoerner \& Sulger GmbH, Schlo{\ss}Platz 8, 68723 Schwetzingen, Germany\\
    $^{52}$ Technical University of Denmark, Elektrovej 327, 2800 Kgs. Lyngby, Denmark\\
    $^{53}$ Cosmic Dawn Center (DAWN), Denmark\\
    $^{54}$ Max-Planck-Institut f\"ur Astronomie, K\"onigstuhl 17, 69117 Heidelberg, Germany\\
    $^{55}$ Universit\"ats-Sternwarte M\"unchen, Fakult\"at f\"ur Physik, Ludwig-Maximilians-Universit\"at M\"unchen, Scheinerstrasse 1, 81679 M\"unchen, Germany\\
    $^{56}$ Aix-Marseille Universit\'e, CNRS/IN2P3, CPPM, Marseille, France\\
    $^{57}$ Universit\'e de Gen\`eve, D\'epartement de Physique Th\'eorique and Centre for Astroparticle Physics, 24 quai Ernest-Ansermet, CH-1211 Gen\`eve 4, Switzerland\\
    $^{58}$ Department of Physics, P.O. Box 64, 00014 University of Helsinki, Finland\\
    $^{59}$ Helsinki Institute of Physics, Gustaf H{\"a}llstr{\"o}min katu 2, University of Helsinki, Helsinki, Finland\\
    $^{60}$ NOVA optical infrared instrumentation group at ASTRON, Oude Hoogeveensedijk 4, 7991PD, Dwingeloo, The Netherlands\\
    $^{61}$ Argelander-Institut f\"ur Astronomie, Universit\"at Bonn, Auf dem H\"ugel 71, 53121 Bonn, Germany\\
    $^{62}$ Department of Physics, Institute for Computational Cosmology, Durham University, South Road, DH1 3LE, UK\\
    $^{63}$ Universit\'e Paris Cit\'e, CNRS, Astroparticule et Cosmologie, 75013 Paris, France\\
    $^{64}$ University of Applied Sciences and Arts of Northwestern Switzerland, School of Engineering, 5210 Windisch, Switzerland\\
    $^{65}$ Institut d'Astrophysique de Paris, 98bis Boulevard Arago, 75014, Paris, France\\
    $^{66}$ CEA Saclay, DFR/IRFU, Service d'Astrophysique, Bat. 709, 91191 Gif-sur-Yvette, France\\
    $^{67}$ European Space Agency/ESTEC, Keplerlaan 1, 2201 AZ Noordwijk, The Netherlands\\
    $^{68}$ Kapteyn Astronomical Institute, University of Groningen, PO Box 800, 9700 AV Groningen, The Netherlands\\
    $^{69}$ Leiden Observatory, Leiden University, Niels Bohrweg 2, 2333 CA Leiden, The Netherlands\\
    $^{70}$ Department of Physics and Astronomy, University of Aarhus, Ny Munkegade 120, DK-8000 Aarhus C, Denmark\\
    $^{71}$ Universit\'e Paris-Saclay, Universit\'e Paris Cit\'e, CEA, CNRS, Astrophysique, Instrumentation et Mod\'elisation Paris-Saclay, 91191 Gif-sur-Yvette, France\\
    $^{72}$ Space Science Data Center, Italian Space Agency, via del Politecnico snc, 00133 Roma, Italy\\
    $^{73}$ Dipartimento di Fisica e Astronomia "G. Galilei", Universit\'a di Padova, Via Marzolo 8, 35131 Padova, Italy\\
    $^{74}$ Departamento de F\'isica, FCFM, Universidad de Chile, Blanco Encalada 2008, Santiago, Chile\\
    $^{75}$ Institut d'Estudis Espacials de Catalunya (IEEC), Carrer Gran Capit\'a 2-4, 08034 Barcelona, Spain\\
    $^{76}$ Institut de Ciencies de l'Espai (IEEC-CSIC), Campus UAB, Carrer de Can Magrans, s/n Cerdanyola del Vall\'es, 08193 Barcelona, Spain\\
    $^{77}$ Centro de Investigaciones Energ\'eticas, Medioambientales y Tecnol\'ogicas (CIEMAT), Avenida Complutense 40, 28040 Madrid, Spain\\
    $^{78}$ Instituto de Astrof\'isica e Ci\^encias do Espa\c{c}o, Faculdade de Ci\^encias, Universidade de Lisboa, Tapada da Ajuda, 1349-018 Lisboa, Portugal\\
    $^{79}$ Universidad Polit\'ecnica de Cartagena, Departamento de Electr\'onica y Tecnolog\'ia de Computadoras,  Plaza del Hospital 1, 30202 Cartagena, Spain\\
    $^{80}$ Institut de Recherche en Astrophysique et Plan\'etologie (IRAP), Universit\'e de Toulouse, CNRS, UPS, CNES, 14 Av. Edouard Belin, 31400 Toulouse, France\\
    $^{81}$ INFN-Bologna, Via Irnerio 46, 40126 Bologna, Italy\\
    $^{82}$ Infrared Processing and Analysis Center, California Institute of Technology, Pasadena, CA 91125, USA\\
    $^{83}$ Instituto de Astrof\'isica de Canarias, Calle V\'ia L\'actea s/n, 38204, San Crist\'obal de La Laguna, Tenerife, Spain\\
    $^{84}$ INAF-Istituto di Astrofisica e Planetologia Spaziali, via del Fosso del Cavaliere, 100, 00100 Roma, Italy\\
    $^{85}$ Department of Physics and Helsinki Institute of Physics, Gustaf H\"allstr\"omin katu 2, 00014 University of Helsinki, Finland\\
    $^{86}$ Junia, EPA department, 41 Bd Vauban, 59800 Lille, France\\
    $^{87}$ Instituto de F\'isica Te\'orica UAM-CSIC, Campus de Cantoblanco, 28049 Madrid, Spain\\
    $^{88}$ CERCA/ISO, Department of Physics, Case Western Reserve University, 10900 Euclid Avenue, Cleveland, OH 44106, USA\\
    $^{89}$ Laboratoire de Physique de l'\'Ecole Normale Sup\'erieure, ENS, Universit\'e PSL, CNRS, Sorbonne Universit\'e, 75005 Paris, France\\
    $^{90}$ Observatoire de Paris, Universit\'e PSL, Sorbonne Universit\'e, LERMA, 750 Paris, France\\
    $^{91}$ Astrophysics Group, Blackett Laboratory, Imperial College London, London SW7 2AZ, UK\\
    $^{92}$ Scuola Normale Superiore, Piazza dei Cavalieri 7, 56126 Pisa, Italy\\
    $^{93}$ SISSA, International School for Advanced Studies, Via Bonomea 265, 34136 Trieste TS, Italy\\
    $^{94}$ IFPU, Institute for Fundamental Physics of the Universe, via Beirut 2, 34151 Trieste, Italy\\
    $^{95}$ INFN, Sezione di Trieste, Via Valerio 2, 34127 Trieste TS, Italy\\
    $^{96}$ Istituto Nazionale di Fisica Nucleare, Sezione di Ferrara, Via Giuseppe Saragat 1, 44122 Ferrara, Italy\\
    $^{97}$ Institut de Physique Th\'eorique, CEA, CNRS, Universit\'e Paris-Saclay 91191 Gif-sur-Yvette Cedex, France\\
    $^{98}$ Dipartimento di Fisica - Sezione di Astronomia, Universit\'a di Trieste, Via Tiepolo 11, 34131 Trieste, Italy\\
    $^{99}$ NASA Ames Research Center, Moffett Field, CA 94035, USA\\
    $^{100}$ Kavli Institute for Particle Astrophysics \& Cosmology (KIPAC), Stanford University, Stanford, CA 94305, USA\\
    $^{101}$ Department of Astronomy and Astrophysics, University of California, Santa Cruz, 1156 High Street, Santa Cruz, CA 95064, USA\\
    $^{102}$ INAF, Istituto di Radioastronomia, Via Piero Gobetti 101, 40129 Bologna, Italy\\
    $^{103}$ Universit\'e C\^{o}te d'Azur, Observatoire de la C\^{o}te d'Azur, CNRS, Laboratoire Lagrange, Bd de l'Observatoire, CS 34229, 06304 Nice cedex 4, France\\
    $^{104}$ Institute for Theoretical Particle Physics and Cosmology (TTK), RWTH Aachen University, 52056 Aachen, Germany\\
    $^{105}$ Institute for Astronomy, University of Hawaii, 2680 Woodlawn Drive, Honolulu, HI 96822, USA\\
    $^{106}$ Department of Physics \& Astronomy, University of California Irvine, Irvine CA 92697, USA\\
    $^{107}$ UCB Lyon 1, CNRS/IN2P3, IUF, IP2I Lyon, 4 rue Enrico Fermi, 69622 Villeurbanne, France\\
    $^{108}$ Department of Astronomy \& Physics and Institute for Computational Astrophysics, Saint Mary's University, 923 Robie Street, Halifax, Nova Scotia, B3H 3C3, Canada\\
    $^{109}$ Dipartimento di Fisica, Universit\'a degli studi di Genova, and INFN-Sezione di Genova, via Dodecaneso 33, 16146, Genova, Italy\\
    $^{110}$ Institute of Space Sciences (ICE, CSIC), Campus UAB, Carrer de Can Magrans, s/n, 08193 Barcelona, Spain\\
    $^{111}$ Department of Computer Science, Aalto University, PO Box 15400, Espoo, FI-00 076, Finland\\
    $^{112}$ Ruhr University Bochum, Faculty of Physics and Astronomy, Astronomical Institute (AIRUB), German Centre for Cosmological Lensing (GCCL), 44780 Bochum, Germany\\
    $^{113}$ Department of Physics, Lancaster University, Lancaster, LA1 4YB, UK\\
    $^{114}$ Instituto de Astrof\'isica de Canarias (IAC); Departamento de Astrof\'isica, Universidad de La Laguna (ULL), 38200, La Laguna, Tenerife, Spain\\
    $^{115}$ Universit\'e Paris-Cit\'e, 5 Rue Thomas Mann, 75013, Paris, France\\
    $^{116}$ Universit\'e PSL, Observatoire de Paris, Sorbonne Universit\'e, CNRS, LERMA, 75014, Paris, France\\
    $^{117}$ Department of Physics and Astronomy, University College London, Gower Street, London WC1E 6BT, UK\\
    $^{118}$ Department of Physics and Astronomy, Vesilinnantie 5, 20014 University of Turku, Finland\\
    $^{119}$ AIM, CEA, CNRS, Universit\'{e} Paris-Saclay, Universit\'{e} de Paris, 91191 Gif-sur-Yvette, France\\
    $^{120}$ Oskar Klein Centre for Cosmoparticle Physics, Department of Physics, Stockholm University, Stockholm, SE-106 91, Sweden\\
    $^{121}$ Centre de Calcul de l'IN2P3/CNRS, 21 avenue Pierre de Coubertin 69627 Villeurbanne Cedex, France\\
    $^{122}$ Dipartimento di Fisica, Sapienza Universit\`a di Roma, Piazzale Aldo Moro 2, 00185 Roma, Italy\\
    $^{123}$ INFN-Sezione di Roma, Piazzale Aldo Moro, 2 - c/o Dipartimento di Fisica, Edificio G. Marconi, 00185 Roma, Italy\\
    $^{124}$ Centro de Astrof\'{\i}sica da Universidade do Porto, Rua das Estrelas, 4150-762 Porto, Portugal\\
    $^{125}$ Department of Mathematics and Physics E. De Giorgi, University of Salento, Via per Arnesano, CP-I93, 73100, Lecce, Italy\\
    $^{126}$ INAF-Sezione di Lecce, c/o Dipartimento Matematica e Fisica, Via per Arnesano, 73100, Lecce, Italy\\
    $^{127}$ INFN, Sezione di Lecce, Via per Arnesano, CP-193, 73100, Lecce, Italy\\
    $^{128}$ Institute of Space Science, Str. Atomistilor, nr. 409 M\u{a}gurele, Ilfov, 077125, Romania\\
    $^{129}$ Institute for Computational Science, University of Zurich, Winterthurerstrasse 190, 8057 Zurich, Switzerland\\
    $^{130}$ Physikalisches Institut, Ruprecht-Karls-Universit\"at Heidelberg, Im Neuenheimer Feld 226, 69120 Heidelberg, Germany\\
    $^{131}$ Universit\'e St Joseph; Faculty of Sciences, Beirut, Lebanon\\
    $^{132}$ Department of Astrophysical Sciences, Peyton Hall, Princeton University, Princeton, NJ 08544, USA\\
    $^{133}$ Centro de Estudios de F\'isica del Cosmos de Arag\'on (CEFCA), Plaza San Juan, 1, planta 2, 44001, Teruel, Spain}

 \date{Received xxx; accepted yyy}

%
%
   \abstract{Forthcoming imaging surveys will increase the number of known galaxy-scale strong lenses by several orders of magnitude. For this to happen, images of billions of galaxies will have to be inspected to identify potential candidates. In this context, deep-learning techniques are particularly suitable for the finding patterns in large data sets, and convolutional neural networks (CNNs) in particular can efficiently process large volumes of images.
   We assess and compare the performance of three network architectures in the classification of strong-lensing systems on the basis of their morphological characteristics. In particular, we implemented a classical CNN architecture, an inception network, and a residual network.
   We trained and tested our networks on different subsamples of a data set of 40 000 mock images whose characteristics were similar to those expected in the wide survey planned  with the ESA mission \Euclid, gradually including larger fractions of faint lenses. We also evaluated the importance of adding information about the color difference between the lens and source galaxies by repeating the same training on single- and multiband images.
   Our models find samples of clear lenses with  $\gtrsim 90\%$ precision and completeness. Nevertheless, when lenses with fainter arcs are included in the training set, the performance of the three models deteriorates with accuracy values of $\sim 0.87$ to $\sim 0.75$, depending on the model. Specifically, the classical CNN and the inception network perform similarly in most of our tests, while the residual network generally produces worse results. Our analysis focuses on the application of CNNs to high-resolution space-like images, such as those that the \Euclid telescope will deliver. Moreover, we investigated the optimal training strategy for this specific survey to fully exploit the scientific potential of the upcoming observations. We suggest that training the networks separately on lenses with different morphology might be needed to identify the faint arcs. We also tested the relevance of the color information for the detection of these systems, and we find that it does not yield a significant improvement. The accuracy ranges from $\sim 0.89$ to $\sim 0.78$ for the different models. The reason might be that the resolution of the \Euclid telescope in the infrared bands is lower than that of the the images in the visual band.}
%
%
%
%

   \maketitle
%
%
%
%
   
\section{Introduction}\label{sec:intro}
Galaxy-galaxy strong-lensing (GGSL) events occur when a foreground galaxy substantially deflects the light emitted by a background galaxy. When the observer, the lens, and the source are nearly aligned and their mutual distances are favorable, the background galaxy appears as a set of multiple images surrounding the lens. These images often have the form of extended arcs or rings.

These events have multiple astrophysical and cosmological applications.
For example, GGSL enables us to probe the total mass of the lens galaxies within the so-called Einstein radius (e.g., \citealt{treu_lensmass, gavazzi_lensmass, nightingale_lensmass}). By independently measuring the stellar mass and combining lensing with other probes of the gravitational potential of the lens (e.g., stellar kinematics), we can distinguish the contributions from dark and baryonic mass and thus study the interplay between these two mass components \citep[e.g.,][]{barnabe_dis, suyu_dis, schuldt_dis}. Accurately measuring the dark matter mass profiles and the substructure content of galaxies also enables us to test the predictions of the standard cold dark matter (CDM) model of structure formation and to shed light on the nature of dark matter \citep[e.g.,][]{grillo_dm, oguri_dm, vegetti_dm, quinn_dm}. Finally, the lensing magnification makes it possible to study very faint and high-redshift sources that would be not observable in the absence of the lensing effects \citep[e.g.,][]{impellizzeri_mag, allison_mag, stacey_mag}. 

The high-mass density in the central regions of galaxy clusters boosts the strong-lensing cross section of individual galaxies \citep{desprez_2018,angora_gc}. Thus, the probability for GGSL is particularly high in cluster fields. \cite{meneghetti_ggsl} suggested that the frequency of GGSL events is a powerful tool for a stress-test of the CDM paradigm \citep[see also][]{2022A&A...668A.188M, ragagnin2022}. Modeling these lensing events helps constraining the cluster mass distribution on the scale of cluster galaxies \citep[e.g.,][]{tu_2008, grillo2014, jauzac_2020, bergamini_2021}.

Fewer than 1000 galaxy-scale lenses have been confirmed so far. They have been discovered, along with more candidates, by employing a variety of methods, including searches for unexpected emission lines in the spectra of elliptical galaxies \citep{bolton_slacs}, sources with anomalously high fluxes at submillimeter wavelengths \citep{2010Sci...330..800N,2017MNRAS.465.3558N}, and sources with unusual shapes \citep{2003MNRAS.341....1M}. Some arc and ring finders have been developed to analyze optical images, and they typically search for blue features around red galaxies \citep[e.g.,][]{2007A&A...461..813C,2007A&A...472..341S,2014ApJ...785..144G,2014A&A...567A.111M,2018PASJ...70S..29S}. 
Assembling extensive catalogs of GGSL systems is arduous because these systems are rare, but this is expected to change in the next decade through upcoming imaging surveys. It has been estimated that the ESA \Euclid space telescope \citep{Laureijs11} and the Legacy Survey of Space and Time \citep[LSST;][]{lsst_paper} performed with the Vera C. Rubin Observatory will observe more than 100 000 strong lenses \citep{collett_sl}, which will significantly increase the number of known systems. Producing large and homogeneous catalogs of GGSL systems like this will be possible because of the significant improvements in spatial resolution, area, and seeing of these surveys compared to previous observations.

Identifying potential candidates will require the examination of hundreds of millions of galaxies; thus, developing reliable methods for analyzing large volumes of data is of fundamental importance. Over the past few years, machine-learning (ML), and specifically, deep-learning (DL), techniques have proven extremely promising in this context. We focus on supervised ML techniques. These automated methods learn to perform a given task in three steps. In the first step, the training, they analyze many labeled examples and extract relevant features from the data. In the second step, the validation, the networks are validated on labeled data whose labels they cannot access to ensure that the learning does not lead to overfitting. The validation occurs at the same time as the training and is used to guide it. In the third step, the architectures are tested on more labeled data that were not used in the previous phases, whose labels are unknown to the models, but that are used to evaluate their performance.

In particular, convolutional neural networks (CNNs; e.g., \citealt{lecun_cnn}) are a DL algorithm that has been successfully applied to several astrophysical problems and is expected to play a key role in the future of astronomical data analysis. Among the many different applications, they have been employed to estimate the photometric redshifts of luminous sources \citep[ e.g.,][]{pasquet_zphot, shuntov_zphot, 2022A&A...666A..85L}, to perform the morphological classification of galaxies \citep[e.g.,][]{huertas_company_morpho, dom_sanchez_morpho, zhu_morphology, ghosh_morphology}, to constrain the cosmological parameters \citep[e.g.,][]{merten_cosmo, fluri_cosmo, pan_cosmo}, to identify cluster members \citep[e.g.,][]{angora_gc}, to find galaxy-scale strong lenses in galaxy clusters \citep[e.g.,][]{angora_ggsl}, to quantify galaxy metallicities \citep[e.g.,][]{wu_metallicity, liewcain_metallicity}, and  to estimate the dynamical masses of galaxy clusters \citep[e.g.,][]{ho_clustermass, gupta_clustermass}. Recently, \cite{oriordan_2023} also tested whether CNNs can be used to detect subhalos in simulated \Euclid-like galaxy-scale strong lenses. 

Several CNN architectures were also used recently to identify strong lenses in ground-based wide-field surveys such as the Kilo Degree Survey \citep[KiDS;][]{kids, 2017MNRAS.472.1129P, petrillo_cnn2, he_cnn, li_cnn, napolitano_cnn, 2021ApJ...923...16L}, the Canada-France-Hawaii Telescope Legacy Survey \citep[CFHTLS;][]{cfhtls, jacobs_cnn3}, the Canada France Imaging Survey \citep[CFIS;][]{savary_cnn}, the Hyper Suprime-Cam Subaru Strategic Program Survey \citep[HSC;][]{aihara_2018, canameras_2021, wong_2022}, and the Dark Energy Survey \citep[DES;][]{des, jacobs_cnn, jacobs_cnn2, rojas_cnn}. Most of them were also employed in two challenges aimed at comparing and quantifying the performance of several methods to find lenses, either based on artificial intelligence or working without it. The first challenge results, presented in  \cite{metcalf_challenge}, showed that DL methods are particularly promising with respect to other traditional techniques, such as visual inspection and classical arcfinders. 

In this work, we investigate the ability of three different network architectures to identify GGSL systems. We test them on different subsamples of a data set of \Euclid-like mock observations. In particular, we evaluate the effect of including faint lenses in the training set on the classification. 

This paper is organized as follows: in Sect. \ref{sec:cnn} we explain how CNNs are implemented and trained to be applied to image-recognition problems, in Sect. \ref{sec:data_set} we introduce the data set of simulated images used for training and testing our networks, and in Sect. \ref{sec:results} we describe our experiments and present and discuss our results. In Sect. \ref{sec:conclusion} we summarize our conclusions.

\section{Convolutional neural networks}\label{sec:cnn}
Artificial neural networks (ANNs; e.g., \citealt{ann, goodfellow}) are an ML algorithm inspired by the biological functioning of the human brain. They consist of artificial neurons, or nodes, that are organized in consecutive layers and linked together through weighted connections. The weights define the sensitivity among individual nodes \citep{hebb_connections} and are adapted to enable the network to carry out a specific task. 

The output of the $k$th layer $\vec{h}^k$ depends on the output of the previous layer $\vec{h}^{k-1}$ \citep{bengio}
\begin{equation}
    \vec{h}^k = f(\vec{b}^k + \tens{W}^k\vec{h}^{k-1}).
    \label{eq:neuron_sum}
\end{equation}

\noindent Here, $\vec{b}^k$ is the vector of offsets (biases), and $\tens{W}^k$ is the weight matrix associated with the layer. The dimension of $\vec{b}^k$ and $\tens{W}^{k}$ corresponds to the number of nodes within the layer, and the symbol $f$ represents the activation function, which introduces nonlinearity in the network that would otherwise only be characterized by linear operations. 

The CNNs are a special class of ANNs that use the convolution operation. Through this property, they perform particularly well on pattern recognition tasks. The basic structure of a CNN can be described as a sequence of convolutional and pooling layers, followed by fully connected layers. Convolutional layers consist of a series of filters, also called kernels, which are matrices of weights with a typical dimension of $3\times3$ to $7\times7$ and act as the weights of a generic ANN. They are convolved with the layer input to produce the feature maps. The feature maps are passed through an activation function that introduces nonlinearity in the network, and they are then fed as input to the subsequent layer. In our networks, we use the leaky rectified linear unit (Leaky ReLU; \citealt{2015arXiv150500853X}) as the activation function. The organization of the filters in multiple layers ensures that the CNN can infer complex mappings between the inputs and outputs by dividing them into simpler functions, each extracting relevant features from the images. The pooling operation downsamples each feature map by dividing it into quadrants with a typical dimension of $2\times2$ or $3\times3$ and substituting them with a summary statistic, such as the maximum \citep{zhou_maxpooling}. This operation has the twofold purpose of reducing the size of the feature maps and therefore the number of parameters of the model, and making the architecture invariant to small modifications of the input \citep{goodfellow}. 

After these layers, the feature maps are flattened into a 1D vector that is processed by fully connected layers and is then passed to the output layer that predicts the output. In classification problems, the activation function used for the output layer is often the softmax, providing an output in the range $[0, 1]$ that can be interpreted \citep{bengio} as an indicator of $P(Y=i\,|\,\vec{x})$, where $Y$ is the class associated with the input $\vec{x}$ of all the possible classes $i$.

The CNNs master the execution of a given task due to a supervised learning process, called training, in which they analyze thousands of known input-output pairs. The weights of the network, which are randomly initialized, are readjusted so that the output predictions of the network are correct for the largest number of possible examples. This step is crucial because the weights are not modified afterward when the final model is applied to other data. The training aims to minimize a loss (or cost) function that estimates the difference between the outputs predicted by the network and the true labels. To do this, the images are passed to the network several times, and at the end of each pass, called epoch, the gradient of the cost function is computed with respect to the weights and is backpropagated \citep{rumelhart_backprop} from the output to the input layer so that the kernels can be adapted accordingly. The magnitude of the variation of the weights is regulated through the learning rate, a hyperparameter that is to be defined at the beginning of the training, whose specific value is fine-tuned by testing different values to find the one that minimizes the loss function.

In addition to showing good performance on the training set, it is essential that the network generalizes to other images. Preventing the model from overfitting (i.e., memorizing peculiar characteristics of the images in the training set that cannot be used to make correct predictions on other data sets) is possible by monitoring the training with a validation step. At the end of each epoch, the network performance is assessed on the validation set, which is a small part of the data set (usually $5-10 \%$) that was excluded from the training set. If the loss function evaluated on these images does not improve for several consecutive epochs, the training should be interrupted or the learning rate reduced. Dropout \citep{srivastava_dropout} is another technique that is used to mitigate overfitting. This method consists of randomly dropping units from the network during training, that is, temporarily removing incoming and outcoming connections from a given node. When the training is completed, the performance of the final model is evaluated on the test set, which is a part of the data set  (about $20-25\%$) that was excluded from the other subsets. The CNN can then be applied to new images.

The CNNs conveniently handle large data sets for several reasons. While the training can take up to a few days to be completed, processing a single image afterward requires a fraction of a second through graphics processing units (GPUs). Moreover, the feature-extraction process during the training is completely automated. The algorithm selects the most significant characteristics for achieving the best results without any previous knowledge of the data. The following subsections provide more information about the specific architectures we test in this work and technical details about our training. 

\subsection{Network architectures}\label{sec:architectures}
We implemented three CNN architectures: a visual geometry group-like network (VGG-like network; \citealt{vggnet}), an inception network (IncNet; \citealt{gnet_1, gnet_2}), and a residual network (ResNet; \citealt{resnet, resnext}).  The definition of the final configuration of the networks that we applied to the images is the result of several trials in which we tested different hyperparameters for the optimization (e.g. the learning rate) and general architectures (e.g., the number of layers and kernels) to find the most suitable arrangement for our classification problem.

\subsubsection{VGG-like network}\label{sec:vggnet}
The visual geometry group network (VGGNet) was first presented by \cite{vggnet}. The most significant innovation introduced with this architecture is the application of small convolutional filters with a receptive field of $3\times3$, which means that the portion of the image that the filter processes at any given moment is $3\times3$ pixels wide. This allowed the construction of deeper models because the introduction of small filters keeps the number of trainable parameters in the CNN smaller than that of networks that use larger filters (e.g., with a dimension of $5\times5$ or $7\times7$). Because the concatenation of multiple kernels with sizes of $3\times3$ has the same resulting receptive field as larger filters \citep{gnet_2}, it is possible to analyze features of larger scales while building deeper architectures.

Our implementation of the VGGNet comprises ten convolutional layers that alternate with five max pooling layers. We define a convolutional-pooling block as two convolutional layers followed by a pooling layer. At the end of each convolutional-pooling block, we perform the batch normalization of the output of the block. Batch normalization consists of the renormalization of the layer inputs \citep{batch_norm} and is employed to accelerate and stabilize the training of deep networks. After five convolutional-pooling blocks, two fully connected layers of 256 nodes each alternate with dropout layers, and finally, a softmax layer as the output layer. The number of parameters for this architecture is about two million.

When training on multiband observations, we add a second branch to process the \Euclid Near Infrared Spectrometer and Photometer (NISP; \citealt{Maciaszek22}) images, passing them to the network through a second input channel. Because they are smaller than the Visual Imager (VIS; \citealt{vis_instrument}) images (see Table \ref{tab:euclid_prop}), this branch of the network is only four convolutional-pooling blocks deep. The outputs of the two branches are flattened and concatenated before they are passed to the output layer. Like in the single-branch version of this architecture, we have two fully connected layers with 256 nodes each, and finally, the output layer. In this configuration, our network uses about three million parameters. In Appendix \ref{app:implementation}, Fig. \ref{fig:my_vgg} shows the VGG-like network configuration we tested on the VIS images (panel a) and on the multiband images (panel b).

\subsubsection{Inception network}\label{sec:incnet}
The reasons for the IncNet architecture were outlined by \cite{gnet_1}, who applied the ideas of \cite{Lin2013NetworkIN} to CNNs. Trying to improve the performance of a CNN by enlarging its depth and width leads to a massive increase in the number of parameters of the model, favoring overfitting and increasing the requirements of computational resources. \cite{gnet_1} suggested applying filters with different sizes to the same input, making the model extract features on different scales in the same feature maps. This is implemented through the inception module.
In the simplest configuration, each module applies filters of several sizes ($1\times1$, $3\times3$, and $5\times5$) and a pooling function to the same input and concatenates their outputs, passing the result of this operation as input to the following layer. However, this implementation can be improved by applying $1\times1$ filters before the $3\times3$ and $5\times5$ filters. Introducing $1\times 1$ filters has the main purpose of reducing the dimensionality of the feature maps, and thus the computational cost of convolutions, while keeping their spatial information. This is possible by reducing the number of channels of the feature maps. An IncNet is a series of such modules stacked upon each other. A further improvement of the original inception module design is presented in  \cite{gnet_2}: The $5\times5$ filters are replaced by two $3\times3$ filters stacked together in order to decrease the number of parameters required by the model. This version of the inception module is used in our network implementation.

Before they are fed to the inception modules, the images are processed through two convolutional layers alternating with two max pooling layers. The network is composed of seven modules, the fifth of which is connected to an additional classifier. The outputs of the two classifiers are taken into account when computing the loss function by computing the individual losses and then taking a weighted sum of them. The intermediate output layer is weighted with weight 0.3, while the final one is weighted with weight 1.0. Dropout is performed before both output layers, while batch normalization is performed on the output of each max pooling layer. The output layers are both softmax layers. The total number of parameters that compose the model is approximately two million.

The configuration used to analyze the multiband images has a secondary branch with one initial convolutional layer and seven inception modules. This branch is characterized by approximately one million parameters, thus leading to a total of around three million parameters. In Appendix \ref{app:implementation}, Fig. \ref{fig:my_incnet} shows the IncNet configuration we tested on the VIS images (panel a) and the multiband images (panel b).  

\subsubsection{Residual network}\label{sec:resnet}
\cite{resnet} introduced residual learning to make the training of deep networks more efficient. 
The basic idea behind the ResNets is that it is easier for a certain layer (or a few stacked layers) to infer a residual function with respect to the input rather than the complete, and more complicated, full mapping. 

In practice, this is implemented using residual blocks with shortcut connections. Let $\vec{x}$ be the input of a given residual block. The input is simultaneously propagated through the layers within the block and stored without being changed, through the shortcut connection. The residual function $\mathcal{F}(\vec{x})$ that the block is expected to infer can be written as
\begin{equation}
    \mathcal{F}(\vec{x}) :=  \mathcal{H}(\vec{x}) - \vec{x},
    \label{eq:residual_mapping}
\end{equation}
where $\mathcal{H}(\vec{x})$ is the function that a convolutional layer would have to learn in the absence of shortcut connections. Thus, the original function can be computed as $\mathcal{F}(\vec{x}) + \vec{x}$. 

This architecture was later improved by \cite{resnext}, who presented the ResNeXt architecture. The main modification introduced in this work is the ResNeXt block, which aggregates a set of transformations, and can be presented as
\begin{equation}
    \mathcal{F}(\vec{x}) = \sum_{i=1}^C \mathcal{T}_i(\vec{x})
    \label{eq:aggregated_transformations}
\end{equation}

\noindent and serves as the residual function in Eq. (\ref{eq:residual_mapping}). Here, $\mathcal{T}_i(\vec{x})$ is an arbitrary function, and $C$ is a hyperparameter called cardinality, which represents the size of the set of transformations to be aggregated. 
 
In our implementation of the ResNet, we use this last ResNeXt block as the fundamental block, with the cardinality set to eight. In particular, the input is initially processed by two convolutional layers alternated with two pooling layers. The resulting feature maps are passed to four residual blocks alternated with two max pooling layers. There follows a dropout layer and finally a softmax layer. Moreover, batch normalization is performed after every max pooling layer. The NISP images are processed by a similar branch, which differs from this one in that it has only one initial convolutional layer.

The parameters of the model are circa one million in the VIS configuration and about two million in the multiband configuration, so they are significantly fewer than those of our implementations of the VGG-like network and of the IncNet. However, we tested different configurations of the ResNet when designing the network architectures, and this specific setup outperformed the others, including those that had a higher number of weights. In Appendix \ref{app:implementation}, Fig. \ref{fig:my_resnext} shows the ResNet configuration we applied to the VIS images (panel a) and the multiband images (panel b).

\section{The data set}\label{sec:data_set}
Training CNNs requires thousands of labeled examples. Because not enough observed galaxy-scale lenses are known to date, simulating the events is necessary for training a classifier to identify them. In some cases, it is possible to include real observations in the training set, but in our case, it is inevitable to adopt a fully simulated data set because no real images have been observed with the \Euclid telescope yet. The realism of the simulations is essential to ensure that the evaluation of the model performance is indicative of the results we may expect from real observations. 

The image simulations were used to produce all the images in the data set, that is, both the lenses and nonlenses. We generated all the images and then divided them into the two classes according to the criteria that we introduced below. The simulations used the galaxy and halo catalogs provided by the Flagship simulation (v1.10.11; Castander et al., in prep.) through the CosmoHub portal\footnote{\url{https://cosmohub.pic.es/home}} \citep{2017ehep.confE.488C,TALLADA2020100391}. 

We constructed the images using the following procedure. We randomly selected a trial lens galaxy from the light cone subject to a magnitude cut of 23 in the VIS band from the \Euclid telescope, that is, the $\IE$ band. After this, we randomly selected a  background source from a catalog of Hubble Ultra Deep Field (UDF; \citealt{2006AJ....132..926C}) sources with known redshift. We decomposed these sources into shapelets for denoising, following the procedure described in \cite{2008AandA...482..403M,2010AandA...514A..93M}. This procedure has its limitations because in regions of high magnification, the finite resolution of the shapelets can be apparent and there can be low surface brightness ringing that is usually not visible above the noise. We investigate the potential impact of these effects on the results of this paper in Sec. \ref{sec:cosmos_lenses}. The mass of the lens is represented by a truncated singular isothermal ellipsoid (TSIE) and a Navarro, Frenk \& White (NFW; \citealt{1996ApJ...462..563N}) halo. The SIE model has been shown to fit existing GGSLs well \citep{2007ApJ...667..176G}.

We used the GLAMER lensing code \citep{glamer_i,glamer_ii} to perform the ray-tracing. Light rays coming from the position of the observer are shot within a $20\arcsecond \times 20\arcsecond$ square centered on the lens object, with an initial resolution of $\ang{;;0.05}$, that is, twice the final resolution of the VIS instrument. We used these rays to compute the deflection angles that trace the path of the light back to the sources. The code detects any caustics in the field and provides some further refinement to characterize them. Specifically, more rays are shot in a region surrounding the caustics to constrain their position with higher resolution. If the area within the largest critical curve is larger than $0.2\, \rm arcsec^2$ and smaller than $20\,\rm arcsec^2$, the object is accepted as a lens of the appropriate size range.  

The lensed image is constructed using the shapelet source and S\'{e}rsic profiles for the lens galaxy and any other galaxy that appears within the field. We took the parameters for the S\'{e}rsic profiles from the Flagship catalog with some randomization. While we placed the lens galaxy at the center of the cutout, the positions of the other galaxies were determined following the Flagship catalogs as well, with some randomization. In this way, the density of galaxies along the line of sight is the same as that of the Flagship simulations, but the sources have a different angular position. We placed the background source galaxy at a random point on the source plane within a circle surrounding the caustic. The radius of the circle was set to one-half of the largest separation between points in the caustic times 2.5. 
    
A model for the point spread function (PSF) is applied to the image which initially has a resolution of 0.025 arcsecs and then downsamples to 0.1 arcsecs for VIS and 0.3 arcsecs for the infrared bands. The VIS PSF was derived from modeling the instrument (Euclid collaboration et al., in prep.). For the infrared bands, a simple Gaussian model with a width of 0.3 arcsecs was used. The noise was simulated with a Gaussian random field to reproduce the noise level expected by the Euclid Wide Survey \citep{Scaramella-EP1}. 

To avoid repeating a particular lens and to increase the number of images at a low computational cost, we randomized each lens. In this step, all the galaxies within a sphere centered on the primary lens are rotated randomly in three dimensions about the primary lens. The sphere radius was set to 30 arcsecs at the distance of the lens. In addition, the galaxies outside this sphere but within the field of view were independently rotated about the primary in the plane of the sky. The mass associated with each galaxy is moved with the galaxy image. The position angles of each galaxy were also randomly resampled.

The final step is the classification of the images as lenses. Some of the images will have low signal-to-noise ratios in some lensed images or are not distorted enough to be recognizable lenses.

This procedure is similar to the one used for the lens-finding challenges that was described in more detail in \cite{metcalf_challenge}. These simulations are currently being improved to provide more realistic representations of lens and source galaxies. This is important both for training the CNNs and for statistical studies (see Sect. \ref{sec:finding_euclid}). A possible improvement that would be relevant in the context of GGSL searches is a better characterization of the blending between the lens and source galaxies in the definition of \texttt{n\_pix\_source} by taking into consideration the fraction of light from lens and source in each pixel. Moreover, the simulations miss some instrumental effects, such as nonlinearity, charge transfer inefficiency, and a more intricate PSF model, which are included in other studies \citep[e.g.,][]{Pires20}.

The result of these simulations are 100 000 \Euclid-like mock images simulated in the $\IE$ band of the VIS instrument and $\HE$, $\YE$ and $\JE$ bands of the NISP instrument \citep{Schirmer-EP18}. The dimensions of the VIS and NISP images are 200$\times$200 and 66$\times$66  pixels, respectively. Given the resolution of the instruments, reported in Table~\ref{tab:euclid_prop}, these correspond to $20'' \times 20''$ images.

\begin{table}[htp]
\centering
\caption{Main characteristics of the \Euclid VIS and NISP \citep{Schirmer-EP18} instruments.} 
\begin{tabular}{@{}cccc@{}}
    Instrument & Capability & $\lambda$ range& Pixel size\\
    &&(nm)&(arcsec)\\ \hline\\ [-1.7ex]
    VIS & Visual imaging & $\IE$ (530--920)
 & 0.1\\
   NISP & NIR imaging  & $\YE$ (949.6--1212.3), & 0.3 \\
    &photometry  & $\JE$ (1167.6--1567.0), &	0.3\\ 
    & & $\HE$ (1521.5--2021.4) & 0.3\\ 
\hline

\end{tabular}
\label{tab:euclid_prop}
\end{table}

When preparing the images for the training, we clean the data set by removing the images with sources at $z > 7$, thus leaving a catalog of $\num{99409}$ objects. We do this because there are just a few hundreds of such objects in the simulated  data set and their number would not be sufficient to grant generalization after training. Moreover the sources at such high redshift are not as reliable as the others used in the simulations. The images in the data set are considered lenses if they meet the following criteria simultaneously:
\begin{equation}
    \begin{cases}
        \texttt{n\_source\_im} > 0; \\  
        \texttt{mag\_eff} > 1.6; \\ 
        \texttt{n\_pix\_source} > 20. 
    \end{cases}
    \label{eq:lens_criteria}
\end{equation} 
Here, \texttt{n\_source\_im} represents the number of images of the background source, \texttt{mag\_eff} is the effective magnification of the source, and \texttt{n\_pix\_source} is the number of pixels in which the surface brightness of the source is 1$\sigma$ above the background noise level. For every image, the magnification is computed as the ratio of the sum of all the pixels with a flux above the noise level in the lensed images on the image plane and the pixels of the unlensed image on the source plane. The most discriminatory parameters seems to be \texttt{n\_pix\_source}.
The same criteria were adopted in the lens-finding challenge 2.0\footnote{\url{http://metcalf1.difa.unibo.it/blf-portal/gg_challenge.html}} (Metcalf et al., in prep.).

In many cases, one or more background sources are present in the nonlenses, but they are too faint or too weakly magnified to be classified as a lens, or both. For this reason, the parameters \texttt{n\_pix\_source} and \texttt{mag\_eff} are also considered in the classification criteria (Eq. \ref{eq:lens_criteria}). Depending on the sensitivity of the model, the classification of the images with a low signal-to-noise ratio might vary, while the clearest images should be immediately assigned to the correct category.

By using these conditions, we divided the images we simulated into $\num{19591}$ lenses and $\num{79816}$ nonlenses, thus obtaining two very unbalanced classes out of the complete data set. It is well known that unbalanced classes result in biased classification \citep{buda2018systematic}. For this reason, we used all the lenses for the training, and we randomly selected only a subsample of $\num{20000}$ nonlenses. As we discuss in Sec. \ref{sec:data_prep}, these numbers were increased by data augmentation. We refer to the nonlenses as class 0 and to the lenses as class 1. More strategies would be possible to deal with the unbalanced data set, such as using different weights for the two classes in the loss function or optimizing our classifiers with respect to purity, but we did not test them.

\begin{figure}
    \centering
    \includegraphics[width = 0.45\textwidth]{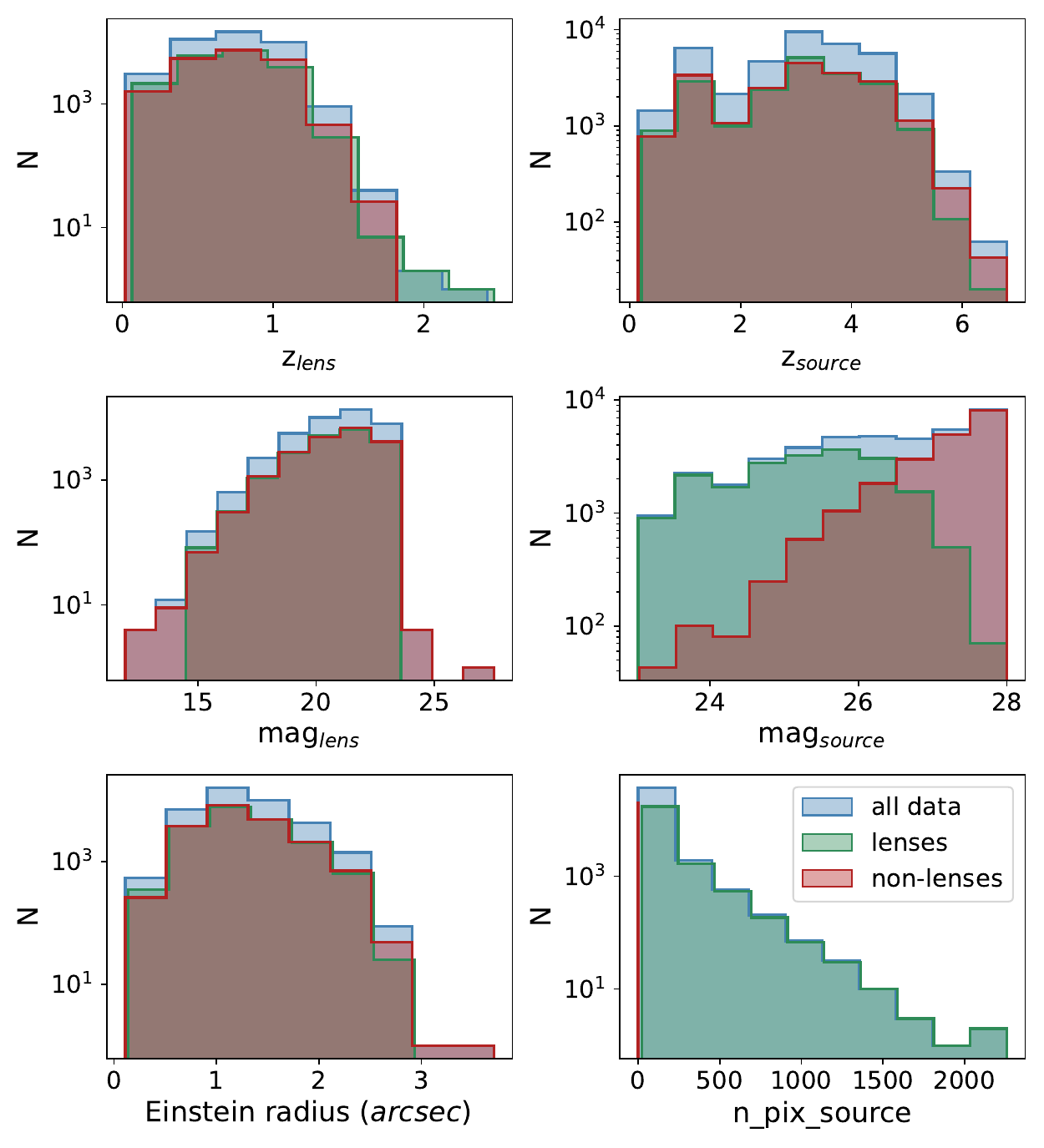}
    \caption{Distribution of several properties of the simulated images in the data set (blue histograms) selected for training, which consisted of $\num{40000}$ mocks in total. The distributions of the same properties in the separate subsets of lenses and nonlenses are given by the green and red histograms, respectively. In the panels in the upper and middle rows, we show the distributions of lens and source redshifts and $\IE$ band magnitudes (in the case of the sources, we refer to the intrinsic magnitude). The bottom panels show the distributions of Einstein radii of the lenses and of the number of pixels for which the source brightness exceeds $1\sigma$ above the background noise level.}
    \label{fig:challenge_hist}
\end{figure}

In Fig. \ref{fig:challenge_hist} we report the distribution of some properties of the images in the data set. From top left to bottom right, we show the distribution of the redshifts of the galaxy lenses and sources, of the magnitudes of the galaxy lenses and sources, of the Einstein radii of the largest critical curve in the lensing system, and of \texttt{n\_pix\_source}. The histograms in each panel refer to the lenses (green) and nonlenses (red) separately and to the complete data set (blue). The galaxy lenses in the two classes share similar distributions of redshift, magnitude, and Einstein radius (top, middle, and bottom left panels, respectively). The redshift distribution of the sources in the top right panel is also similar for the two subsets. On the other hand, the simulated sources (middle right panel) in the nonlenses class are fainter on average than that of the sources in the lenses. This is intuitive because sources with lower magnitudes (i.e., brighter sources) will be more evident in the images, and it will be more likely that they produce a clear lensing event. A similar argument can be made about \texttt{n\_pix\_source} (bottom right panel): the higher the value of this parameter, the clearer the distortion of the source images, hence the lensing system.

\section{Results and discussion}\label{sec:results}
\subsection{Data preprocessing}\label{sec:data_prep}
The data preparation consists of a sequence of several steps. We divided the entire data set into three subsets: the training set (70\%), the validation set (5\%), and the test set (25\%). The images in the data set were randomly assigned to one of these subsets, but we checked that all subsets (training, validation, and testing) were representative of the entire data set. We did this by inspecting the distributions of several parameters that define the characteristics of the lenses and sources in the data set, such as their redshift, magnitude, and Einstein radius.

After the data set was split, we randomly selected 20\% of the images in the training set for augmentation. We performed five augmentations: We rotated these images by $90 \degree$, $180 \degree$, and $270 \degree$ and flipped them with respect to the horizontal and vertical axes. After performing these operations, we doubled the size of the training set. Neither the test set nor the validation set were augmented.

Afterward, we proceeded with the normalization of the images in the data set. We subtracted the mean and divided it by the standard deviation of the mean image of the training set. The mean image of the training set is the image that has for every pixel $i,j$ the mean value of the pixel $i,j$ of all images in the training set. The reason for this type of normalization is that the computation of the gradients in the training stage of the networks is easier when the features in the training set are in a similar range. Moreover, scaling the inputs in this way makes the parameter sharing more efficient \citep{goodfellow}.

\subsection{Training procedure}\label{sec:training}
We implemented, trained, and tested our networks using the library \texttt{Keras}\footnote{\url{https://keras.io/}} \citep{chollet_keras} 2.4.3 with the \texttt{TensorFlow}\footnote{\url{https://www.tensorflow.org/}} \citep{abaldi_tf} 2.2.0 backend on an NVIDIA Titan Xp GPU.

We used the adaptive moment estimation (Adam; \citealt{kingma_adam, reddi_adam}) optimizer with an initial learning rate of $10^{-4}$. We employed the binary cross-entropy $\mathcal{L}$ to estimate the loss at the end of each epoch,

\begin{equation} 
    \mathcal{L} = -\frac{1}{N}\sum_{i=1}^N {y(\textbf{x}_i)\;\ln[y_p(\textbf{x}_i)] + [1-y(\textbf{x}_i)]\;\ln[1-y_p(\textbf{x}_i)]},
    \label{eq:binary_crossentropy}
\end{equation}

\noindent where $N$ is the number of training examples, $\textbf{x}_i$ is  the batch of images used to compute the loss, $y$ is the ground truth, and $y_p$ is the probability that the $i$th example has the label 1, as predicted by the network, so that $1 - y_p$ is the probability that the $i$th example has the label 0. 

The performance of the network on the validation set is estimated at the end of every epoch and is used to monitor the training process. If the loss function evaluated on this independent subset does not decrease for 20 consecutive epochs, the training will be stopped with the \texttt{EarlyStopping}\footnote{\url{https://keras.io/api/callbacks/early_stopping/}} class from \texttt{Keras}. This step is particularly useful to avoid overfitting. At the end of training, we used the best models, that is, those with the lowest value of the loss function on the validation set, for our tests. 

\subsection{Performance evaluation}\label{sec:metrics}
We assessed the performance of our trained networks by examining the properties of the catalogs produced by the classification of the images in the test set. In particular, we considered four statistical metrics that were immediately derived from the confusion matrix \citep{stehman_cm}. A generic element of the confusion matrix $\tens{C}_{ij}$ is given by the number of images belonging to the class $i$ and classified as members of the class $j$. In a binary classification problem like the one considered here, the diagonal elements indicate the number of correctly classified objects, that is, the number of true positives (TP) and the number of true negatives (TN), while the off-diagonal terms show the number of misclassified objects, that is, the number of false positives (FP) and the number of false negatives (FN).

Considering the class of Positives, the combination of these quantities leads to the definition of the following metrics:
\begin{itemize}
    \item The precision (\textit{P}) can be computed as
    \begin{equation}
        P = \frac{\rm TP}{\rm TP + \rm FP},
        \label{eq:precision}
    \end{equation}
    
    which measures the level of purity of the retrieved catalog.
    
    \item The recall (\textit{R}) can be computed as
    \begin{equation}
        R = \frac{\rm TP}{\rm TP + \rm FN},
        \label{eq:recall}
    \end{equation}
    
    which measures the level of completeness of the retrieved catalog.
    
    \item The F1-score (F1) is the harmonic average of \textit{P} and \textit{R},
    \begin{equation}
        {\rm F1} = 2\;\frac{P\;R}{P + R}.
        \label{eq:fbeta_score}
    \end{equation}
    
    \item The accuracy (\textit{A}) is the ratio of the number of correctly classified objects and the total number of objects,
    \begin{equation}
        A = \rm \frac{TP + TN}{TP + TN + FP + FN}.
        \label{eq:accuracy}
    \end{equation}

\end{itemize}

The first three indicators can be similarly computed for the class of the Negatives, while the accuracy is a global indicator of the performance.

In addition, we computed the receiver operating characteristic (ROC; \citealt{hanley_roc}) curve, which visually represents the variation of the true-positive rate (TPR) and false-positive rate (FPR) with the detection threshold $t \in (0, 1)$, which was used to discriminate whether an image contains a lens. The area under the ROC curve (AUC) summarizes the information conveyed by the ROC: while 1.0 would be the score of a perfect classifier, 0.5 indicates that the classification is equivalent to a random choice and hence worthless.

\subsection{Experiment setup}\label{sec:experiments}
The identification of GGSL events is primarily based on their distinctive morphological characteristics, namely on the distortion of the images of the background source into arcs and rings, as well as on the color difference between the foreground and background galaxies. However, real lenses can show complex configurations and might not be so easily recognizable. Our experiments aimed at evaluating the ability of CNNs to detect the less clear lenses and at assessing their performance on a diversified data set.

We did this by training the three networks we presented on four selections of images, labeled S1 to S4, which gradually include a greater fraction of objects that present challenging visual identification, as we discuss below for nonlenses and lenses separately. These samples consist of approximately 2000, 10 000, 20 000, and 40 000 images, respectively. They were built to have an approximately equal number of lenses and nonlenses (see Table \ref{tab:data_selection}). The criteria we adopted to progressively broaden our selections took the features into account that might be employed by the networks to classify the objects as members of the correct category. 

In the case of the nonlenses, the lack of a background source, or the absence of its images, makes the classification more likely to be correct. Therefore, we initially considered a sample of the approximately 10 000 nonlenses without a background source. Specifically, we selected 1000 of them in S1, 5000 in S2, and 10 000 in S3.  In S4, we broadened our sample by including the images to which a background source was added, but that do not correspond to a visible image, extending our selection to the other objects that are classified as nonlenses according to the criteria in Eq.~(\ref{eq:lens_criteria}). 

\begin{figure}
    \centering
    \includegraphics[width = 0.48\textwidth]{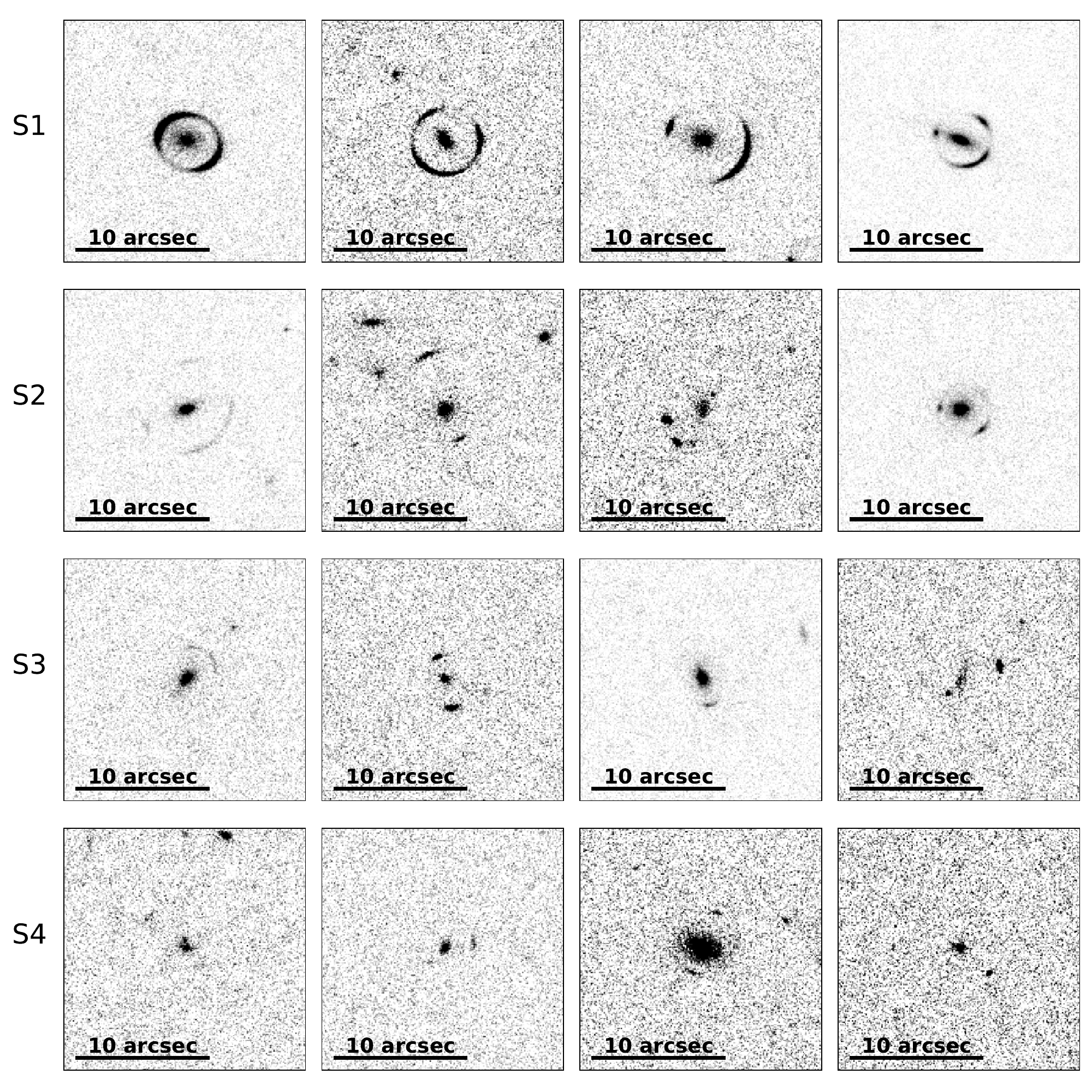}
    \caption{Examples of the kind of lenses included in all the selections used for training. From top to bottom row, we show four random lenses that were extracted from data sets S1, S2, S3, and S4, as simulated in the $\IE$ band.}
    \label{fig:lenses_per_selection}
\end{figure}

In the case of the lenses, the definition of an effective criterion to identify the clearest examples in the data set is more important and also more challenging. The mere presence of an image of the source does not guarantee a straightforward classification of the system because several factors contribute to the actual clarity of the observable features. They include the magnitude of the source and the extension of the image produced by the lensing effect. After several tests involving these parameters and others (e.g., the Einstein area and the magnification of the sources), we selected \texttt{n\_pix\_source} as an appropriate parameter to distinguish between clear and faint lenses. The complete sample of lenses is characterized by the minimum value $ \texttt{n\_pix\_source} > 20$.  From S4 to S1, we increased this threshold to different levels, which depended on the number of images we sought to isolate: the higher the value, the smaller the number of selected images and the clearer the lenses. The thresholds established for the creation of the selections described so far also take into account the necessity to have a comparable number of images of each class, so that the examples passed to the networks in the training phase are balanced. In Table~\ref{tab:data_selection} we summarize the criteria we used to identify the images to include in each selection. We also show in Fig. \ref{fig:lenses_per_selection} some randomly chosen examples of lenses that are characteristic of each selection to better illustrate which kind of selection we introduce by considering different thresholds for \texttt{n\_pix\_source} in the definition of the training sets. 

\begin{table*}[htpb]
\caption{Summary of the criteria we adopted to choose the images included in the different selections of lenses and nonlenses for our experiments. While the identification of the lenses is solely based on the variation of a threshold value for the parameter \texttt{n\_pix\_source}, the identification of the nonlenses is primarily based on the possible presence and visibility of a background source.}
\label{tab:data_selection}
\begin{center}
\begin{tabular}{cccccc}
\hline\hline\\ [-1.8ex]
\multirow{3}{*}{Selection} & \multicolumn{2}{c}{\textbf{Lenses}} & \multicolumn{2}{c}{\textbf{nonlenses}} & \multirow{3}{*}{Total} \\ \cline{2-5}\\ [-1.8ex]
 & \multirow{2}{*}{Criterion} & Number & \multirow{2}{*}{Criterion} & Number &  \\
 &  & \multicolumn{1}{l}{of images} &  & \multicolumn{1}{l}{of images} &  \\ \hline\\ [-1.8ex]
\multirow{3}{*}{S1} & \multirow{3}{*}{\texttt{n\_pix\_source} \textgreater 430} & \multirow{3}{*}{$\num{1001}$} & Randomly selected & \multirow{3}{*}{$\num{1000}$} & \multirow{3}{*}{$\num{2001}$} \\
 &  &  & objects with &  &  \\
 &  &  & \texttt{n\_sources} = 0 &  &  \\ \hline\\ [-1.8ex]\\ [-1.8ex]
\multirow{3}{*}{S2} & \multirow{3}{*}{\texttt{n\_pix\_source} \textgreater 140} & \multirow{3}{*}{$\num{5083}$} & Randomly selected & \multirow{3}{*}{$\num{5000}$} & \multirow{3}{*}{$\num{10083}$} \\
 &  &  & objects with &  &  \\
 &  &  & \texttt{n\_sources} = 0 &  &  \\ \hline\\ [-1.8ex]
\multirow{3}{*}{S3} & \multirow{3}{*}{\texttt{n\_pix\_source} \textgreater 70} & \multirow{3}{*}{$\num{9709}$} & Randomly selected & \multirow{3}{*}{$\num{10000}$} & \multirow{3}{*}{$\num{19709}$} \\
 &  &  & objects with &  &  \\
 &  &  & \texttt{n\_sources} = 0 &  &  \\ \hline\\ [-1.8ex]
\multirow{3}{*}{S4} & \multirow{3}{*}{\texttt{n\_pix\_source} \textgreater 20} & \multirow{3}{*}{$\num{19591}$} & Randomly selected & \multirow{3}{*}{$\num{20000}$} & \multirow{3}{*}{$\num{39591}$} \\
 &  &  & objects with &  &  \\
 &  &  & \texttt{n\_source\_im} = 0 &  &  \\ \hline
\end{tabular}
\end{center}
\end{table*}

We trained and tested on these selections of the data set the three architectures we discussed above: a VGG-like network \citep{vggnet}, an IncNet (\citealt{gnet_1, gnet_2}), and a ResNet (\citealt{resnet, resnext}). We conducted 24 training sessions in total because we trained each architecture on each selection of data. Twelve of them used the VIS images, and the other 12 used the NISP bands in addition to the VIS one. Every training was carried out for 100 epochs because the EarlyStopping method we had set up to prevent overfitting did not interrupt any of them. The best results of each architecture and each classification experiment, which were conducted using the $\IE$ band images, are summarized in Table \ref{tab:summary_challenge}, where the precision, recall, F1-score, accuracy, and AUC obtained from the application of our models are reported. An anologous summary for the training on the multiband images is provided in Table \ref{tab:summary_challenge_mw}.

\subsection{Discussion}
\begin{figure}
    \centering
    \includegraphics[width = 0.45\textwidth]{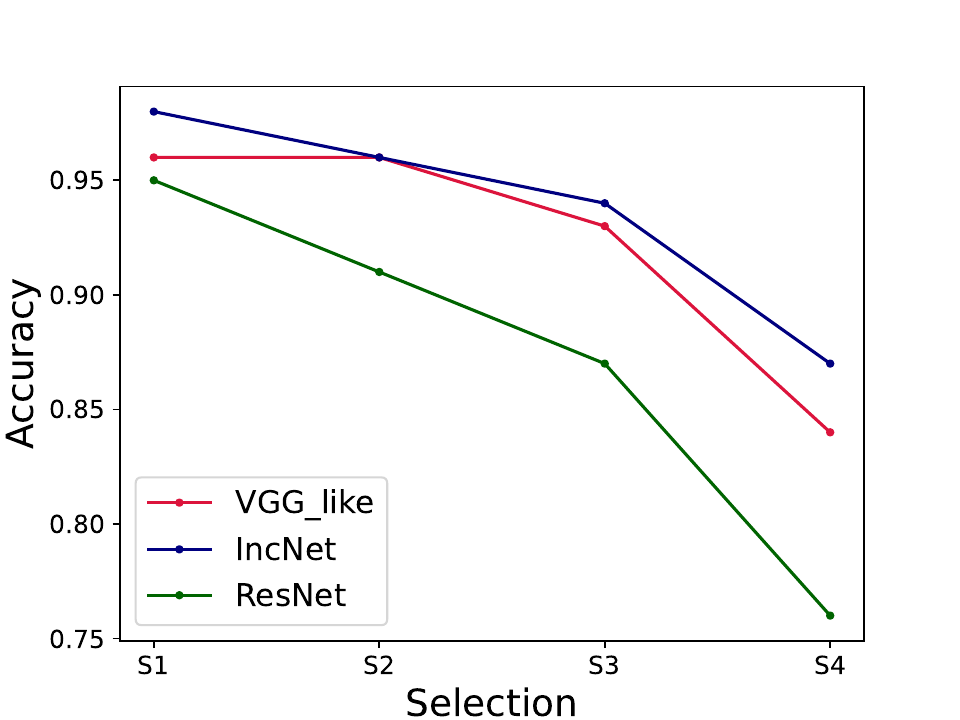}
    \caption{Trend of the classification accuracy of the single-branch versions of the VGG-like network (red), the IncNet (blue) and the ResNet (green) tested on the four data selections.}
    \label{fig:acc_challenge}
\end{figure}

By studying how the metrics depend on the selections, we find that the ability of our networks to correctly classify the images tends to deteriorate as the fraction of included lenses with a low signal-to-noise ratio increases. All the results described in the paper were found by considering a classification threshold of 0.5. The trend of the accuracy is shown in Fig.~\ref{fig:acc_challenge}. Our three models succeed in the classification of the objects in the selections S1 and S2, where the accuracy is in the range $\sim 0.9$ to $\sim 0.96$. The IncNet and VGG-like network also perform similarly on S3, while they reach an accuracy level of $\sim 0.87$ on S4. On the other hand, ResNet performs worst, with an accuracy of $\sim 0.75$ on the complete data set.

\begin{figure}
    \centering
    \includegraphics[width = 0.45 \textwidth]{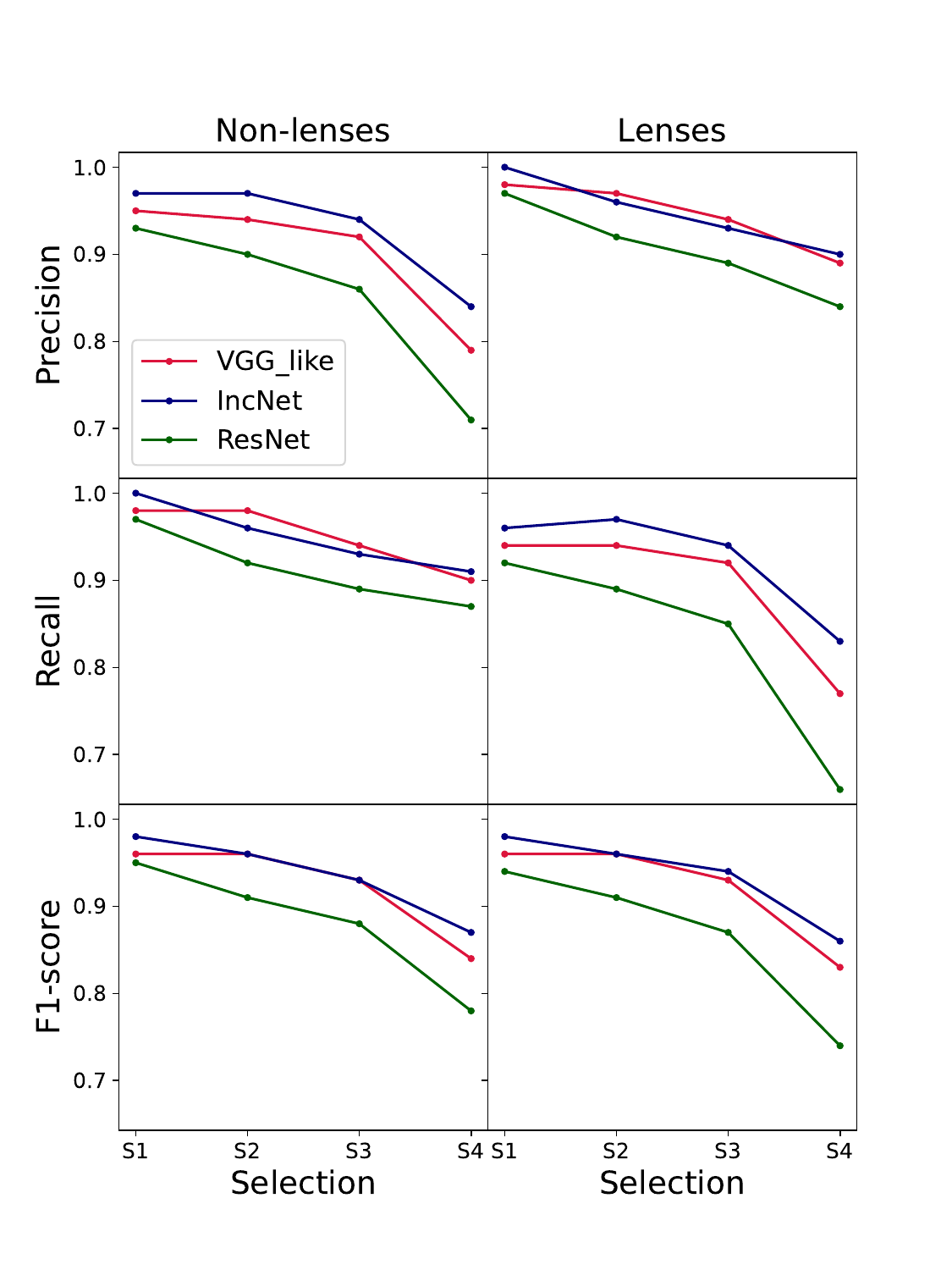}
    \caption{Trend of the precision (first row), recall (second row), and F1-score (third row) in the classification of the nonlenses (left column) and of the lenses (right column) in the different selections. Differently colored lines refer to different networks, as labeled, in the single-branch configuration.}
    \label{fig:metrics_challenge}
\end{figure}

The global trends of precision, recall, and F1-score are also similar to that of the accuracy. They are shown in the top, middle, and bottom panels of Fig. \ref{fig:metrics_challenge}, respectively. These metrics were evaluated separately on the nonlenses (left panels) and on the lenses (right panels), but the same consideration applies to both classes. This suggests that the degradation of the performance does not only affect the identification of the lenses, but affects the classification of the two categories. In particular, the F1-score, which depends on precision and completeness, peaks at $\sim 0.96$ on S1 and decreases to $\sim 0.87$ on S4, and ResNet is again the worst-performing network.


\begin{figure}
    \centering
    \includegraphics[width = 0.4 \textwidth]{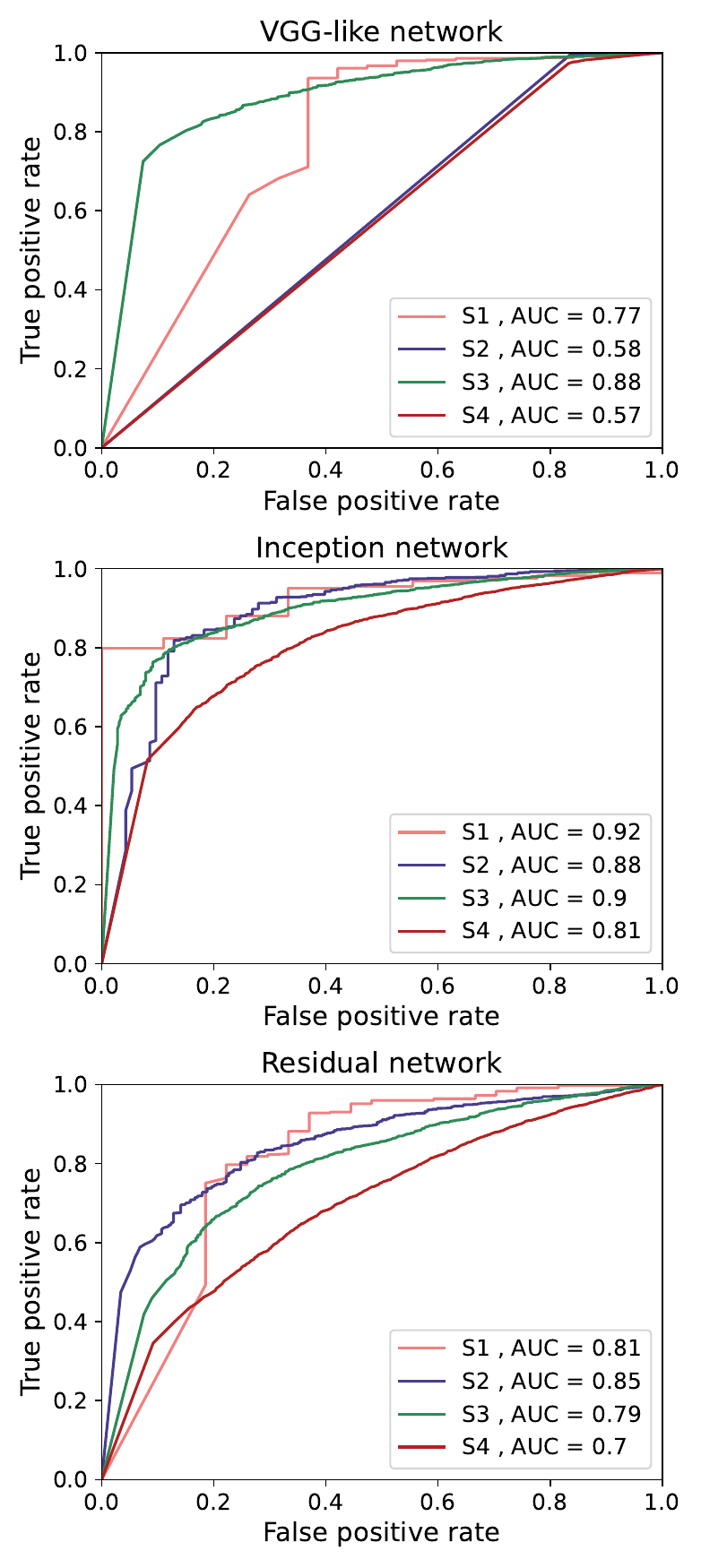}
    \caption{ROC curves as obtained from the tests of the single-branch versions of our architectures. From top to bottom, each panel of this image shows the ROC curves of the VGG-like network, the IncNet, and the ResNet to the test sets of the different selections S1 (pink line), S2 (blue line), S3 (green line) and S4 (red line) of the data set.}
    \label{fig:roc_challenge}
\end{figure}

In each panel of Fig. \ref{fig:roc_challenge}, we show the ROC curves of one of our networks, evaluated on the test sets of the selections S1, S2, S3, and S4. Their trends for the IncNet (middle panel) and the ResNet (bottom panel) are similar, and the AUC decreases by $\sim 10\%$ from S1 to S4. It should, however, be pointed out that IncNet performs systematically better than ResNet: while the AUC of the former is 0.92 on S1 and 0.81 on S4, the AUC of the latter ranges from 0.81 on S1 to 0.7 on S4. On the other hand, the ROC of the VGG-like network on S2 and S4 has a lower AUC, of $\sim 0.57$, compared to the other models, and higher AUC values only for the selections S1 and S3. After studying the predictions of this network on the different selections, we think that this is due to a significant difference in the number of objects that is predicted in the two classes when a high threshold is applied to the output probabilities. 

\begin{figure}
    \centering
    \includegraphics[width = 0.45 \textwidth]{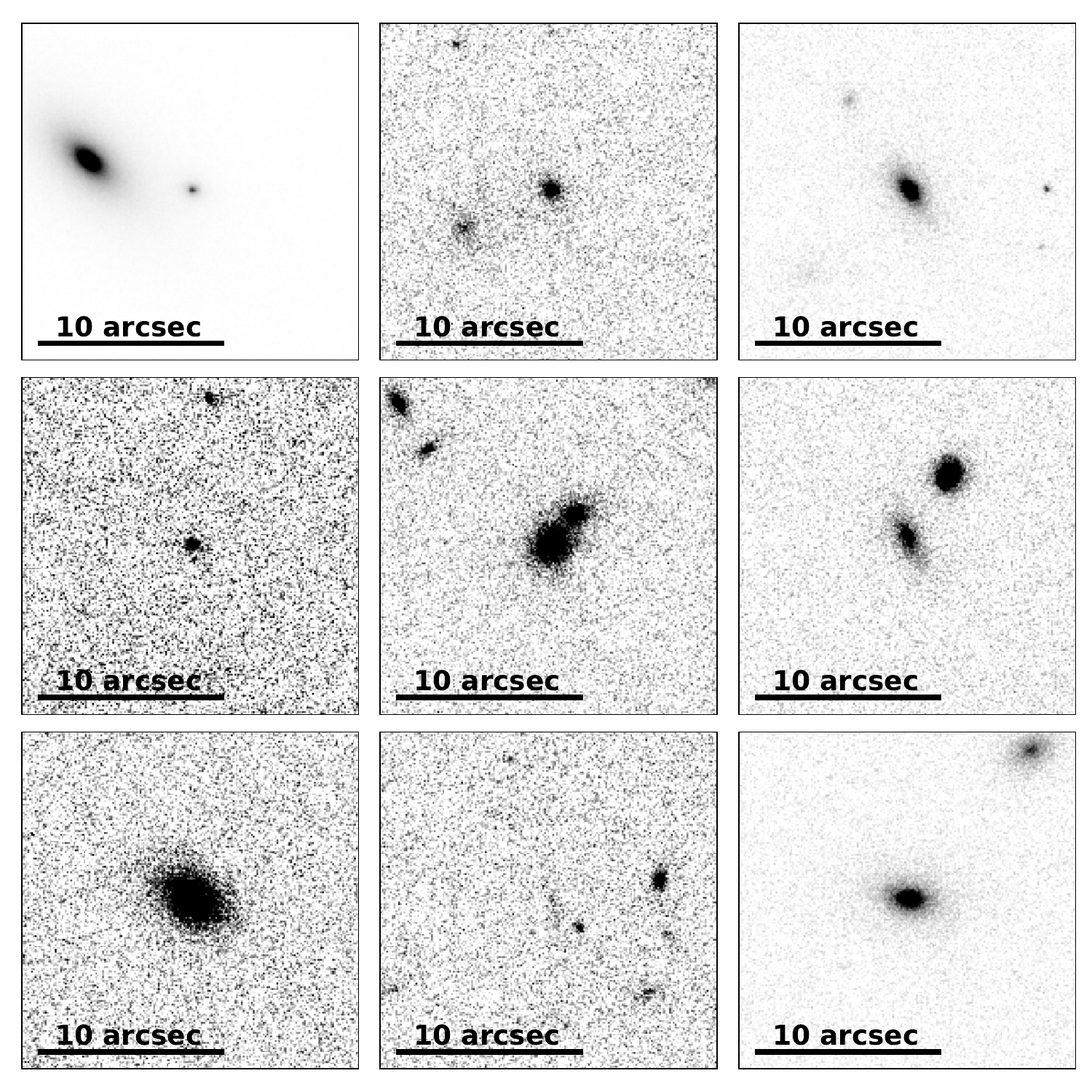}
    \caption{Example of false positives produced by the three networks in the single-branch configuration when applied to the selection S4, here pictured in the $\IE$ band.}
    \label{fig:wrong_non_lenses_challenge}
\end{figure}

\begin{figure}
    \centering
    \includegraphics[width = 0.45 \textwidth]{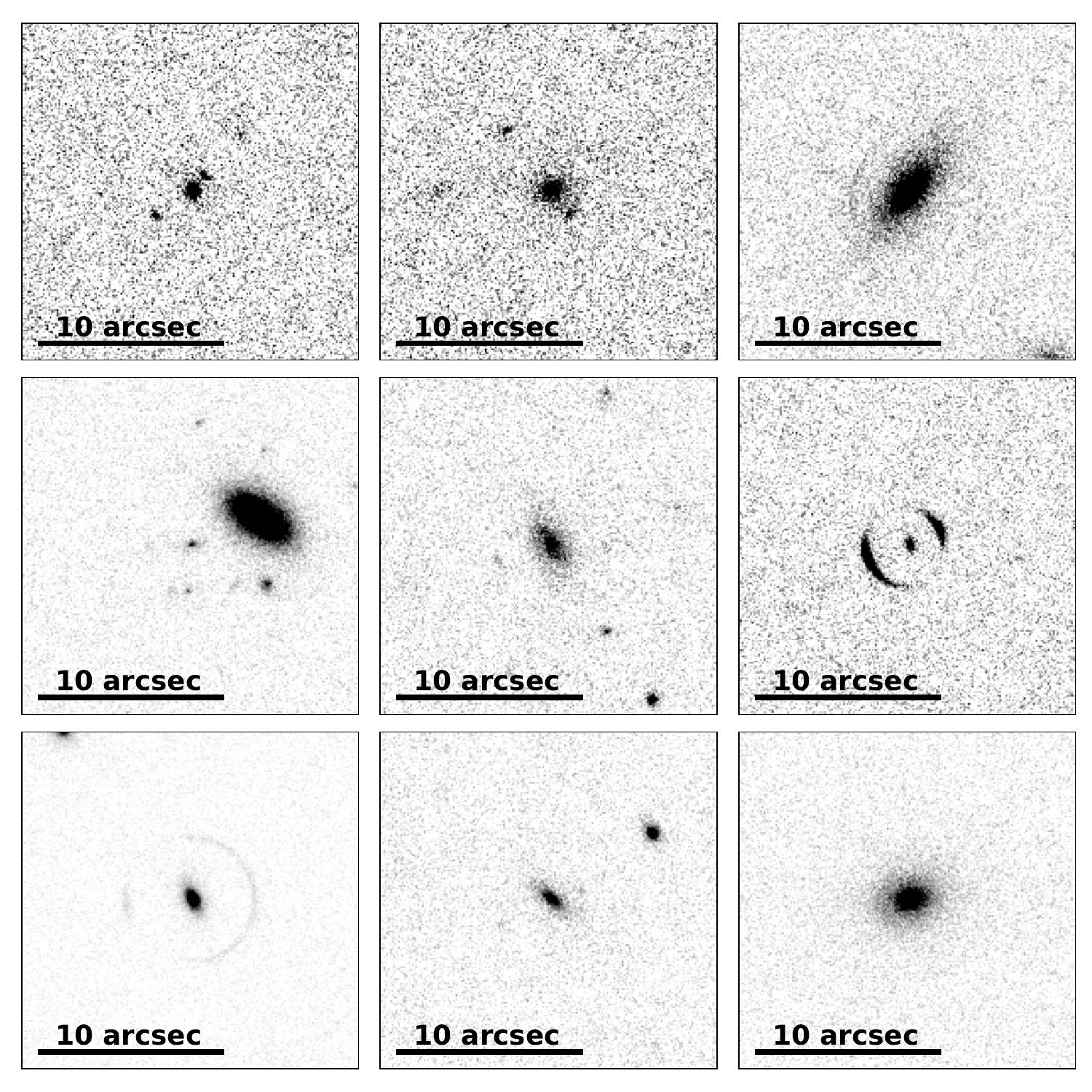}
    \caption{Example of false negatives produced by the three networks in the single-branch configuration when applied to the selection S4, here pictured in the $\IE$ band.}
    \label{fig:wrong_lenses_challenge}
\end{figure}

We focus on the selection S4, that is, on the performance of our models on the complete data set. Fig. \ref{fig:wrong_non_lenses_challenge} shows nine misclassified nonlenses, and Fig. \ref{fig:wrong_lenses_challenge} shows nine misclassified lenses. The images reported in these figures were selected from those that were misclassified by all three models, and therefore, they should be characterized by the features that the networks generally find harder to attribute to the correct class.

The false positives in Fig. \ref{fig:wrong_non_lenses_challenge} are mostly characterized by the coexistence of more than one source in addition to the lens galaxy, which might be mistaken for multiple images of the same source. The misinterpretation of these objects might be exacerbated by the inclusion of several low \texttt{n\_pix\_source} lenses in the training set. Many of the lenses in the labeled examples do not present clear arcs or rings, and the faint distortions encountered in the feature-extraction process are likely to resemble specific morphological features of nonlensed galaxies, such as spiral arms, or isolated, but elongated galaxies. One possible way to mitigate the misclassification of nonlenses with a background source could be to train the networks on multiband images to benefit from the color information. We investigate this possibility in Sec.~\ref{sec:mw_training}.

The false negatives in Fig. \ref{fig:wrong_lenses_challenge} are partly not even recognizable as lenses by visual inspection. Although they were classified as lenses according to the criteria in Eq.  (\ref{eq:lens_criteria}), many of these objects do not show evident lensing features. Therefore, if the classification were to be carried out on unlabeled observations, we would not expect the models to be able to identify them as lenses. An approach to solving the issue of nondetectable lenses might be to complement the use of the aforementioned criteria with the visual inspection of the images in the training set. In addition to this, we might include an additional criterion to ensure that the arc is detectable with respect to the other sources in the image. In this case, we would only accept systems as lenses in which the flux of the brightest pixel of the background source is greater than the flux of the other objects along the line of sight at the same pixel \citep[see][]{shu_2022, canameras_2023}. However, in some of the images, the arc-shaped and ring-shaped sources are evident. Nevertheless, their classification is incorrect, which signals that some clear lenses might also be missed by our classifiers.

\begin{figure}
    \centering
    \includegraphics[width = 0.45\textwidth]{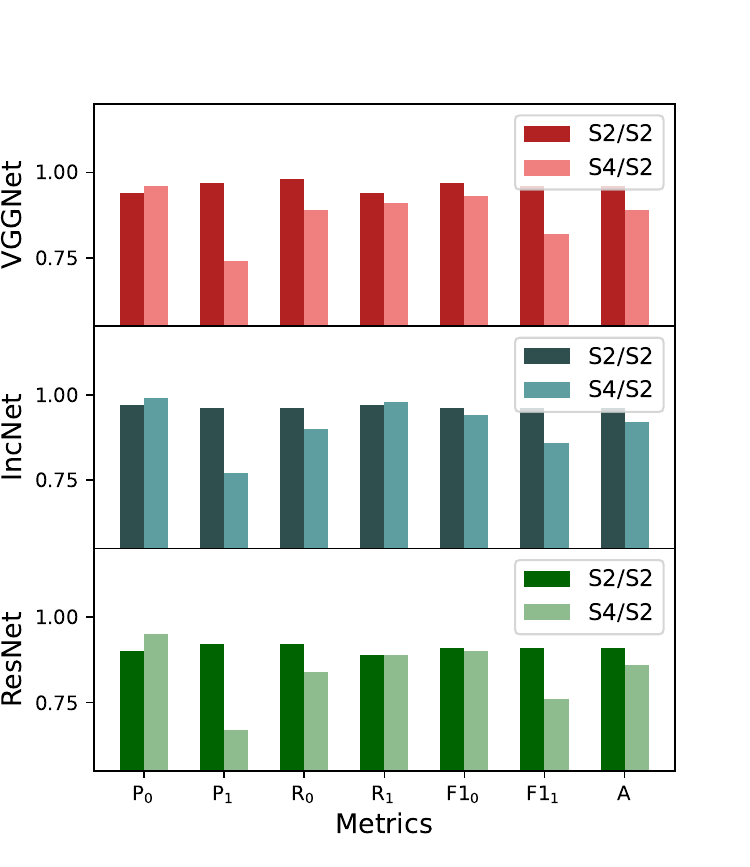}
    \caption{Comparison of the tests S2/S2 and S4/S2 (darker and lighter histograms) run with the VGG-like network (top), the IncNet (center), and the ResNet (bottom). In each panel, we show the results for the different metrics. From left to right, we show the precision on the class of the nonlenses ($P_0$) and lenses ($P_1$), the recall on the class of the nonlenses ($R_0$) and lenses ($R_1$), the F1-score on the class of the nonlenses (F1$_0$) and lenses (F1$_1$), and the overall accuracy ($A$).}
    \label{fig:S3vsS6metrics}
\end{figure}

In order to further investigate the ability of the networks trained on S4 to identify clear lenses, we tested them on the images in S2 (test S4/S2). The networks trained on S4 have analyzed during training and validation some of the images that are part of S2. We removed these images from our test set S4/S2, because otherwise the network performance would be biased to a better performance than can be achieved on unseen data. We compared the result of this test with results obtained from training and testing the networks on S2 (test S2/S2). The results of this comparison are shown in Fig.~\ref{fig:S3vsS6metrics}, and  more details can be found in Table~\ref{tab:S3vsS6metrics}.

The performance of the models trained on S4 in identifying the lenses in S2 is generally worse than that of the models trained on S2, even though the images that are part of S2 are also inevitably part of S4 because S4 consists of the complete data set. One reason for this is that the networks we used in the test S2/S2 were specifically trained to identify the lenses in S2, while the networks trained on the larger data set S4 were exposed to a larger variety of systems and are not as specialized on the S2 lenses. We examine the results in Table \ref{tab:S3vsS6metrics}, however. While the completeness of the retrieved catalog of lenses is constant in the two tests, the precision decreases by $\sim 20\%$, passing from $\sim 0.95$ in the test S2/S2 to $\sim 0.73$ in S4/S2, with only minor differences between the different architectures. Even though the magnitude of the overall deterioration is not large per se (the accuracy decreases by $\sim 5\%$ for the three networks), this is problematic because it is also due to the misclassification of clear lenses, which are also the most useful for scientific purposes.

This result suggests that the performance of the models trained on S4 is worse in general because a significant fraction of this selection is composed of nonobvious lenses that are intrinsically harder to classify. Moreover, the ability of the models to recognize the clearest GGSL events in the data set that are also present in S2 deteriorates.

This effect might result from a combination of two complementary factors regarding the characteristics of the images in the data set. First, the fraction of clear images in the training set of S4 is smaller than in the other selections because of the relevant fraction of low \texttt{n\_pix\_source} lenses included. This is reflected in the fact that the networks might not learn how to properly distinguish them. Wide arcs and rings are recognizable only in a moderate number of images, and they are therefore not as significant as they are in S2 for the classification of the lenses. Second, the most frequently recurring features in the training set are those that occur in images with a low signal-to-noise ratio, and they thus contribute to explaining the misinterpretation of some of the images that present evident lensing features.

As shown in Fig.~ \ref{fig:wrong_lenses_challenge}, a large fraction of the lenses that were classified as nonlenses by the networks trained on S4 do not present clear lensing features. However, a non-negligible fraction of evident lenses might also be missed if the training set were extended to include a significant number of fainter arcs because the evident systems might become under-represented. In addition to this, the architecture of the network appears to be influential in the outcome of the classification only to a certain degree. In particular, when trained and tested on the same selections, the IncNet and VGG-like networks generally perform similarly when the metrics in Figs.~\ref{fig:acc_challenge} and \ref{fig:metrics_challenge} are compared. The ResNet, on the other hand, performs significantly worse than the others, especially on S4. 

\subsection{Additional tests}\label{sec:tests}

We tested the models trained on S2 on the wider selections S3 and S4 (tests S2/S3 and S2/S4, respectively) after removing the parts of these samples that were also included in the training set of S2. This test had the purpose of assessing whether the networks trained on clear examples are flexible enough to detect fainter systems. A lower performance from S2/S3 to S2/S4 was also expected because CNNs mostly generalize to the images that are similar to those in the data set they were trained with. Consequently, they might perform the same task poorly for images that are characterized by features they never saw before. In the present case, most images in the training set of S2 show clear lensing features, while the test sets progressively include a greater fraction of images with new features.

The general performance of the networks trained on S2 deteriorates on the other broader selections. The accuracy of the classification varies from $\sim 0.85$ in the case S2/S3 to $\sim 0.7$ in the case S2/S4. By comparing these results with those of the test S4/S4 in Figs. \ref{fig:acc_challenge} and \ref{fig:metrics_challenge}, we observe several differences in the precision, recall, and F1-score, computed separately for the nonlenses and lenses, as well as in the accuracy.  We report the results of these tests in Table \ref{tab:s3_test}.

The purity of the nonlenses decreases when broader selections are used as test sets. The precision reaches $\sim 0.64$ with S4. On the other hand, the recall is approximately constant at values of $\sim 0.96$ independently of the considered selection, meaning that the largest fraction of the objects in this class is correctly identified. In the case of the lenses, the trend is roughly reversed. The precision of the classification is roughly constant at $\sim 0.94$, while the recall decreases drastically from $\sim 0.7$ in S3 to $\sim 0.38$ in S4. These values suggest that the networks trained on the S2 sample cannot recognize a large fraction of the lenses in the complete data set.

These trends can be interpreted by considering the impact of including the fainter features in the test sets. In particular, the training set of S2 mostly includes clear lenses and images of isolated nonlenses that are not surrounded by other sources. When processing the images in S3 and S4, the absence of clear arcs and rings, and more generally the faintness of the lensing features induce a growing fraction of lenses to be classified as nonlenses. Our results highlight the inability of our models to recover a considerable fraction of lenses that are not similar to those in S2, leading to a decrease of more than $\sim20\%$ in the recall of the lenses from S2/S2 to S2/S3 and of $\sim 30\% $ from S2/S3 to S2/S4 (see Table \ref{tab:s3_test} for more details). 

\subsection{The impact of the shapelet decomposition} \label{sec:cosmos_lenses}
In the simulation of the images in our data set, we used the galaxies observed in the UDF as background sources. For the purpose of denoising them, we decomposed the galaxies with a shapelet-based approach. The shapelet technique is a very powerful mathematical tool for describing astrophysical objects, and its limitations have been investigated in some works \citep[see e.g.,][]{melchior_2007, melchior_2010}. In this section, we investigate the impact of these limitations on the performance of our networks.

We assessed this by testing our networks on a sample of 134 real lenses mainly found in the Sloan Lens ACS Survey \citep[SLACS;][]{bolton_slacs} and in the BOSS Emission-Line Lens Survey \citep[BELLS;][]{bells_survey} and on 300 nonlensed galaxies of the UDF. The purpose of this test was not to evaluate the performance of our networks on a realistic sample, which would require including a larger number of nonlenses in the test set. We wished to estimate whether the shapelet decomposition prevents the networks from being applied to real observations. The failure of the networks to identify the observed lenses as lenses would indicate that the simulations are not descriptive enough for the characteristics of real galaxies.

We used the networks trained on S2 to carry out this test. We preprocessed all the images by normalizing them with a procedure similar to the one we applied to the simulations as described in Sec. \ref{sec:data_prep}. In the case of the galaxies of the UDF, we also reshaped the images to the size expected by the networks. 

The results of this test are that we recovered 129 of the lenses with the IncNet and 126 lenses with the VGG-like network and with the ResNet. In the case of the nonlensed UDF galaxies, all the three networks correctly classified 296 of them. Based on these recovery rates, the shapelet decomposition does not introduce significant limitations in our simulations.

\subsection{Training with multiband images}\label{sec:mw_training}
The correct identification of GGSL events may benefit significantly from color information emerging from the analysis of multiband data. Lenses and sources typically have different colors because their spectral energy distributions (and redshifts) are different. For example, the most common sources are star-forming galaxies that appear bluer than the lenses, which in contrast are often early-type passive galaxies. Moreover, the color similarity of multiple images of the same source can be leveraged to identify strongly lensed sources. This is particularly useful in systems that do not present evident morphological distortions. 

For example, \cite{gentile_2021} reported that training CNNs on multiband images resulted in an improved classification of systems with small Einstein radii, while training on single-band images was more efficient for finding lenses with large radii. \cite{metcalf_challenge} also found that using multiband images for the training substantially improved the performance of the classifiers for mock ground-based data, even though the color information came from observations with poorer spatial resolution. 

\begin{figure}
    \centering
    \includegraphics[width = 0.45 \textwidth]{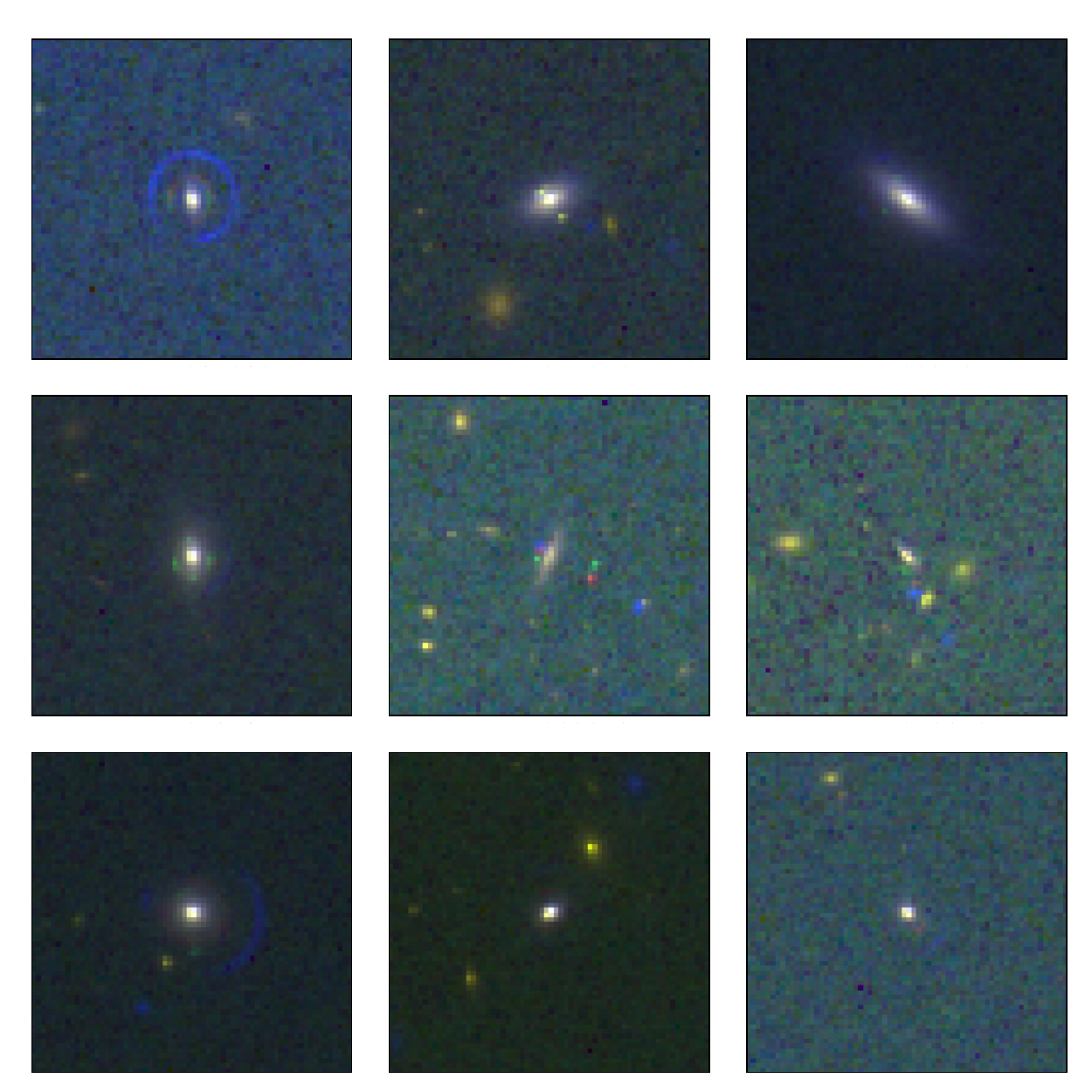}
    \caption{Example of randomly chosen lenses in the configuration used for multiband training. For visualization pruposes, the images simulated in the $\IE$ band were downgraded to the resolution of the NISP bands in these examples.}
    \label{fig:rgb_images}
\end{figure}

We evaluated the importance of color information for the identification of the low \texttt{n\_pix\_source} lenses in \Euclid-like data by repeating the same training as before, but this time included the NIR images that are also available from the simulations. We show in Fig. \ref{fig:rgb_images} some randomly chosen examples of lenses obtained by combining the VIS and NIR bands. We changed the architecture of our models to take the different sizes of the VIS and NISP images into account, as explained in Sect. \ref{sec:cnn} and represented in panels (b) of Figs. \ref{fig:my_vgg}, \ref{fig:my_incnet} and \ref{fig:my_resnext}, but otherwise, we used the same setup as in our previous experiments. We report the results of these tests in Table \ref{tab:summary_challenge_mw}.

By comparing these values to those of the VIS training (see Table \ref{tab:summary_challenge}), we do not observe a significant improvement in the model performances for a training with multiband data. This is expected for the smaller selections, which are limited to the clearest lenses, whose correct identification through their morphology is relatively easy. In these cases, the color information is therefore expected to be less relevant. However, for broader selections, in which the morphology of the lenses is less clear, we might expect to see some improvement in the classification performance when the models are fed with color information. Surprisingly, we do not note any significant variations in the metrics that quantify the model performance. 

We interpret this result as follows. First, the wavelength range covered by the VIS instrument (see Table \ref{tab:euclid_prop}) does not include the wavelengths at which the color difference between the background and foreground galaxies is particularly evident, that is, the blue wavelengths of the optical spectrum. Second, the images in the NIR bands are characterized by lower resolution than those in the $\IE$ band (also see Table \ref{tab:euclid_prop}), which means that the morphological information is degraded in these channels. This also suggests that morphological information is more important than color for identifying lenses, at least in this wavelength range.

\subsection{Finding lenses in unbalanced data sets}\label{sec:real_prop_test}
As we discussed in Sec. \ref{sec:data_set}, training on a balanced training set is important for the networks to learn how to assign the images to the correct class, but a balanced test set is not a requirement. While in all the previous tests we used a balanced test set, with a ratio of about 1:1 between lenses and nonlenses, this is very different from reality, where we reasonably expect to observe less than one lens for 1000 nonlenses \citep{marshall2009}. In this scenario, even very efficient classifiers will produce a large number of false positives \citep{savary_cnn, jacobs_cnn2, jacobs_cnn}, and the visual inspection of thousands of candidates is required to find definite samples of strong lenses. While training on simulations instead of real observations plays a role in this because it is possible that the images present irregular features or shapes that were not included in the training, the high imbalance between the two populations is a major factor to consider.

For this reason, we ran an additional test with realistic proportions in the number of images of the two subsamples. We focused on the networks trained on S1, which globally have the best performances (Figs. \ref{fig:acc_challenge} and \ref{fig:metrics_challenge}). We applied the networks trained on this selection on a test set that has the same lenses as in the original test set of S1, that is, 240 lenses, and used the $\sim \num{80000}$ nonlenses that were excluded from the training (as discussed in Sect. \ref{sec:data_set}). While most of the metrics have similar values to those we found in the test with balanced classes, the precision drops to $\sim 0.15$ for the VGG-like network, to $\sim 0.45$ for the IncNet and to $\sim 0.13$ for the ResNet. This is expected and due to the larger number of false positives predicted by the networks. To reduce the occurrence of false positives, we combined the results of the three networks by averaging their predictions, as this has shown to benefit the rate of correct predictions \citep[e.g.,][]{andika_2023}. We find that the ensemble prediction indeed has a higher precision (with a precision of $\sim 0.46$) than those of the VGG-like network and of the ResNet, while it is comparable to that of the IncNet. More details for this test are given in Table \ref{tab:test_trueprop}.

Even though it is difficult to design a method that will produce a highly pure and complete sample of strong lenses, different strategies are possible to mitigate the issue of many false positives. A common way to reduce their number is to use a high threshold for the classification of the lenses \citep{petrillo_cnn2, gentile_2021} and perform a visual inspection of the candidates that are most likely to be lenses to further refine the selection. The drawback of this method is that the completeness of the sample decreases because the systems that are classified with a lower probability are missed. Another possible strategy is increasing the number of images with misleading features in the negative class of the training set \citep{canameras_2020}. This should make the networks more familiar with these objects and thus more efficient in recognizing them when applied to real data. Moreover, methods such as transfer learning and domain adaptation might improve the classification performance with real data \citep{dominguez_tl, domainadaptation}. These techniques would require retraining networks that were trained on simulations on a small sample (a few hundred) of observed lenses and might lead to a significant improvement of the network performances.

\subsection{Finding lenses in Euclid}\label{sec:finding_euclid}
Future \Euclid observations will offer the opportunity to increase the number of known GGSL events by orders of magnitude as long as potential candidates
are efficiently identified. The optimization of the lens-finding strategy, especially in the first year after the launch, is also essential for efficient follow-up observations. For example, the 4-meter Multi-Object Spectroscopic Telescope \citep[4MOST;][]{4mostsurvey} Strong Lens Spectroscopic Legacy Survey\footnote{\url{https://www.4most.eu/cms/science/extragalactic-community-surveys/}} will observe about $\num{10000}$ lens candidates observed by \Euclid and LSST, providing spectroscopic redshifts for them.

The strategy currently planned for finding lenses in the survey relies both on fully simulated images and data-driven simulations. Training CNNs on simulated images is inevitable in the initial phase of the \Euclid observations because only so few galaxy-galaxy lenses are known at the moment. As the data accumulate, more sophisticated simulations will be made, in which the lenses are real galaxies observed by \Euclid. The networks will be retrained with images that include realistic properties of both lenses and sources, thus improving the performance of the classifiers in the next step of the data analysis. The addition of information about photometric redshifts of the sources might also yield some improvement, but this comes with the challenge of measuring them with good accuracy. A large enough separation between the lens galaxy and the source or efficient deblending techniques are decisive in this context.

The greatest advantage of searching for lenses with \Euclid is that it will resolve faint Einstein rings with small radii ($\sim 0.5 \arcsecond$), mostly lensed by bulges of spiral galaxies, in addition to lenses on a larger angular scale. These systems are usually unresolved by ground-based facilities, but will be found with the high resolution of \Euclid. Moreover, they will be most common according to forecasts \citep{collett_sl}. \Euclid observations could also be combined with and complemented by those of other surveys. The LSST, for instance, will observe a comparable number of lenses that will likely be skewed to larger radii because of the lower resolution of ground-based observations. A complementary data set of lenses in the radio band with high resolution will be produced by Square Kilometer Array \citep{ska}. They are complementary to the others because the parent population of the systems observed in radio is different from that of the systems observed in optical and infrared bands \citep{koopmans_2004}.

The fully simulated data sets are also critical for studying the selection functions of the algorithms that will be used for finding lenses in the survey. An accurate characterization of the selection function is necessary for the scientific exploitation of the GGSLs found by \Euclid. For example, \cite{sonnenfeld_selectionf1} discussed the importance of characterizing the selection function for inferring the properties of the population of galaxies of which the strong lenses are a biased subsample. Moreover, they showed how the information about the number of nondetections can be used to further constrain models of galaxy structure. More recently, \cite{sonnenfeld_selectionf2} investigated the difference between lens galaxies and lensed sources from their parent population, that is, the strong-lensing bias. Because \Euclid will provide the largest sample of homogeneously discovered strong lenses ever gathered, this type of study will be more significant than in the past.

\section{Conclusions}\label{sec:conclusion} 
In this work, we have presented a detailed analysis of the performance of three CNN architectures in identifying GGSL events. We used a data set of 40 000 images simulated by the Bologna Lens Factory to mimic the data quality expected from the \Euclid space mission. The classification was primarily based on the morphology of the systems because we mainly conducted our experiments with the images simulated in the $\IE$ band. Still, we evaluated the importance of color information using multiband images. We trained and tested our CNNs on four data-set selections that gradually included a greater fraction of objects characterized by faint lensing features and that will be more difficult to recognize. We evaluated the outcome of the classification by estimating the precision, recall, and F1 score of the lens catalogs we obtained.

We found that the morphological characteristics of the lenses included in the training set influence the ability of our CNNs to identify the lenses in a separate test set in a critical way, whether they show clear or faint lensing features. We found that the inclusion of a large fraction of images deteriorates the performance of our models, causing a decrease in the overall accuracy of $\sim 10\%$, from $\sim 0.95$ to $\sim 0.85$ for the IncNet and VGG-like network, and an even greater decrease for the ResNet, which reaches an accuracy of $\sim 0.74$. 
Moreover, we also found that it impacts the ability of our models to identify the most evident lenses because they become under-represented in the training set. 

These results emphasize the importance of building realistic training sets for DL models. This is particularly relevant for the first searches because we will not have real lensing systems at our disposal, and the simulations of large data sets will be the only option for training. In this phase, the inclusion of the real galaxies observed by \Euclid in the simulation will make the mocks more realistic than those used so far to train the networks. In particular, they suggest that identifying lenses with different morphologies might require specific training focused on the type of lenses of interest for a certain purpose. Alternatively, the classification of the lenses might be considered and solved a multiclass classification problem, distinguishing the clear and probable lenses from the probable and evident nonlenses. In this last case, however, the distinction between obvious and nonobvious objects should be further investigated and quantified.

We also retrained our models on the same selections of the data set, including a separate channel for processing the near-IR images in addition to those in the $\IE$ band, thus assessing how relevant the color information is for identifying lenses with a low signal-to-noise ratio. We found no significant improvement in the performance of any of our networks. We suggest that this might depend on a combination of two factors. First, the images in the $\IE$ band have a higher resolution than those in the near-IR bands. Second, the $\IE$ band covers a wavelength range in which the color difference between lens and source galaxies might not be important (see Table \ref{tab:euclid_prop}).

Finally, we highlight that the three architectures retrieve catalogs with similar characteristics in terms of completeness and precision when they are applied to the same selections of images. The only exception is ResNet, whose accuracy on the full data set is lower by $\sim10\%$ than the others. Because of the higher precision of IncNet in the test with an unbalanced number of images, we would conclude that this is the best-performing network of those we tested. The results of this test are indeed the closest to what we might expect from real data, and they are therefore particularly relevant to evaluate the performance of our models.

In the future, we could improve our selection method by testing a combination of physical parameters to differentiate between faint and clear lenses instead of using \texttt{n\_pix\_source}, which we have as a result of our simulations, but is not a physical property of the galaxies. It would also be useful to study whether there is a bias in the properties of the lenses found by our models to characterize the type of systems better that are most likely to be found or missed.

\begin{acknowledgements}
\AckECon We acknowledge support from the grants PRIN-MIUR 2017 WSCC32, PRIN-MIUR 2020 SKSTHZ and ASI n.2018-23-HH.0. MM was supported by INAF Grant ``The Big-Data era of cluster lensing". This work has made use of CosmoHub.  CosmoHub has been developed by the Port d'Informaci\'{o} Científica (PIC), maintained through a collaboration of the Institut de F\'{i}sica d'Altes Energies (IFAE) and the Centro de Investigaciones Energ\'{e}ticas, Medioambientales y Tecnológicas (CIEMAT) and the Institute of Space Sciences (CSIC \& IEEC), and was partially funded by the "Plan Estatal de Investigaci\'{o}n Científica y T\'{e}cnica y de Innovación" program of the Spanish government.
\end{acknowledgements}

\bibliography{z_biblio, Euclid}

\begin{thebibliography}{121}
\expandafter\ifx\csname natexlab\endcsname\relax\def\natexlab#1{#1}\fi

\bibitem[{Abadi {et~al.}(2016)Abadi, Barham, Chen, Chen, Davis, Dean, Devin,
  Ghemawat, Irving, Isard, Kudlur, Levenberg, Monga, Moore, Murray, Steiner,
  Tucker, Vasudevan, Warden, Wicke, Yu, \& Zheng}]{abaldi_tf}
Abadi, M., Barham, P., Chen, J., {et~al.} 2016, 12th {USENIX} Symposium on
  Operating Systems Design and Implementation ({OSDI} 16), 265

\bibitem[{{Aihara} {et~al.}(2018){Aihara}, {Arimoto}, {Armstrong}, {Arnouts},
  {Bahcall}, {Bickerton}, {Bosch}, {Bundy}, {Capak}, {Chan}, {Chiba}, {Coupon},
  {Egami}, {Enoki}, {Finet}, {Fujimori}, {Fujimoto}, {Furusawa}, {Furusawa},
  {Goto}, {Goulding}, {Greco}, {Greene}, {Gunn}, {Hamana}, {Harikane},
  {Hashimoto}, {Hattori}, {Hayashi}, {Hayashi}, {He{\l}miniak}, {Higuchi},
  {Hikage}, {Ho}, {Hsieh}, {Huang}, {Huang}, {Ikeda}, {Imanishi}, {Inoue},
  {Iwasawa}, {Iwata}, {Jaelani}, {Jian}, {Kamata}, {Karoji}, {Kashikawa},
  {Katayama}, {Kawanomoto}, {Kayo}, {Koda}, {Koike}, {Kojima}, {Komiyama},
  {Konno}, {Koshida}, {Koyama}, {Kusakabe}, {Leauthaud}, {Lee}, {Lin}, {Lin},
  {Lupton}, {Mandelbaum}, {Matsuoka}, {Medezinski}, {Mineo}, {Miyama},
  {Miyatake}, {Miyazaki}, {Momose}, {More}, {More}, {Moritani}, {Moriya},
  {Morokuma}, {Mukae}, {Murata}, {Murayama}, {Nagao}, {Nakata}, {Niida},
  {Niikura}, {Nishizawa}, {Obuchi}, {Oguri}, {Oishi}, {Okabe}, {Okamoto},
  {Okura}, {Ono}, {Onodera}, {Onoue}, {Osato}, {Ouchi}, {Price}, {Pyo}, {Sako},
  {Sawicki}, {Shibuya}, {Shimasaku}, {Shimono}, {Shirasaki}, {Silverman},
  {Simet}, {Speagle}, {Spergel}, {Strauss}, {Sugahara}, {Sugiyama}, {Suto},
  {Suyu}, {Suzuki}, {Tait}, {Takada}, {Takata}, {Tamura}, {Tanaka}, {Tanaka},
  {Tanaka}, {Tanaka}, {Terai}, {Terashima}, {Toba}, {Tominaga}, {Toshikawa},
  {Turner}, {Uchida}, {Uchiyama}, {Umetsu}, {Uraguchi}, {Urata}, {Usuda},
  {Utsumi}, {Wang}, {Wang}, {Wong}, {Yabe}, {Yamada}, {Yamanoi}, {Yasuda},
  {Yeh}, {Yonehara}, \& {Yuma}}]{aihara_2018}
{Aihara}, H., {Arimoto}, N., {Armstrong}, R., {et~al.} 2018, \pasj, 70, S4

\bibitem[{{Allison} {et~al.}(2017){Allison}, {Moss}, {Macquart}, {Curran},
  {Duchesne}, {Mahony}, {Sadler}, {Whiting}, {Bannister}, {Chippendale},
  {Edwards}, {Harvey-Smith}, {Heywood}, {Indermuehle}, {Lenc}, {Marvil},
  {McConnell}, \& {Sault}}]{allison_mag}
{Allison}, J.~R., {Moss}, V.~A., {Macquart}, J.~P., {et~al.} 2017, \mnras, 465,
  4450

\bibitem[{{Angora} {et~al.}(2020){Angora}, {Rosati}, {Brescia}, {Mercurio},
  {Grillo}, {Caminha}, {Meneghetti}, {Nonino}, {Vanzella}, {Bergamini},
  {Biviano}, \& {Lombardi}}]{angora_gc}
{Angora}, G., {Rosati}, P., {Brescia}, M., {et~al.} 2020, \aap, 643, A177

\bibitem[{{Angora} {et~al.}(2023){Angora}, {Rosati}, {Meneghetti}, {Brescia},
  {Mercurio}, {Grillo}, {Bergamini}, {Acebron}, {Caminha}, {Nonino},
  {Tortorelli}, {Bazzanini}, \& {Vanzella}}]{angora_ggsl}
{Angora}, G., {Rosati}, P., {Meneghetti}, M., {et~al.} 2023, arXiv:2303.00769

\bibitem[{{Barnab{\`e}} {et~al.}(2011){Barnab{\`e}}, {Czoske}, {Koopmans},
  {Treu}, \& {Bolton}}]{barnabe_dis}
{Barnab{\`e}}, M., {Czoske}, O., {Koopmans}, L. V.~E., {Treu}, T., \& {Bolton},
  A.~S. 2011, \mnras, 415, 2215

\bibitem[{Bengio(2009)}]{bengio}
Bengio, Y. 2009, {Foundation and Trends in Machine Learning}, vol. 2, 1

\bibitem[{{Bergamini} {et~al.}(2021){Bergamini}, {Rosati}, {Vanzella},
  {Caminha}, {Grillo}, {Mercurio}, {Meneghetti}, {Angora}, {Calura}, {Nonino},
  \& {Tozzi}}]{bergamini_2021}
{Bergamini}, P., {Rosati}, P., {Vanzella}, E., {et~al.} 2021, \aap, 645, A140

\bibitem[{{Bolton} {et~al.}(2006){Bolton}, {Burles}, {Koopmans}, {Treu}, \&
  {Moustakas}}]{bolton_slacs}
{Bolton}, A.~S., {Burles}, S., {Koopmans}, L. V.~E., {Treu}, T., \&
  {Moustakas}, L.~A. 2006, \apj, 638, 703

\bibitem[{{Brownstein} {et~al.}(2012){Brownstein}, {Bolton}, {Schlegel},
  {Eisenstein}, {Kochanek}, {Connolly}, {Maraston}, {Pandey}, {Seitz}, {Wake},
  {Wood-Vasey}, {Brinkmann}, {Schneider}, \& {Weaver}}]{bells_survey}
{Brownstein}, J.~R., {Bolton}, A.~S., {Schlegel}, D.~J., {et~al.} 2012, \apj,
  744, 41

\bibitem[{Buda {et~al.}(2018)Buda, Maki, \& Mazurowski}]{buda2018systematic}
Buda, M., Maki, A., \& Mazurowski, M.~A. 2018, Neural networks, 106, 249

\bibitem[{{Ca{\~n}ameras} {et~al.}(2021){Ca{\~n}ameras}, {Schuldt}, {Shu},
  {Suyu}, {Taubenberger}, {Meinhardt}, {Leal-Taix{\'e}}, {Chao}, {Inoue},
  {Jaelani}, \& {More}}]{canameras_2021}
{Ca{\~n}ameras}, R., {Schuldt}, S., {Shu}, Y., {et~al.} 2021, \aap, 653, L6

\bibitem[{{Ca{\~n}ameras} {et~al.}(2020){Ca{\~n}ameras}, {Schuldt}, {Suyu},
  {Taubenberger}, {Meinhardt}, {Leal-Taix{\'e}}, {Lemon}, {Rojas}, \&
  {Savary}}]{canameras_2020}
{Ca{\~n}ameras}, R., {Schuldt}, S., {Suyu}, S.~H., {et~al.} 2020, \aap, 644,
  A163

\bibitem[{{Cabanac} {et~al.}(2007){Cabanac}, {Alard}, {Dantel-Fort}, {Fort},
  {Gavazzi}, {Gomez}, {Kneib}, {Le F{\`e}vre}, {Mellier}, {Pello}, {Soucail},
  {Sygnet}, \& {Valls-Gabaud}}]{2007A&A...461..813C}
{Cabanac}, R.~A., {Alard}, C., {Dantel-Fort}, M., {et~al.} 2007, \aap, 461, 813

\bibitem[{{Carretero} {et~al.}(2017){Carretero}, {Tallada}, {Casals}, {Caubet},
  {Castander}, {Blot}, {Alarc{\'o}n}, {Serrano}, {Fosalba}, {Acosta-Silva},
  {Tonello}, {Torradeflot}, {Eriksen}, {Neissner}, \&
  {Delfino}}]{2017ehep.confE.488C}
{Carretero}, J., {Tallada}, P., {Casals}, J., {et~al.} 2017, in Proceedings of
  the European Physical Society Conference on High Energy Physics. 5-12 July,
  488

\bibitem[{{Cañameras} {et~al.}(2023){Cañameras}, {Schuldt}, {Shu}, {Suyu},
  {Taubenberger}, {Andika}, {Bag}, {Inoue}, {Jaelani}, {Leal-Taixe},
  {Meinhardt}, {Melo}, \& {More}}]{canameras_2023}
{Cañameras}, R., {Schuldt}, S., {Shu}, Y., {et~al.} 2023, arXiv e-prints,
  arXiv:2306.03136

\bibitem[{Chollet(2015)}]{chollet_keras}
Chollet, F. 2015, keras, \url{https://github.com/fchollet/keras}

\bibitem[{{{\'C}iprijanovi{\'c}} {et~al.}(2022){{\'C}iprijanovi{\'c}},
  {Kafkes}, {Snyder}, {S{\'a}nchez}, {Perdue}, {Pedro}, {Nord}, {Madireddy}, \&
  {Wild}}]{domainadaptation}
{{\'C}iprijanovi{\'c}}, A., {Kafkes}, D., {Snyder}, G., {et~al.} 2022, Machine
  Learning: Science and Technology, 3, 035007

\bibitem[{{Coe} {et~al.}(2006){Coe}, {Ben{\'\i}tez}, {S{\'a}nchez}, {Jee},
  {Bouwens}, \& {Ford}}]{2006AJ....132..926C}
{Coe}, D., {Ben{\'\i}tez}, N., {S{\'a}nchez}, S.~F., {et~al.} 2006, \aj, 132,
  926

\bibitem[{{Collett}(2015)}]{collett_sl}
{Collett}, T.~E. 2015, \apj, 811, 20

\bibitem[{{Cropper} {et~al.}(2012){Cropper}, {Cole}, {James}, {Mellier},
  {Martignac}, {Di Giorgio}, {Paltani}, {Genolet}, {Fourmond}, {Cara},
  {Amiaux}, {Guttridge}, {Walton}, {Thomas}, {Rees}, {Pool}, {Endicott},
  {Holland}, {Gow}, {Murray}, {Duvet}, {Augueres}, {Laureijs}, {Gondoin},
  {Kitching}, {Massey}, \& {Hoekstra}}]{vis_instrument}
{Cropper}, M., {Cole}, R., {James}, A., {et~al.} 2012, in Society of
  Photo-Optical Instrumentation Engineers (SPIE) Conference Series, Vol. 8442,
  Space Telescopes and Instrumentation 2012: Optical, Infrared, and Millimeter
  Wave, ed. M.~C. {Clampin}, G.~G. {Fazio}, H.~A. {MacEwen}, \& J.~{Oschmann},
  Jacobus~M., 84420V

\bibitem[{{de Jong} {et~al.}(2015){de Jong}, {Verdoes Kleijn}, {Boxhoorn},
  {Buddelmeijer}, {Capaccioli}, {Getman}, {Grado}, {Helmich}, {Huang},
  {Irisarri}, {Kuijken}, {La Barbera}, {McFarland}, {Napolitano}, {Radovich},
  {Sikkema}, {Valentijn}, {Begeman}, {Brescia}, {Cavuoti}, {Choi}, {Cordes},
  {Covone}, {Dall'Ora}, {Hildebrandt}, {Longo}, {Nakajima}, {Paolillo},
  {Puddu}, {Rifatto}, {Tortora}, {van Uitert}, {Buddendiek},
  {Harnois-D{\'e}raps}, {Erben}, {Eriksen}, {Heymans}, {Hoekstra}, {Joachimi},
  {Kitching}, {Klaes}, {Koopmans}, {K{\"o}hlinger}, {Roy}, {Sif{\'o}n},
  {Schneider}, {Sutherland}, {Viola}, \& {Vriend}}]{kids}
{de Jong}, J. T.~A., {Verdoes Kleijn}, G.~A., {Boxhoorn}, D.~R., {et~al.} 2015,
  \aap, 582, A62

\bibitem[{{de Jong} {et~al.}(2019){de Jong}, {Agertz}, {Berbel}, {Aird},
  {Alexander}, {Amarsi}, {Anders}, {Andrae}, {Ansarinejad}, {Ansorge},
  {Antilogus}, {Anwand-Heerwart}, {Arentsen}, {Arnadottir}, {Asplund}, {Auger},
  {Azais}, {Baade}, {Baker}, {Baker}, {Balbinot}, {Baldry}, {Banerji},
  {Barden}, {Barklem}, {Barth{\'e}l{\'e}my-Mazot}, {Battistini}, {Bauer},
  {Bell}, {Bellido-Tirado}, {Bellstedt}, {Belokurov}, {Bensby}, {Bergemann},
  {Bestenlehner}, {Bielby}, {Bilicki}, {Blake}, {Bland-Hawthorn}, {Boeche},
  {Boland}, {Boller}, {Bongard}, {Bongiorno}, {Bonifacio}, {Boudon}, {Brooks},
  {Brown}, {Brown}, {Br{\"u}ggen}, {Brynnel}, {Brzeski}, {Buchert},
  {Buschkamp}, {Caffau}, {Caillier}, {Carrick}, {Casagrande}, {Case}, {Casey},
  {Cesarini}, {Cescutti}, {Chapuis}, {Chiappini}, {Childress}, {Christlieb},
  {Church}, {Cioni}, {Cluver}, {Colless}, {Collett}, {Comparat}, {Cooper},
  {Couch}, {Courbin}, {Croom}, {Croton}, {Daguis{\'e}}, {Dalton}, {Davies},
  {Davis}, {de Laverny}, {Deason}, {Dionies}, {Disseau}, {Doel}, {D{\"o}scher},
  {Driver}, {Dwelly}, {Eckert}, {Edge}, {Edvardsson}, {Youssoufi}, {Elhaddad},
  {Enke}, {Erfanianfar}, {Farrell}, {Fechner}, {Feiz}, {Feltzing}, {Ferreras},
  {Feuerstein}, {Feuillet}, {Finoguenov}, {Ford}, {Fotopoulou}, {Fouesneau},
  {Frenk}, {Frey}, {Gaessler}, {Geier}, {Gentile Fusillo}, {Gerhard},
  {Giannantonio}, {Giannone}, {Gibson}, {Gillingham},
  {Gonz{\'a}lez-Fern{\'a}ndez}, {Gonzalez-Solares}, {Gottloeber}, {Gould},
  {Grebel}, {Gueguen}, {Guiglion}, {Haehnelt}, {Hahn}, {Hansen}, {Hartman},
  {Hauptner}, {Hawkins}, {Haynes}, {Haynes}, {Heiter}, {Helmi}, {Aguayo},
  {Hewett}, {Hinton}, {Hobbs}, {Hoenig}, {Hofman}, {Hook}, {Hopgood},
  {Hopkins}, {Hourihane}, {Howes}, {Howlett}, {Huet}, {Irwin}, {Iwert},
  {Jablonka}, {Jahn}, {Jahnke}, {Jarno}, {Jin}, {Jofre}, {Johl}, {Jones},
  {J{\"o}nsson}, {Jordan}, {Karovicova}, {Khalatyan}, {Kelz}, {Kennicutt},
  {King}, {Kitaura}, {Klar}, {Klauser}, {Kneib}, {Koch}, {Koposov},
  {Kordopatis}, {Korn}, {Kosmalski}, {Kotak}, {Kovalev}, {Kreckel}, {Kripak},
  {Krumpe}, {Kuijken}, {Kunder}, {Kushniruk}, {Lam}, {Lamer}, {Laurent},
  {Lawrence}, {Lehmitz}, {Lemasle}, {Lewis}, {Li}, {Lidman}, {Lind}, {Liske},
  {Lizon}, {Loveday}, {Ludwig}, {McDermid}, {Maguire}, {Mainieri}, {Mali},
  {Mandel}, {Mandel}, {Mannering}, {Martell}, {Martinez Delgado}, {Matijevic},
  {McGregor}, {McMahon}, {McMillan}, {Mena}, {Merloni}, {Meyer}, {Michel},
  {Micheva}, {Migniau}, {Minchev}, {Monari}, {Muller}, {Murphy},
  {Muthukrishna}, {Nandra}, {Navarro}, {Ness}, {Nichani}, {Nichol}, {Nicklas},
  {Niederhofer}, {Norberg}, {Obreschkow}, {Oliver}, {Owers}, {Pai},
  {Pankratow}, {Parkinson}, {Paschke}, {Paterson}, {Pecontal}, {Parry},
  {Phillips}, {Pillepich}, {Pinard}, {Pirard}, {Piskunov}, {Plank},
  {Pl{\"u}schke}, {Pons}, {Popesso}, {Power}, {Pragt}, {Pramskiy}, {Pryer},
  {Quattri}, {Queiroz}, {Quirrenbach}, {Rahurkar}, {Raichoor}, {Ramstedt},
  {Rau}, {Recio-Blanco}, {Reiss}, {Renaud}, {Revaz}, {Rhode}, {Richard},
  {Richter}, {Rix}, {Robotham}, {Roelfsema}, {Romaniello}, {Rosario},
  {Rothmaier}, {Roukema}, {Ruchti}, {Rupprecht}, {Rybizki}, {Ryde}, {Saar},
  {Sadler}, {Sahl{\'e}n}, {Salvato}, {Sassolas}, {Saunders}, {Saviauk},
  {Sbordone}, {Schmidt}, {Schnurr}, {Scholz}, {Schwope}, {Seifert}, {Shanks},
  {Sheinis}, {Sivov}, {Sk{\'u}lad{\'o}ttir}, {Smartt}, {Smedley}, {Smith},
  {Smith}, {Sorce}, {Spitler}, {Starkenburg}, {Steinmetz}, {Stilz}, {Storm},
  {Sullivan}, {Sutherland}, {Swann}, {Tamone}, {Taylor}, {Teillon}, {Tempel},
  {ter Horst}, {Thi}, {Tolstoy}, {Trager}, {Traven}, {Tremblay}, {Tresse},
  {Valentini}, {van de Weygaert}, {van den Ancker}, {Veljanoski}, {Venkatesan},
  {Wagner}, {Wagner}, {Walcher}, {Waller}, {Walton}, {Wang}, {Winkler},
  {Wisotzki}, {Worley}, {Worseck}, {Xiang}, {Xu}, {Yong}, {Zhao}, {Zheng},
  {Zscheyge}, \& {Zucker}}]{4mostsurvey}
{de Jong}, R.~S., {Agertz}, O., {Berbel}, A.~A., {et~al.} 2019, The Messenger,
  175, 3

\bibitem[{{Desprez} {et~al.}(2018){Desprez}, {Richard}, {Jauzac}, {Martinez},
  {Siana}, \& {Cl{\'e}ment}}]{desprez_2018}
{Desprez}, G., {Richard}, J., {Jauzac}, M., {et~al.} 2018, \mnras, 479, 2630

\bibitem[{{Dewdney} {et~al.}(2009){Dewdney}, {Hall}, {Schilizzi}, \&
  {Lazio}}]{ska}
{Dewdney}, P.~E., {Hall}, P.~J., {Schilizzi}, R.~T., \& {Lazio}, T.~J.~L.~W.
  2009, IEEE Proceedings, 97, 1482

\bibitem[{{Dom{\'\i}nguez S{\'a}nchez} {et~al.}(2019){Dom{\'\i}nguez
  S{\'a}nchez}, {Huertas-Company}, {Bernardi}, {Kaviraj}, {Fischer}, {Abbott},
  {Abdalla}, {Annis}, {Avila}, {Brooks}, {Buckley-Geer}, {Carnero Rosell},
  {Carrasco Kind}, {Carretero}, {Cunha}, {D'Andrea}, {da Costa}, {Davis}, {De
  Vicente}, {Doel}, {Evrard}, {Fosalba}, {Frieman}, {Garc{\'\i}a-Bellido},
  {Gaztanaga}, {Gerdes}, {Gruen}, {Gruendl}, {Gschwend}, {Gutierrez},
  {Hartley}, {Hollowood}, {Honscheid}, {Hoyle}, {James}, {Kuehn}, {Kuropatkin},
  {Lahav}, {Maia}, {March}, {Melchior}, {Menanteau}, {Miquel}, {Nord},
  {Plazas}, {Sanchez}, {Scarpine}, {Schindler}, {Schubnell}, {Smith}, {Smith},
  {Soares-Santos}, {Sobreira}, {Suchyta}, {Swanson}, {Tarle}, {Thomas},
  {Walker}, \& {Zuntz}}]{dominguez_tl}
{Dom{\'\i}nguez S{\'a}nchez}, H., {Huertas-Company}, M., {Bernardi}, M.,
  {et~al.} 2019, \mnras, 484, 93

\bibitem[{{Dom{\'\i}nguez S{\'a}nchez} {et~al.}(2018){Dom{\'\i}nguez
  S{\'a}nchez}, {Huertas-Company}, {Bernardi}, {Tuccillo}, \&
  {Fischer}}]{dom_sanchez_morpho}
{Dom{\'\i}nguez S{\'a}nchez}, H., {Huertas-Company}, M., {Bernardi}, M.,
  {Tuccillo}, D., \& {Fischer}, J.~L. 2018, \mnras, 476, 3661

\bibitem[{{Euclid Collaboration: Scaramella} {et~al.}(2022){Euclid
  Collaboration: Scaramella}, {Amiaux}, {Mellier}, {Burigana}, {Carvalho},
  {Cuillandre}, {Da Silva}, {Derosa}, {Dinis}, {Maiorano}, {Maris}, {Tereno},
  {Laureijs}, {Boenke}, {Buenadicha}, {Dupac}, {Gaspar Venancio},
  {G{\'o}mez-{\'A}lvarez}, {Hoar}, {Lorenzo Alvarez}, {Racca},
  {Saavedra-Criado}, {Schwartz}, {Vavrek}, {Schirmer}, {Aussel}, {Azzollini},
  {Cardone}, {Cropper}, {Ealet}, {Garilli}, {Gillard}, {Granett}, {Guzzo},
  {Hoekstra}, {Jahnke}, {Kitching}, {Maciaszek}, {Meneghetti}, {Miller},
  {Nakajima}, {Niemi}, {Pasian}, {Percival}, {Pottinger}, {Sauvage},
  {Scodeggio}, {Wachter}, {Zacchei}, {Aghanim}, {Amara}, {Auphan}, {Auricchio},
  {Awan}, {Balestra}, {Bender}, {Bodendorf}, {Bonino}, {Branchini},
  {Brau-Nogue}, {Brescia}, {Candini}, {Capobianco}, {Carbone}, {Carlberg},
  {Carretero}, {Casas}, {Castander}, {Castellano}, {Cavuoti}, {Cimatti},
  {Cledassou}, {Congedo}, {Conselice}, {Conversi}, {Copin}, {Corcione},
  {Costille}, {Courbin}, {Degaudenzi}, {Douspis}, {Dubath}, {Duncan}, {Dusini},
  {Farrens}, {Ferriol}, {Fosalba}, {Fourmanoit}, {Frailis}, {Franceschi},
  {Franzetti}, {Fumana}, {Gillis}, {Giocoli}, {Grazian}, {Grupp}, {Haugan},
  {Holmes}, {Hormuth}, {Hudelot}, {Kermiche}, {Kiessling}, {Kilbinger},
  {Kohley}, {Kubik}, {K{\"u}mmel}, {Kunz}, {Kurki-Suonio}, {Lahav}, {Ligori},
  {Lilje}, {Lloro}, {Mansutti}, {Marggraf}, {Markovic}, {Marulli}, {Massey},
  {Maurogordato}, {Melchior}, {Merlin}, {Meylan}, {Mohr}, {Moresco}, {Morin},
  {Moscardini}, {Munari}, {Nichol}, {Padilla}, {Paltani}, {Peacock},
  {Pedersen}, {Pettorino}, {Pires}, {Poncet}, {Popa}, {Pozzetti}, {Raison},
  {Rebolo}, {Rhodes}, {Rix}, {Roncarelli}, {Rossetti}, {Saglia}, {Schneider},
  {Schrabback}, {Secroun}, {Seidel}, {Serrano}, {Sirignano}, {Sirri},
  {Skottfelt}, {Stanco}, {Starck}, {Tallada-Cresp{\'\i}}, {Tavagnacco},
  {Taylor}, {Teplitz}, {Toledo-Moreo}, {Torradeflot}, {Trifoglio}, {Valentijn},
  {Valenziano}, {Verdoes Kleijn}, {Wang}, {Welikala}, {Weller}, {Wetzstein},
  {Zamorani}, {Zoubian}, {Andreon}, {Baldi}, {Bardelli}, {Boucaud}, {Camera},
  {Di Ferdinando}, {Fabbian}, {Farinelli}, {Galeotta}, {Graci{\'a}-Carpio},
  {Maino}, {Medinaceli}, {Mei}, {Neissner}, {Polenta}, {Renzi}, {Romelli},
  {Rosset}, {Sureau}, {Tenti}, {Vassallo}, {Zucca}, {Baccigalupi},
  {Balaguera-Antol{\'\i}nez}, {Battaglia}, {Biviano}, {Borgani}, {Bozzo},
  {Cabanac}, {Cappi}, {Casas}, {Castignani}, {Colodro-Conde}, {Coupon},
  {Courtois}, {Cuby}, {de la Torre}, {Desai}, {Dole}, {Fabricius}, {Farina},
  {Ferreira}, {Finelli}, {Flose-Reimberg}, {Fotopoulou}, {Ganga}, {Gozaliasl},
  {Hook}, {Keihanen}, {Kirkpatrick}, {Liebing}, {Lindholm}, {Mainetti},
  {Martinelli}, {Martinet}, {Maturi}, {McCracken}, {Metcalf}, {Morgante},
  {Nightingale}, {Nucita}, {Patrizii}, {Potter}, {Riccio}, {S{\'a}nchez},
  {Sapone}, {Schewtschenko}, {Schultheis}, {Scottez}, {Teyssier}, {Tutusaus},
  {Valiviita}, {Viel}, {Vriend}, \& {Whittaker}}]{Scaramella-EP1}
{Euclid Collaboration: Scaramella}, R., {Amiaux}, J., {Mellier}, Y., {et~al.}
  2022, \aap, 662, A112

\bibitem[{{Euclid Collaboration: Schirmer} {et~al.}(2022){Euclid Collaboration:
  Schirmer}, {Jahnke}, {Seidel}, {Aussel}, {Bodendorf}, {Grupp}, {Hormuth},
  {Wachter}, {Appleton}, {Barbier}, {Brinchmann}, {Carrasco}, {Castander},
  {Coupon}, {De Paolis}, {Franco}, {Ganga}, {Hudelot}, {Jullo}, {Lan{\c{c}}on},
  {Nucita}, {Paltani}, {Smadja}, {Strafella}, {Venancio}, {Weiler}, {Amara},
  {Auphan}, {Auricchio}, {Balestra}, {Bender}, {Bonino}, {Branchini},
  {Brescia}, {Capobianco}, {Carbone}, {Carretero}, {Casas}, {Castellano},
  {Cavuoti}, {Cimatti}, {Cledassou}, {Congedo}, {Conselice}, {Conversi},
  {Copin}, {Corcione}, {Costille}, {Courbin}, {Da Silva}, {Degaudenzi},
  {Douspis}, {Dubath}, {Dupac}, {Dusini}, {Ealet}, {Farrens}, {Ferriol},
  {Fosalba}, {Frailis}, {Franceschi}, {Franzetti}, {Fumana}, {Garilli},
  {Gillard}, {Gillis}, {Giocoli}, {Grazian}, {Guzzo}, {Haugan}, {Hoekstra},
  {Holmes}, {Hornstrup}, {K{\"u}mmel}, {Kermiche}, {Kiessling}, {Kilbinger},
  {Kitching}, {Kohley}, {Kunz}, {Kurki-Suonio}, {Laureijs}, {Ligori}, {Lilje},
  {Lloro}, {Maciaszek}, {Maiorano}, {Mansutti}, {Marggraf}, {Markovic},
  {Marulli}, {Massey}, {Maurogordato}, {Mellier}, {Meneghetti}, {Merlin},
  {Meylan}, {Moresco}, {Moscardini}, {Munari}, {Nakajima}, {Nichol}, {Niemi},
  {Padilla}, {Pasian}, {Pedersen}, {Percival}, {Pettorino}, {Pires}, {Poncet},
  {Popa}, {Pozzetti}, {Prieto}, {Raison}, {Rhodes}, {Rix}, {Roncarelli},
  {Rossetti}, {Saglia}, {Sartoris}, {Scaramella}, {Schneider}, {Secroun},
  {Serrano}, {Sirignano}, {Sirri}, {Stanco}, {Tallada-Cresp{\'\i}}, {Taylor},
  {Teplitz}, {Tereno}, {Toledo-Moreo}, {Torradeflot}, {Trifoglio}, {Valentijn},
  {Valenziano}, {Wang}, {Weller}, {Zamorani}, {Zoubian}, {Andreon}, {Bardelli},
  {Boucaud}, {Camera}, {Farinelli}, {Graci{\'a}-Carpio}, {Maino}, {Medinaceli},
  {Mei}, {Morisset}, {Polenta}, {Renzi}, {Romelli}, {Tenti}, {Vassallo},
  {Zacchei}, {Zucca}, {Baccigalupi}, {Balaguera-Antol{\'\i}nez}, {Biviano},
  {Blanchard}, {Borgani}, {Bozzo}, {Burigana}, {Cabanac}, {Cappi}, {Carvalho},
  {Casas}, {Castignani}, {Colodro-Conde}, {Cooray}, {Courtois}, {Crocce},
  {Cuby}, {Davini}, {de la Torre}, {Di Ferdinando}, {Escartin}, {Farina},
  {Ferreira}, {Finelli}, {Fotopoulou}, {Galeotta}, {Garcia-Bellido},
  {Gaztanaga}, {George}, {Gozaliasl}, {Hook}, {Ili{\'c}}, {Kansal},
  {Kashlinsky}, {Keihanen}, {Kirkpatrick}, {Lindholm}, {Mainetti}, {Maoli},
  {Martinelli}, {Martinet}, {Maturi}, {Mauri}, {McCracken}, {Metcalf},
  {Monaco}, {Morgante}, {Nightingale}, {Patrizii}, {Peel}, {Popa}, {Porciani},
  {Potter}, {Reimberg}, {Riccio}, {S{\'a}nchez}, {Sapone}, {Scottez},
  {Sefusatti}, {Teyssier}, {Tutusaus}, {Valieri}, {Valiviita}, {Viel}, \&
  {Hildebrandt}}]{Schirmer-EP18}
{Euclid Collaboration: Schirmer}, M., {Jahnke}, K., {Seidel}, G., {et~al.}
  2022, \aap, 662, A92

\bibitem[{{Fluri} {et~al.}(2019){Fluri}, {Kacprzak}, {Lucchi}, {Refregier},
  {Amara}, {Hofmann}, \& {Schneider}}]{fluri_cosmo}
{Fluri}, J., {Kacprzak}, T., {Lucchi}, A., {et~al.} 2019, \prd, 100, 063514

\bibitem[{{Gavazzi} {et~al.}(2014){Gavazzi}, {Marshall}, {Treu}, \&
  {Sonnenfeld}}]{2014ApJ...785..144G}
{Gavazzi}, R., {Marshall}, P.~J., {Treu}, T., \& {Sonnenfeld}, A. 2014, \apj,
  785, 144

\bibitem[{{Gavazzi} {et~al.}(2012){Gavazzi}, {Treu}, {Marshall}, {Brault}, \&
  {Ruff}}]{gavazzi_lensmass}
{Gavazzi}, R., {Treu}, T., {Marshall}, P.~J., {Brault}, F., \& {Ruff}, A. 2012,
  \apj, 761, 170

\bibitem[{{Gavazzi} {et~al.}(2007){Gavazzi}, {Treu}, {Rhodes}, {Koopmans},
  {Bolton}, {Burles}, {Massey}, \& {Moustakas}}]{2007ApJ...667..176G}
{Gavazzi}, R., {Treu}, T., {Rhodes}, J.~D., {et~al.} 2007, \apj, 667, 176

\bibitem[{{Gentile} {et~al.}(2022){Gentile}, {Tortora}, {Covone}, {Koopmans},
  {Spiniello}, {Fan}, {Li}, {Liu}, {Napolitano}, {Vaccari}, \&
  {Fu}}]{gentile_2021}
{Gentile}, F., {Tortora}, C., {Covone}, G., {et~al.} 2022, \mnras, 510, 500

\bibitem[{{Ghosh} {et~al.}(2020){Ghosh}, {Urry}, {Wang}, {Schawinski}, {Turp},
  \& {Powell}}]{ghosh_morphology}
{Ghosh}, A., {Urry}, C.~M., {Wang}, Z., {et~al.} 2020, \apj, 895, 112

\bibitem[{Goodfellow {et~al.}(2016)Goodfellow, Bengio, \&
  Courville}]{goodfellow}
Goodfellow, I., Bengio, Y., \& Courville, A. 2016, Deep Learning (The MIT
  Press)

\bibitem[{{Grillo}(2012)}]{grillo_dm}
{Grillo}, C. 2012, \apjl, 747, L15

\bibitem[{{Grillo} {et~al.}(2014){Grillo}, {Gobat}, {Presotto}, {Balestra},
  {Mercurio}, {Rosati}, {Nonino}, {Vanzella}, {Christensen}, {Graves},
  {Biviano}, {Lemze}, {Bartelmann}, {Benitez}, {Bouwens}, {Bradley},
  {Broadhurst}, {Coe}, {Donahue}, {Ford}, {Infante}, {Jouvel}, {Kelson},
  {Koekemoer}, {Lahav}, {Medezinski}, {Melchior}, {Meneghetti}, {Merten},
  {Molino}, {Monna}, {Moustakas}, {Moustakas}, {Postman}, {Seitz}, {Umetsu},
  {Zheng}, \& {Zitrin}}]{grillo2014}
{Grillo}, C., {Gobat}, R., {Presotto}, V., {et~al.} 2014, \apj, 786, 11

\bibitem[{{Gupta} \& {Reichardt}(2020)}]{gupta_clustermass}
{Gupta}, N. \& {Reichardt}, C.~L. 2020, \apj, 900, 110

\bibitem[{{Gwyn}(2012)}]{cfhtls}
{Gwyn}, S. D.~J. 2012, \aj, 143, 38

\bibitem[{Hanley(1982)}]{hanley_roc}
Hanley, J. V. \&~McNeil, B. 1982, Radiology, 143, 29

\bibitem[{{He} {et~al.}(2016){He}, {Zhang}, {Ren}, \& {Sun}}]{resnet}
{He}, K., {Zhang}, X., {Ren}, S., \& {Sun}, J. 2016, 2016 IEEE Conference on
  Computer Vision and Pattern Recognition (CVPR), 770

\bibitem[{{He} {et~al.}(2020){He}, {Er}, {Long}, {Liu}, {Liu}, {Li}, {Liu},
  {Deng}, \& {Fan}}]{he_cnn}
{He}, Z., {Er}, X., {Long}, Q., {et~al.} 2020, \mnras, 497, 556

\bibitem[{Hebb(1949)}]{hebb_connections}
Hebb, D.~O. 1949, The organization of behavior: {A} neuropsychological theory
  (Wiley)

\bibitem[{{Ho} {et~al.}(2019){Ho}, {Rau}, {Ntampaka}, {Farahi}, {Trac}, \&
  {P{\'o}czos}}]{ho_clustermass}
{Ho}, M., {Rau}, M.~M., {Ntampaka}, M., {et~al.} 2019, \apj, 887, 25

\bibitem[{{Huertas-Company} {et~al.}(2015){Huertas-Company}, {Gravet},
  {Cabrera-Vives}, {P{\'e}rez-Gonz{\'a}lez}, {Kartaltepe}, {Barro}, {Bernardi},
  {Mei}, {Shankar}, {Dimauro}, {Bell}, {Kocevski}, {Koo}, {Faber}, \&
  {Mcintosh}}]{huertas_company_morpho}
{Huertas-Company}, M., {Gravet}, R., {Cabrera-Vives}, G., {et~al.} 2015, \apjs,
  221, 8

\bibitem[{{Impellizzeri} {et~al.}(2008){Impellizzeri}, {McKean}, {Castangia},
  {Roy}, {Henkel}, {Brunthaler}, \& {Wucknitz}}]{impellizzeri_mag}
{Impellizzeri}, C.~M.~V., {McKean}, J.~P., {Castangia}, P., {et~al.} 2008,
  \nat, 456, 927

\bibitem[{Ioffe \& Szegedy(2015)}]{batch_norm}
Ioffe, S. \& Szegedy, C. 2015, in Proceedings of Machine Learning Research,
  Vol.~37, Proceedings of the 32nd International Conference on Machine
  Learning, ed. F.~Bach \& D.~Blei (Lille, France: PMLR), 448

\bibitem[{{Jacobs} {et~al.}(2019{\natexlab{a}}){Jacobs}, {Collett},
  {Glazebrook}, {Buckley-Geer}, {Diehl}, {Lin}, {McCarthy}, {Qin}, {Odden},
  {Caso Escudero}, {Dial}, {Yung}, {Gaitsch}, {Pellico}, {Lindgren}, {Abbott},
  {Annis}, {Avila}, {Brooks}, {Burke}, {Carnero Rosell}, {Carrasco Kind},
  {Carretero}, {da Costa}, {De Vicente}, {Fosalba}, {Frieman},
  {Garc{\'\i}a-Bellido}, {Gaztanaga}, {Goldstein}, {Gruen}, {Gruendl},
  {Gschwend}, {Hollowood}, {Honscheid}, {Hoyle}, {James}, {Krause},
  {Kuropatkin}, {Lahav}, {Lima}, {Maia}, {Marshall}, {Miquel}, {Plazas},
  {Roodman}, {Sanchez}, {Scarpine}, {Serrano}, {Sevilla-Noarbe}, {Smith},
  {Sobreira}, {Suchyta}, {Swanson}, {Tarle}, {Vikram}, {Walker}, {Zhang}, \&
  {DES Collaboration}}]{jacobs_cnn2}
{Jacobs}, C., {Collett}, T., {Glazebrook}, K., {et~al.} 2019{\natexlab{a}},
  \apjs, 243, 17

\bibitem[{{Jacobs} {et~al.}(2019{\natexlab{b}}){Jacobs}, {Collett},
  {Glazebrook}, {McCarthy}, {Qin}, {Abbott}, {Abdalla}, {Annis}, {Avila},
  {Bechtol}, {Bertin}, {Brooks}, {Buckley-Geer}, {Burke}, {Carnero Rosell},
  {Carrasco Kind}, {Carretero}, {da Costa}, {Davis}, {De Vicente}, {Desai},
  {Diehl}, {Doel}, {Eifler}, {Flaugher}, {Frieman}, {Garc{\'\i}a-Bellido},
  {Gaztanaga}, {Gerdes}, {Goldstein}, {Gruen}, {Gruendl}, {Gschwend},
  {Gutierrez}, {Hartley}, {Hollowood}, {Honscheid}, {Hoyle}, {James}, {Kuehn},
  {Kuropatkin}, {Lahav}, {Li}, {Lima}, {Lin}, {Maia}, {Martini}, {Miller},
  {Miquel}, {Nord}, {Plazas}, {Sanchez}, {Scarpine}, {Schubnell}, {Serrano},
  {Sevilla-Noarbe}, {Smith}, {Soares-Santos}, {Sobreira}, {Suchyta}, {Swanson},
  {Tarle}, {Vikram}, {Walker}, {Zhang}, {Zuntz}, \& {DES
  Collaboration}}]{jacobs_cnn}
{Jacobs}, C., {Collett}, T., {Glazebrook}, K., {et~al.} 2019{\natexlab{b}},
  \mnras, 484, 5330

\bibitem[{{Jacobs} {et~al.}(2017){Jacobs}, {Glazebrook}, {Collett}, {More}, \&
  {McCarthy}}]{jacobs_cnn3}
{Jacobs}, C., {Glazebrook}, K., {Collett}, T., {More}, A., \& {McCarthy}, C.
  2017, \mnras, 471, 167

\bibitem[{{Jauzac} {et~al.}(2021){Jauzac}, {Klein}, {Kneib}, {Richard},
  {Rexroth}, {Sch{\"a}fer}, \& {Verdier}}]{jauzac_2020}
{Jauzac}, M., {Klein}, B., {Kneib}, J.-P., {et~al.} 2021, \mnras, 508, 1206

\bibitem[{{Kingma} \& {Ba}(2017)}]{kingma_adam}
{Kingma}, D.~P. \& {Ba}, J. 2017, arXiv:1412.6980

\bibitem[{{Koopmans} {et~al.}(2004){Koopmans}, {Browne}, \&
  {Jackson}}]{koopmans_2004}
{Koopmans}, L.~V.~E., {Browne}, I.~W.~A., \& {Jackson}, N.~J. 2004, \nar, 48,
  1085

\bibitem[{{Laureijs} {et~al.}(2011){Laureijs}, {Amiaux}, {Arduini},
  {Augu{\`e}res}, {Brinchmann}, {Cole}, {Cropper}, {Dabin}, {Duvet}, {Ealet},
  {Garilli}, {Gondoin}, {Guzzo}, {Hoar}, {Hoekstra}, {Holmes}, {Kitching},
  {Maciaszek}, {Mellier}, {Pasian}, {Percival}, {Rhodes}, {Saavedra Criado},
  {Sauvage}, {Scaramella}, {Valenziano}, {Warren}, {Bender}, {Castander},
  {Cimatti}, {Le F{\`e}vre}, {Kurki-Suonio}, {Levi}, {Lilje}, {Meylan},
  {Nichol}, {Pedersen}, {Popa}, {Rebolo Lopez}, {Rix}, {Rottgering},
  {Zeilinger}, {Grupp}, {Hudelot}, {Massey}, {Meneghetti}, {Miller}, {Paltani},
  {Paulin-Henriksson}, {Pires}, {Saxton}, {Schrabback}, {Seidel}, {Walsh},
  {Aghanim}, {Amendola}, {Bartlett}, {Baccigalupi}, {Beaulieu}, {Benabed},
  {Cuby}, {Elbaz}, {Fosalba}, {Gavazzi}, {Helmi}, {Hook}, {Irwin}, {Kneib},
  {Kunz}, {Mannucci}, {Moscardini}, {Tao}, {Teyssier}, {Weller}, {Zamorani},
  {Zapatero Osorio}, {Boulade}, {Foumond}, {Di Giorgio}, {Guttridge}, {James},
  {Kemp}, {Martignac}, {Spencer}, {Walton}, {Bl{\"u}mchen}, {Bonoli},
  {Bortoletto}, {Cerna}, {Corcione}, {Fabron}, {Jahnke}, {Ligori}, {Madrid},
  {Martin}, {Morgante}, {Pamplona}, {Prieto}, {Riva}, {Toledo}, {Trifoglio},
  {Zerbi}, {Abdalla}, {Douspis}, {Grenet}, {Borgani}, {Bouwens}, {Courbin},
  {Delouis}, {Dubath}, {Fontana}, {Frailis}, {Grazian}, {Koppenh{\"o}fer},
  {Mansutti}, {Melchior}, {Mignoli}, {Mohr}, {Neissner}, {Noddle}, {Poncet},
  {Scodeggio}, {Serrano}, {Shane}, {Starck}, {Surace}, {Taylor},
  {Verdoes-Kleijn}, {Vuerli}, {Williams}, {Zacchei}, {Altieri}, {Escudero
  Sanz}, {Kohley}, {Oosterbroek}, {Astier}, {Bacon}, {Bardelli}, {Baugh},
  {Bellagamba}, {Benoist}, {Bianchi}, {Biviano}, {Branchini}, {Carbone},
  {Cardone}, {Clements}, {Colombi}, {Conselice}, {Cresci}, {Deacon}, {Dunlop},
  {Fedeli}, {Fontanot}, {Franzetti}, {Giocoli}, {Garcia-Bellido}, {Gow},
  {Heavens}, {Hewett}, {Heymans}, {Holland}, {Huang}, {Ilbert}, {Joachimi},
  {Jennins}, {Kerins}, {Kiessling}, {Kirk}, {Kotak}, {Krause}, {Lahav}, {van
  Leeuwen}, {Lesgourgues}, {Lombardi}, {Magliocchetti}, {Maguire}, {Majerotto},
  {Maoli}, {Marulli}, {Maurogordato}, {McCracken}, {McLure}, {Melchiorri},
  {Merson}, {Moresco}, {Nonino}, {Norberg}, {Peacock}, {Pello}, {Penny},
  {Pettorino}, {Di Porto}, {Pozzetti}, {Quercellini}, {Radovich}, {Rassat},
  {Roche}, {Ronayette}, {Rossetti}, {Sartoris}, {Schneider}, {Semboloni},
  {Serjeant}, {Simpson}, {Skordis}, {Smadja}, {Smartt}, {Spano}, {Spiro},
  {Sullivan}, {Tilquin}, {Trotta}, {Verde}, {Wang}, {Williger}, {Zhao},
  {Zoubian}, \& {Zucca}}]{Laureijs11}
{Laureijs}, R., {Amiaux}, J., {Arduini}, S., {et~al.} 2011, arXiv:1110.3193

\bibitem[{LeCun {et~al.}(1989)LeCun, Boser, Denker, Henderson, Howard, Hubbard,
  \& Jackel}]{lecun_cnn}
LeCun, Y., Boser, B., Denker, J.~S., {et~al.} 1989, Neural Computation, 1, 541

\bibitem[{{Li} {et~al.}(2022){Li}, {Napolitano}, {Feng}, {Li}, {Amaro}, {Xie},
  {Tortora}, {Bilicki}, {Brescia}, {Cavuoti}, \&
  {Radovich}}]{2022A&A...666A..85L}
{Li}, R., {Napolitano}, N.~R., {Feng}, H., {et~al.} 2022, \aap, 666, A85

\bibitem[{{Li} {et~al.}(2021){Li}, {Napolitano}, {Spiniello}, {Tortora},
  {Kuijken}, {Koopmans}, {Schneider}, {Getman}, {Xie}, {Long}, {Shu},
  {Vernardos}, {Huang}, {Covone}, {Dvornik}, {Heymans}, {Hildebrandt},
  {Radovich}, \& {Wright}}]{2021ApJ...923...16L}
{Li}, R., {Napolitano}, N.~R., {Spiniello}, C., {et~al.} 2021, \apj, 923, 16

\bibitem[{{Li} {et~al.}(2020){Li}, {Napolitano}, {Tortora}, {Spiniello},
  {Koopmans}, {Huang}, {Roy}, {Vernardos}, {Chatterjee}, {Giblin}, {Getman},
  {Radovich}, {Covone}, \& {Kuijken}}]{li_cnn}
{Li}, R., {Napolitano}, N.~R., {Tortora}, C., {et~al.} 2020, \apj, 899, 30

\bibitem[{{Liew-Cain} {et~al.}(2021){Liew-Cain}, {Kawata},
  {S{\'a}nchez-Bl{\'a}zquez}, {Ferreras}, \&
  {Symeonidis}}]{liewcain_metallicity}
{Liew-Cain}, C.~L., {Kawata}, D., {S{\'a}nchez-Bl{\'a}zquez}, P., {Ferreras},
  I., \& {Symeonidis}, M. 2021, \mnras, 502, 1355

\bibitem[{Lin {et~al.}(2013)Lin, Chen, \& Yan}]{Lin2013NetworkIN}
Lin, M., Chen, Q., \& Yan, S. 2013, arXiv:1312.4400

\bibitem[{{LSST Science Collaboration} {et~al.}(2009){LSST Science
  Collaboration}, {Abell}, {Allison}, {Anderson}, {Andrew}, {Angel}, {Armus},
  {Arnett}, {Asztalos}, {Axelrod}, {Bailey}, {Ballantyne}, {Bankert},
  {Barkhouse}, {Barr}, {Barrientos}, {Barth}, {Bartlett}, {Becker}, {Becla},
  {Beers}, {Bernstein}, {Biswas}, {Blanton}, {Bloom}, {Bochanski}, {Boeshaar},
  {Borne}, {Bradac}, {Brandt}, {Bridge}, {Brown}, {Brunner}, {Bullock},
  {Burgasser}, {Burge}, {Burke}, {Cargile}, {Chandrasekharan}, {Chartas},
  {Chesley}, {Chu}, {Cinabro}, {Claire}, {Claver}, {Clowe}, {Connolly}, {Cook},
  {Cooke}, {Cooray}, {Covey}, {Culliton}, {de Jong}, {de Vries}, {Debattista},
  {Delgado}, {Dell'Antonio}, {Dhital}, {Di Stefano}, {Dickinson}, {Dilday},
  {Djorgovski}, {Dobler}, {Donalek}, {Dubois-Felsmann}, {Durech},
  {Eliasdottir}, {Eracleous}, {Eyer}, {Falco}, {Fan}, {Fassnacht}, {Ferguson},
  {Fernandez}, {Fields}, {Finkbeiner}, {Figueroa}, {Fox}, {Francke}, {Frank},
  {Frieman}, {Fromenteau}, {Furqan}, {Galaz}, {Gal-Yam}, {Garnavich},
  {Gawiser}, {Geary}, {Gee}, {Gibson}, {Gilmore}, {Grace}, {Green}, {Gressler},
  {Grillmair}, {Habib}, {Haggerty}, {Hamuy}, {Harris}, {Hawley}, {Heavens},
  {Hebb}, {Henry}, {Hileman}, {Hilton}, {Hoadley}, {Holberg}, {Holman},
  {Howell}, {Infante}, {Ivezic}, {Jacoby}, {Jain}, {R}, {Jedicke}, {Jee},
  {Garrett Jernigan}, {Jha}, {Johnston}, {Jones}, {Juric}, {Kaasalainen},
  {Styliani}, {Kafka}, {Kahn}, {Kaib}, {Kalirai}, {Kantor}, {Kasliwal},
  {Keeton}, {Kessler}, {Knezevic}, {Kowalski}, {Krabbendam}, {Krughoff},
  {Kulkarni}, {Kuhlman}, {Lacy}, {Lepine}, {Liang}, {Lien}, {Lira}, {Long},
  {Lorenz}, {Lotz}, {Lupton}, {Lutz}, {Macri}, {Mahabal}, {Mandelbaum},
  {Marshall}, {May}, {McGehee}, {Meadows}, {Meert}, {Milani}, {Miller},
  {Miller}, {Mills}, {Minniti}, {Monet}, {Mukadam}, {Nakar}, {Neill}, {Newman},
  {Nikolaev}, {Nordby}, {O'Connor}, {Oguri}, {Oliver}, {Olivier}, {Olsen},
  {Olsen}, {Olszewski}, {Oluseyi}, {Padilla}, {Parker}, {Pepper}, {Peterson},
  {Petry}, {Pinto}, {Pizagno}, {Popescu}, {Prsa}, {Radcka}, {Raddick},
  {Rasmussen}, {Rau}, {Rho}, {Rhoads}, {Richards}, {Ridgway}, {Robertson},
  {Roskar}, {Saha}, {Sarajedini}, {Scannapieco}, {Schalk}, {Schindler},
  {Schmidt}, {Schmidt}, {Schneider}, {Schumacher}, {Scranton}, {Sebag},
  {Seppala}, {Shemmer}, {Simon}, {Sivertz}, {Smith}, {Allyn Smith}, {Smith},
  {Spitz}, {Stanford}, {Stassun}, {Strader}, {Strauss}, {Stubbs}, {Sweeney},
  {Szalay}, {Szkody}, {Takada}, {Thorman}, {Trilling}, {Trimble}, {Tyson}, {Van
  Berg}, {Vanden Berk}, {VanderPlas}, {Verde}, {Vrsnak}, {Walkowicz},
  {Wandelt}, {Wang}, {Wang}, {Warner}, {Wechsler}, {West}, {Wiecha},
  {Williams}, {Willman}, {Wittman}, {Wolff}, {Wood-Vasey}, {Wozniak}, {Young},
  {Zentner}, \& {Zhan}}]{lsst_paper}
{LSST Science Collaboration}, {Abell}, P.~A., {Allison}, J., {et~al.} 2009,
  arXiv:0912.0201

\bibitem[{{Maciaszek} {et~al.}(2022){Maciaszek}, {Ealet}, {Gillard}, {Jahnke},
  {Barbier}, {Prieto}, {Bon}, {Bonnefoi}, {Caillat}, {Carle}, {Costille},
  {Ducret}, {Fabron}, {Foulon}, {Gimenez}, {Grassi}, {Jaquet}, {Le Mignant},
  {Martin}, {Pamplona}, {Sanchez}, {Cl{\'e}mens}, {Caillat}, {Niclas},
  {Secroun}, {Kubik}, {Ferriol}, {Berthe}, {Barri{\`e}re}, {Fontignie},
  {Valenziano}, {Auricchio}, {Battaglia}, {De Rosa}, {Farinelli}, {Franceschi},
  {Medinaceli}, {Morgante}, {Sortino}, {Trifoglio}, {Corcione}, {Capobianco},
  {Ligori}, {Dusini}, {Borsato}, {Dal Corso}, {Laudisio}, {Sirignano},
  {Stanco}, {Ventura}, {Patrizii}, {Chiarusi}, {Fornari}, {Giacomini},
  {Margiotta}, {Mauri}, {Pasqualini}, {Sirri}, {Spurio}, {Tenti}, {Travaglini},
  {Bonoli}, {Bortoletto}, {Balestra}, {Dalessandro}, {Grupp}, {Penka},
  {Steinwagner}, {Hormuth}, {Schirmer}, {Seidel}, {Padilla}, {Casas}, {Lloro},
  {Toledo-Moreo}, {Gomez}, {Colodro-Conde}, {Liz{\'a}n}, {Diaz}, {Lilje},
  {Andersen}, {Andersen}, {S{\o}rensen}, {Hornstrup}, {Jessen}, {Thizy},
  {Holmes}, {Pniel}, {Jhabvala}, {Pravdo}, {Seiffert}, {Waczynski}, {Laureij},
  {Racca}, {Salvignol}, {Boenke}, {Strada}, \& {Mellier}}]{Maciaszek22}
{Maciaszek}, T., {Ealet}, A., {Gillard}, W., {et~al.} 2022, in Society of
  Photo-Optical Instrumentation Engineers (SPIE) Conference Series, Vol. 12180,
  Space Telescopes and Instrumentation 2022: Optical, Infrared, and Millimeter
  Wave, ed. L.~E. {Coyle}, S.~{Matsuura}, \& M.~D. {Perrin}, arXiv:2210.10112

\bibitem[{{Marshall} {et~al.}(2009){Marshall}, {Hogg}, {Moustakas},
  {Fassnacht}, {Brada{\v{c}}}, {Schrabback}, \& {Blandford}}]{marshall2009}
{Marshall}, P.~J., {Hogg}, D.~W., {Moustakas}, L.~A., {et~al.} 2009, \apj, 694,
  924

\bibitem[{{Maturi} {et~al.}(2014){Maturi}, {Mizera}, \&
  {Seidel}}]{2014A&A...567A.111M}
{Maturi}, M., {Mizera}, S., \& {Seidel}, G. 2014, \aap, 567, A111

\bibitem[{{McCulloch} \& {Pitts}(1943)}]{ann}
{McCulloch}, W. \& {Pitts}, W. 1943, Bulletin of Mathematical Biophysics, 5,
  115

\bibitem[{{Melchior} {et~al.}(2010){Melchior}, {B{\"o}hnert}, {Lombardi}, \&
  {Bartelmann}}]{melchior_2010}
{Melchior}, P., {B{\"o}hnert}, A., {Lombardi}, M., \& {Bartelmann}, M. 2010,
  \aap, 510, A75

\bibitem[{{Melchior} {et~al.}(2007){Melchior}, {Meneghetti}, \&
  {Bartelmann}}]{melchior_2007}
{Melchior}, P., {Meneghetti}, M., \& {Bartelmann}, M. 2007, \aap, 463, 1215

\bibitem[{{Meneghetti} {et~al.}(2020){Meneghetti}, {Davoli}, {Bergamini},
  {Rosati}, {Natarajan}, {Giocoli}, {Caminha}, {Metcalf}, {Rasia}, {Borgani},
  {Calura}, {Grillo}, {Mercurio}, \& {Vanzella}}]{meneghetti_ggsl}
{Meneghetti}, M., {Davoli}, G., {Bergamini}, P., {et~al.} 2020, Science, 369,
  1347

\bibitem[{{Meneghetti} {et~al.}(2008){Meneghetti}, {Melchior}, {Grazian}, {De
  Lucia}, {Dolag}, {Bartelmann}, {Heymans}, {Moscardini}, \&
  {Radovich}}]{2008AandA...482..403M}
{Meneghetti}, M., {Melchior}, P., {Grazian}, A., {et~al.} 2008, \aap, 482, 403

\bibitem[{{Meneghetti} {et~al.}(2022){Meneghetti}, {Ragagnin}, {Borgani},
  {Calura}, {Despali}, {Giocoli}, {Granato}, {Grillo}, {Moscardini}, {Rasia},
  {Rosati}, {Angora}, {Bassini}, {Bergamini}, {Caminha}, {Granata}, {Mercurio},
  {Metcalf}, {Natarajan}, {Nonino}, {Pignataro}, {Ragone-Figueroa}, {Vanzella},
  {Acebron}, {Dolag}, {Murante}, {Taffoni}, {Tornatore}, {Tortorelli}, \&
  {Valentini}}]{2022A&A...668A.188M}
{Meneghetti}, M., {Ragagnin}, A., {Borgani}, S., {et~al.} 2022, \aap, 668, A188

\bibitem[{{Meneghetti} {et~al.}(2010){Meneghetti}, {Rasia}, {Merten},
  {Bellagamba}, {Ettori}, {Mazzotta}, {Dolag}, \&
  {Marri}}]{2010AandA...514A..93M}
{Meneghetti}, M., {Rasia}, E., {Merten}, J., {et~al.} 2010, \aap, 514, A93

\bibitem[{{Merten} {et~al.}(2019){Merten}, {Giocoli}, {Baldi}, {Meneghetti},
  {Peel}, {Lalande}, {Starck}, \& {Pettorino}}]{merten_cosmo}
{Merten}, J., {Giocoli}, C., {Baldi}, M., {et~al.} 2019, \mnras, 487, 104

\bibitem[{{Metcalf} {et~al.}(2019){Metcalf}, {Meneghetti}, {Avestruz},
  {Bellagamba}, {Bom}, {Bertin}, {Cabanac}, {Courbin}, {Davies},
  {Decenci{\`e}re}, {Flamary}, {Gavazzi}, {Geiger}, {Hartley},
  {Huertas-Company}, {Jackson}, {Jacobs}, {Jullo}, {Kneib}, {Koopmans},
  {Lanusse}, {Li}, {Ma}, {Makler}, {Li}, {Lightman}, {Petrillo}, {Serjeant},
  {Sch{\"a}fer}, {Sonnenfeld}, {Tagore}, {Tortora}, {Tuccillo},
  {Valent{\'\i}n}, {Velasco-Forero}, {Verdoes Kleijn}, \&
  {Vernardos}}]{metcalf_challenge}
{Metcalf}, R.~B., {Meneghetti}, M., {Avestruz}, C., {et~al.} 2019, \aap, 625,
  A119

\bibitem[{{Metcalf} \& {Petkova}(2014)}]{glamer_i}
{Metcalf}, R.~B. \& {Petkova}, M. 2014, \mnras, 445, 1942

\bibitem[{{Minor} {et~al.}(2021){Minor}, {Gad-Nasr}, {Kaplinghat}, \&
  {Vegetti}}]{quinn_dm}
{Minor}, Q., {Gad-Nasr}, S., {Kaplinghat}, M., \& {Vegetti}, S. 2021, \mnras,
  507, 1662

\bibitem[{{Myers} {et~al.}(2003){Myers}, {Jackson}, {Browne}, {de Bruyn},
  {Pearson}, {Readhead}, {Wilkinson}, {Biggs}, {Blandford}, {Fassnacht},
  {Koopmans}, {Marlow}, {McKean}, {Norbury}, {Phillips}, {Rusin}, {Shepherd},
  \& {Sykes}}]{2003MNRAS.341....1M}
{Myers}, S.~T., {Jackson}, N.~J., {Browne}, I.~W.~A., {et~al.} 2003, \mnras,
  341, 1

\bibitem[{{Napolitano} {et~al.}(2020){Napolitano}, {Li}, {Spiniello},
  {Tortora}, {Sergeyev}, {D'Ago}, {Guo}, {Xie}, {Radovich}, {Roy}, {Koopmans},
  {Kuijken}, {Bilicki}, {Erben}, {Getman}, {Heymans}, {Hildebrandt}, {Moya},
  {Shan}, {Vernardos}, \& {Wright}}]{napolitano_cnn}
{Napolitano}, N.~R., {Li}, R., {Spiniello}, C., {et~al.} 2020, \apjl, 904, L31

\bibitem[{{Navarro} {et~al.}(1996){Navarro}, {Frenk}, \&
  {White}}]{1996ApJ...462..563N}
{Navarro}, J.~F., {Frenk}, C.~S., \& {White}, S.~D.~M. 1996, \apj, 462, 563

\bibitem[{{Negrello} {et~al.}(2017){Negrello}, {Amber}, {Amvrosiadis}, {Cai},
  {Lapi}, {Gonzalez-Nuevo}, {De Zotti}, {Furlanetto}, {Maddox}, {Allen},
  {Bakx}, {Bussmann}, {Cooray}, {Covone}, {Danese}, {Dannerbauer}, {Fu},
  {Greenslade}, {Gurwell}, {Hopwood}, {Koopmans}, {Napolitano}, {Nayyeri},
  {Omont}, {Petrillo}, {Riechers}, {Serjeant}, {Tortora}, {Valiante}, {Verdoes
  Kleijn}, {Vernardos}, {Wardlow}, {Baes}, {Baker}, {Bourne}, {Clements},
  {Crawford}, {Dye}, {Dunne}, {Eales}, {Ivison}, {Marchetti}, {Micha{\l}owski},
  {Smith}, {Vaccari}, \& {van der Werf}}]{2017MNRAS.465.3558N}
{Negrello}, M., {Amber}, S., {Amvrosiadis}, A., {et~al.} 2017, \mnras, 465,
  3558

\bibitem[{{Negrello} {et~al.}(2010){Negrello}, {Hopwood}, {De Zotti}, {Cooray},
  {Verma}, {Bock}, {Frayer}, {Gurwell}, {Omont}, {Neri}, {Dannerbauer},
  {Leeuw}, {Barton}, {Cooke}, {Kim}, {da Cunha}, {Rodighiero}, {Cox},
  {Bonfield}, {Jarvis}, {Serjeant}, {Ivison}, {Dye}, {Aretxaga}, {Hughes},
  {Ibar}, {Bertoldi}, {Valtchanov}, {Eales}, {Dunne}, {Driver}, {Auld},
  {Buttiglione}, {Cava}, {Grady}, {Clements}, {Dariush}, {Fritz}, {Hill},
  {Hornbeck}, {Kelvin}, {Lagache}, {Lopez-Caniego}, {Gonzalez-Nuevo}, {Maddox},
  {Pascale}, {Pohlen}, {Rigby}, {Robotham}, {Simpson}, {Smith}, {Temi},
  {Thompson}, {Woodgate}, {York}, {Aguirre}, {Beelen}, {Blain}, {Baker},
  {Birkinshaw}, {Blundell}, {Bradford}, {Burgarella}, {Danese}, {Dunlop},
  {Fleuren}, {Glenn}, {Harris}, {Kamenetzky}, {Lupu}, {Maddalena}, {Madore},
  {Maloney}, {Matsuhara}, {Micha{\l}owski}, {Murphy}, {Naylor}, {Nguyen},
  {Popescu}, {Rawlings}, {Rigopoulou}, {Scott}, {Scott}, {Seibert}, {Smail},
  {Tuffs}, {Vieira}, {van der Werf}, \& {Zmuidzinas}}]{2010Sci...330..800N}
{Negrello}, M., {Hopwood}, R., {De Zotti}, G., {et~al.} 2010, Science, 330, 800

\bibitem[{{Nightingale} {et~al.}(2019){Nightingale}, {Massey}, {Harvey},
  {Cooper}, {Etherington}, {Tam}, \& {Hayes}}]{nightingale_lensmass}
{Nightingale}, J.~W., {Massey}, R.~J., {Harvey}, D.~R., {et~al.} 2019, \mnras,
  489, 2049

\bibitem[{{Oguri} {et~al.}(2014){Oguri}, {Rusu}, \& {Falco}}]{oguri_dm}
{Oguri}, M., {Rusu}, C.~E., \& {Falco}, E.~E. 2014, \mnras, 439, 2494

\bibitem[{{O'Riordan} {et~al.}(2023){O'Riordan}, {Despali}, {Vegetti},
  {Lovell}, \& {Molin{\'e}}}]{oriordan_2023}
{O'Riordan}, C.~M., {Despali}, G., {Vegetti}, S., {Lovell}, M.~R., \&
  {Molin{\'e}}, {\'A}. 2023, \mnras, 521, 2342

\bibitem[{{Pan} {et~al.}(2020){Pan}, {Liu}, {Forero-Romero}, {Sabiu}, {Li},
  {Miao}, \& {Li}}]{pan_cosmo}
{Pan}, S., {Liu}, M., {Forero-Romero}, J., {et~al.} 2020, Science China
  Physics, Mechanics, and Astronomy, 63, 110412

\bibitem[{{Pasquet} {et~al.}(2019){Pasquet}, {Bertin}, {Treyer}, {Arnouts}, \&
  {Fouchez}}]{pasquet_zphot}
{Pasquet}, J., {Bertin}, E., {Treyer}, M., {Arnouts}, S., \& {Fouchez}, D.
  2019, \aap, 621, A26

\bibitem[{{Petkova} {et~al.}(2014){Petkova}, {Metcalf}, \&
  {Giocoli}}]{glamer_ii}
{Petkova}, M., {Metcalf}, R.~B., \& {Giocoli}, C. 2014, \mnras, 445, 1954

\bibitem[{{Petrillo} {et~al.}(2017){Petrillo}, {Tortora}, {Chatterjee},
  {Vernardos}, {Koopmans}, {Verdoes Kleijn}, {Napolitano}, {Covone},
  {Schneider}, {Grado}, \& {McFarland}}]{2017MNRAS.472.1129P}
{Petrillo}, C.~E., {Tortora}, C., {Chatterjee}, S., {et~al.} 2017, \mnras, 472,
  1129

\bibitem[{{Petrillo} {et~al.}(2019){Petrillo}, {Tortora}, {Vernardos},
  {Koopmans}, {Verdoes Kleijn}, {Bilicki}, {Napolitano}, {Chatterjee},
  {Covone}, {Dvornik}, {Erben}, {Getman}, {Giblin}, {Heymans}, {de Jong},
  {Kuijken}, {Schneider}, {Shan}, {Spiniello}, \& {Wright}}]{petrillo_cnn2}
{Petrillo}, C.~E., {Tortora}, C., {Vernardos}, G., {et~al.} 2019, \mnras, 484,
  3879

\bibitem[{{Pires} {et~al.}(2020){Pires}, {Vandenbussche}, {Kansal}, {Bender},
  {Blot}, {Bonino}, {Boucaud}, {Brinchmann}, {Capobianco}, {Carretero},
  {Castellano}, {Cavuoti}, {Cl{\'e}dassou}, {Congedo}, {Conversi}, {Corcione},
  {Dubath}, {Fosalba}, {Frailis}, {Franceschi}, {Fumana}, {Grupp}, {Hormuth},
  {Kermiche}, {Knabenhans}, {Kohley}, {Kubik}, {Kunz}, {Ligori}, {Lilje},
  {Lloro}, {Maiorano}, {Marggraf}, {Massey}, {Meylan}, {Padilla}, {Paltani},
  {Pasian}, {Poncet}, {Potter}, {Raison}, {Rhodes}, {Roncarelli}, {Saglia},
  {Schneider}, {Secroun}, {Serrano}, {Stadel}, {Tallada Cresp{\'\i}}, {Tereno},
  {Toledo-Moreo}, \& {Wang}}]{Pires20}
{Pires}, S., {Vandenbussche}, V., {Kansal}, V., {et~al.} 2020, \aap, 638, A141

\bibitem[{{Ragagnin} {et~al.}(2022){Ragagnin}, {Meneghetti}, {Bassini},
  {Ragone-Figueroa}, {Granato}, {Despali}, {Giocoli}, {Granata}, {Moscardini},
  {Bergamini}, {Rasia}, {Valentini}, {Borgani}, {Calura}, {Dolag}, {Grillo},
  {Mercurio}, {Murante}, {Natarajan}, {Rosati}, {Taffoni}, {Tornatore}, \&
  {Tortorelli}}]{ragagnin2022}
{Ragagnin}, A., {Meneghetti}, M., {Bassini}, L., {et~al.} 2022, \aap, 665, A16

\bibitem[{{Reddi} {et~al.}(2019){Reddi}, {Kale}, \& S.}]{reddi_adam}
{Reddi}, S.~J., {Kale}, S., \& S., K. 2019, On the Convergence of Adam and
  Beyond, arXiv:1904.09237

\bibitem[{{Rojas} {et~al.}(2022){Rojas}, {Savary}, {Cl{\'e}ment}, {Maus},
  {Courbin}, {Lemon}, {Chan}, {Vernardos}, {Joseph}, {Ca{\~n}ameras}, \&
  {Galan}}]{rojas_cnn}
{Rojas}, K., {Savary}, E., {Cl{\'e}ment}, B., {et~al.} 2022, \aap, 668, A73

\bibitem[{Rumelhart {et~al.}(1986)Rumelhart, Hinton, \&
  Williams}]{rumelhart_backprop}
Rumelhart, D., Hinton, G.~E., \& Williams, R.~J. 1986, Nature, 323, 533

\bibitem[{{Savary} {et~al.}(2022){Savary}, {Rojas}, {Maus}, {Cl{\'e}ment},
  {Courbin}, {Gavazzi}, {Chan}, {Lemon}, {Vernardos}, {Ca{\~n}ameras},
  {Schuldt}, {Suyu}, {Cuillandre}, {Fabbro}, {Gwyn}, {Hudson}, {Kilbinger},
  {Scott}, \& {Stone}}]{savary_cnn}
{Savary}, E., {Rojas}, K., {Maus}, M., {et~al.} 2022, \aap, 666, A1

\bibitem[{{Schuldt} {et~al.}(2019){Schuldt}, {Chiriv{\`\i}}, {Suyu},
  {Y{\i}ld{\i}r{\i}m}, {Sonnenfeld}, {Halkola}, \& {Lewis}}]{schuldt_dis}
{Schuldt}, S., {Chiriv{\`\i}}, G., {Suyu}, S.~H., {et~al.} 2019, \aap, 631, A40

\bibitem[{{Seidel} \& {Bartelmann}(2007)}]{2007A&A...472..341S}
{Seidel}, G. \& {Bartelmann}, M. 2007, \aap, 472, 341

\bibitem[{{Shu} {et~al.}(2022){Shu}, {Ca{\~n}ameras}, {Schuldt}, {Suyu},
  {Taubenberger}, {Inoue}, \& {Jaelani}}]{shu_2022}
{Shu}, Y., {Ca{\~n}ameras}, R., {Schuldt}, S., {et~al.} 2022, \aap, 662, A4

\bibitem[{{Shuntov} {et~al.}(2020){Shuntov}, {Pasquet}, {Arnouts}, {Ilbert},
  {Treyer}, {Bertin}, {de la Torre}, {Dubois}, {Fouchez}, {Kraljic}, {Laigle},
  {Pichon}, \& {Vibert}}]{shuntov_zphot}
{Shuntov}, M., {Pasquet}, J., {Arnouts}, S., {et~al.} 2020, \aap, 636, A90

\bibitem[{Simonyan \& Zisserman(2015)}]{vggnet}
Simonyan, K. \& Zisserman, A. 2015, 3rd International Conference on Learning
  Representations, {ICLR} 2015, San Diego, CA, USA, May 7-9, 2015, Conference
  Track Proceedings

\bibitem[{{Sonnenfeld}(2022)}]{sonnenfeld_selectionf1}
{Sonnenfeld}, A. 2022, \aap, 659, A132

\bibitem[{{Sonnenfeld} {et~al.}(2018){Sonnenfeld}, {Chan}, {Shu}, {More},
  {Oguri}, {Suyu}, {Wong}, {Lee}, {Coupon}, {Yonehara}, {Bolton}, {Jaelani},
  {Tanaka}, {Miyazaki}, \& {Komiyama}}]{2018PASJ...70S..29S}
{Sonnenfeld}, A., {Chan}, J. H.~H., {Shu}, Y., {et~al.} 2018, \pasj, 70, S29

\bibitem[{{Sonnenfeld} {et~al.}(2023){Sonnenfeld}, {Li}, {Despali}, {Shajib},
  \& {Taylor}}]{sonnenfeld_selectionf2}
{Sonnenfeld}, A., {Li}, S.-S., {Despali}, G., {Shajib}, A.~J., \& {Taylor},
  E.~N. 2023, arXiv:2301.13230

\bibitem[{Srivastava {et~al.}(2014)Srivastava, Hinton, Krizhevsky, Sutskever,
  \& Salakhutdinov}]{srivastava_dropout}
Srivastava, N., Hinton, G., Krizhevsky, A., Sutskever, I., \& Salakhutdinov, R.
  2014, Journal of Machine Learning Research, 15, 1929\textendash 1958

\bibitem[{{Stacey} {et~al.}(2018){Stacey}, {McKean}, {Robertson}, {Ivison},
  {Isaak}, {Schleicher}, {van der Werf}, {Baan}, {Berciano Alba}, {Garrett}, \&
  {Loenen}}]{stacey_mag}
{Stacey}, H.~R., {McKean}, J.~P., {Robertson}, N.~C., {et~al.} 2018, \mnras,
  476, 5075

\bibitem[{{Stehman}(1997)}]{stehman_cm}
{Stehman}, S.~V. 1997, Remote Sensing of Environment, 62, 77

\bibitem[{{Suyu} {et~al.}(2012){Suyu}, {Hensel}, {McKean}, {Fassnacht}, {Treu},
  {Halkola}, {Norbury}, {Jackson}, {Schneider}, {Thompson}, {Auger},
  {Koopmans}, \& {Matthews}}]{suyu_dis}
{Suyu}, S.~H., {Hensel}, S.~W., {McKean}, J.~P., {et~al.} 2012, \apj, 750, 10

\bibitem[{{Szegedy} {et~al.}(2016){Szegedy}, {Vanhoucke}, {Ioffe}, {Shlens}, \&
  {Wojna}}]{gnet_2}
{Szegedy}, C., {Vanhoucke}, V., {Ioffe}, S., {Shlens}, J., \& {Wojna}, Z. 2016,
  2016 IEEE Conference on Computer Vision and Pattern Recognition (CVPR), 2818

\bibitem[{{Szegedy} {et~al.}(2015){Szegedy}, {Wei Liu}, {Yangqing Jia},
  {Sermanet}, {Reed}, {Anguelov}, {Erhan}, {Vanhoucke}, \&
  {Rabinovich}}]{gnet_1}
{Szegedy}, C., {Wei Liu}, {Yangqing Jia}, {et~al.} 2015, 2015 IEEE Conference
  on Computer Vision and Pattern Recognition (CVPR), 1

\bibitem[{Tallada {et~al.}(2020)Tallada, Carretero, Casals, Acosta-Silva,
  Serrano, Caubet, Castander, César, Crocce, Delfino, Eriksen, Fosalba,
  Gaztañaga, Merino, Neissner, \& Tonello}]{TALLADA2020100391}
Tallada, P., Carretero, J., Casals, J., {et~al.} 2020, Astronomy and Computing,
  32, 100391

\bibitem[{{Taufik Andika} {et~al.}(2023){Taufik Andika}, {Suyu},
  {Ca{\~n}ameras}, {Melo}, {Schuldt}, {Shu}, {Eilers}, {Timur Jaelani}, \&
  {Yue}}]{andika_2023}
{Taufik Andika}, I., {Suyu}, S.~H., {Ca{\~n}ameras}, R., {et~al.} 2023, arXiv
  e-prints, arXiv:2307.01090

\bibitem[{{The Dark Energy Survey Collaboration}(2005)}]{des}
{The Dark Energy Survey Collaboration}. 2005, arXiv:0510346

\bibitem[{{Treu} \& {Koopmans}(2004)}]{treu_lensmass}
{Treu}, T. \& {Koopmans}, L. V.~E. 2004, \apj, 611, 739

\bibitem[{{Tu} {et~al.}(2008){Tu}, {Limousin}, {Fort}, {Shu}, {Sygnet},
  {Jullo}, {Kneib}, \& {Richard}}]{tu_2008}
{Tu}, H., {Limousin}, M., {Fort}, B., {et~al.} 2008, \mnras, 386, 1169

\bibitem[{{Vegetti} {et~al.}(2018){Vegetti}, {Despali}, {Lovell}, \&
  {Enzi}}]{vegetti_dm}
{Vegetti}, S., {Despali}, G., {Lovell}, M.~R., \& {Enzi}, W. 2018, \mnras, 481,
  3661

\bibitem[{{Wong} {et~al.}(2022){Wong}, {Chan}, {Chao}, {Jaelani}, {Kayo},
  {Lee}, {More}, \& {Oguri}}]{wong_2022}
{Wong}, K.~C., {Chan}, J. H.~H., {Chao}, D. C.~Y., {et~al.} 2022, \pasj, 74,
  1209

\bibitem[{{Wu} \& {Boada}(2019)}]{wu_metallicity}
{Wu}, J.~F. \& {Boada}, S. 2019, \mnras, 484, 4683

\bibitem[{{Xie} {et~al.}(2017){Xie}, {Girshick}, {Dollár}, {Tu}, \&
  {He}}]{resnext}
{Xie}, S., {Girshick}, R., {Dollár}, P., {Tu}, Z., \& {He}, K. 2017, 2017 IEEE
  Conference on Computer Vision and Pattern Recognition (CVPR), 5987

\bibitem[{{Xu} {et~al.}(2015){Xu}, {Wang}, {Chen}, \&
  {Li}}]{2015arXiv150500853X}
{Xu}, B., {Wang}, N., {Chen}, T., \& {Li}, M. 2015, arXiv:1505.00853

\bibitem[{Zhou \& Chellappa(1988)}]{zhou_maxpooling}
Zhou, Y.-T. \& Chellappa, R. 1988, in IEEE 1988 International Conference on
  Neural Networks, Vol.~2, 71

\bibitem[{{Zhu} {et~al.}(2019){Zhu}, {Dai}, {Bian}, {Chen}, {Chen}, \&
  {Hu}}]{zhu_morphology}
{Zhu}, X.-P., {Dai}, J.-M., {Bian}, C.-J., {et~al.} 2019, \apss, 364, 55

\end{thebibliography}

\begin{appendix}
\onecolumn
\section{Network architectures}\label{app:implementation} 
\begin{figure*}[h!]
    \centering
\includegraphics[width=.75\textwidth,height=.75\textheight,keepaspectratio]{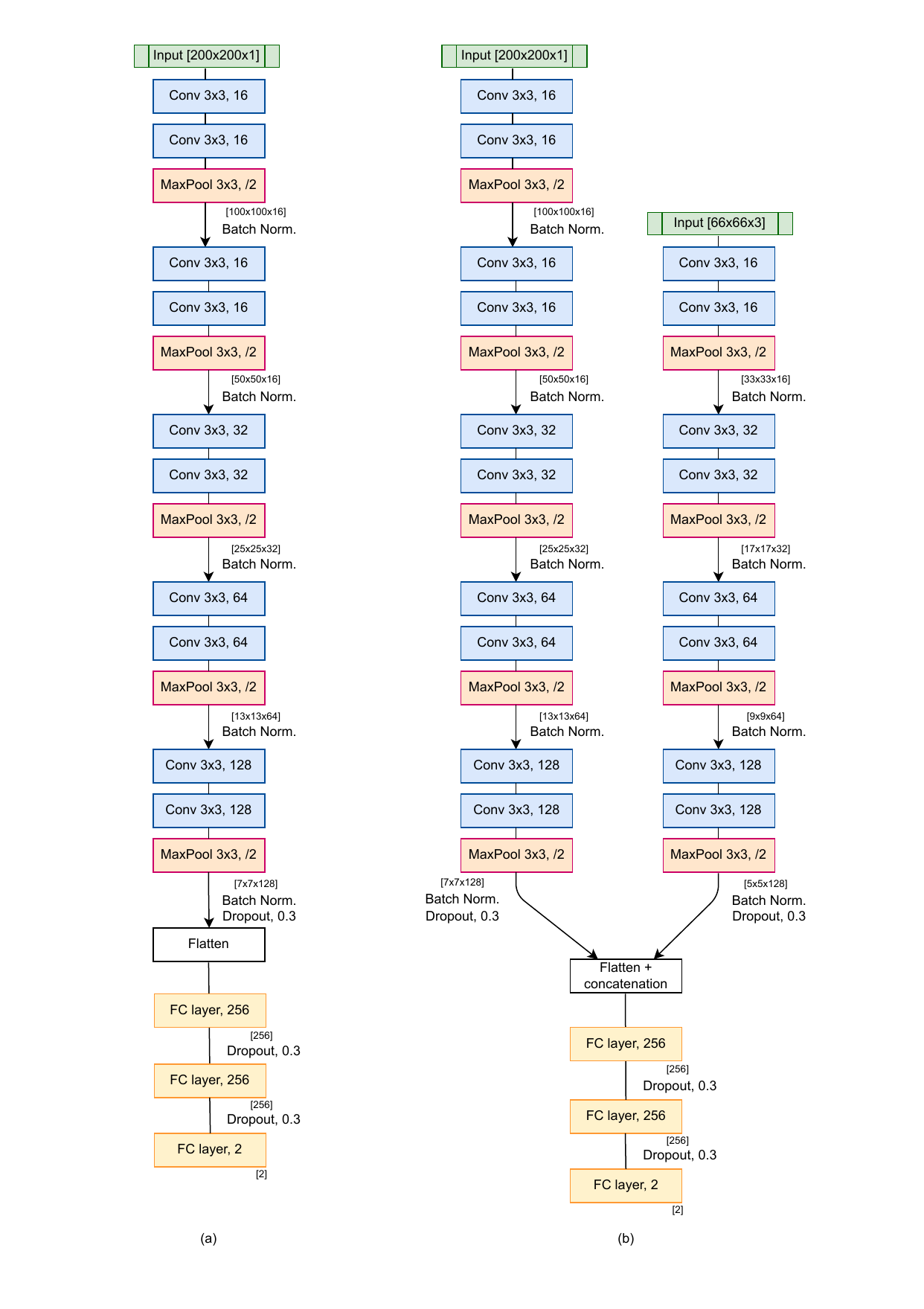}
    \caption{VGG-like network configurations tested on (a) VIS images and (b) multiband images. We report the dimension (\textit{D}) and number (\textit{F}) of the filters used in the convolutional layers in the format \textit{D}$\times$\textit{D}, \textit{F}. We also indicate the pooling region (\textit{R}) and the strides (\textit{S}) in the pooling layers in the format \textit{R}$\times$\textit{R}, /\textit{S}. The numbers in square brackets indicate the dimension and number of the feature maps obtained as the output of the layers in the format [\textit{D}$\times$\textit{D}$\times$\textit{F}] in the case of the convolutional layers, and the number of nodes in the format [N] in the case of the fully connected layers.}
    \label{fig:my_vgg}
\end{figure*}

\begin{figure*}[h!]
    \centering
    \includegraphics[width = 0.8\textwidth]{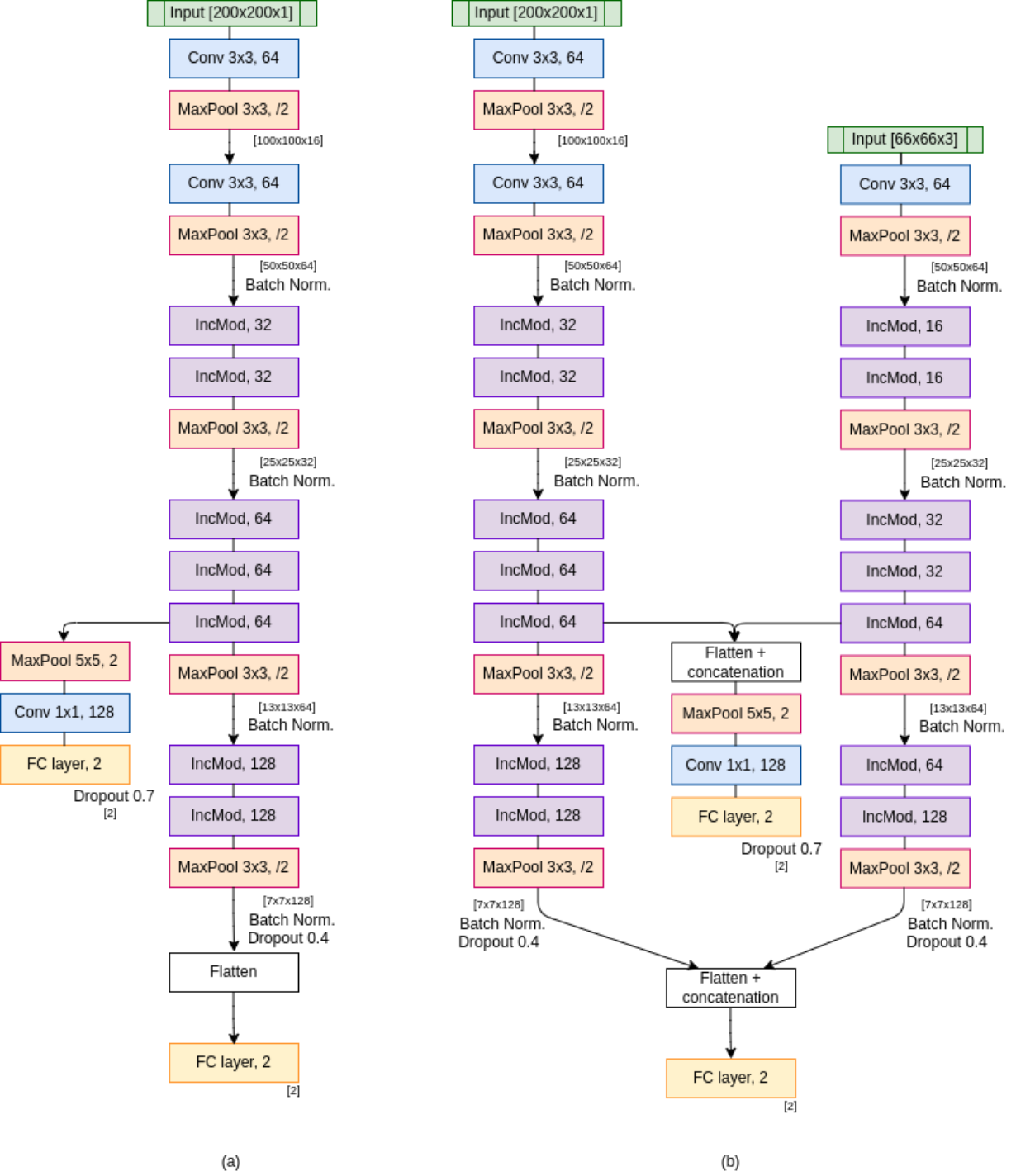}
    \caption{Inception network configurations tested on (a) VIS images and (b) multiband images. These diagrams use the same notation as in Fig. \ref{fig:my_vgg}. Every inception module (IncMod) is built as described in subsection \ref{sec:incnet}.}
    \label{fig:my_incnet}
\end{figure*}

\begin{figure*}[h!]
    \centering
    \includegraphics[width = 0.8\textwidth]{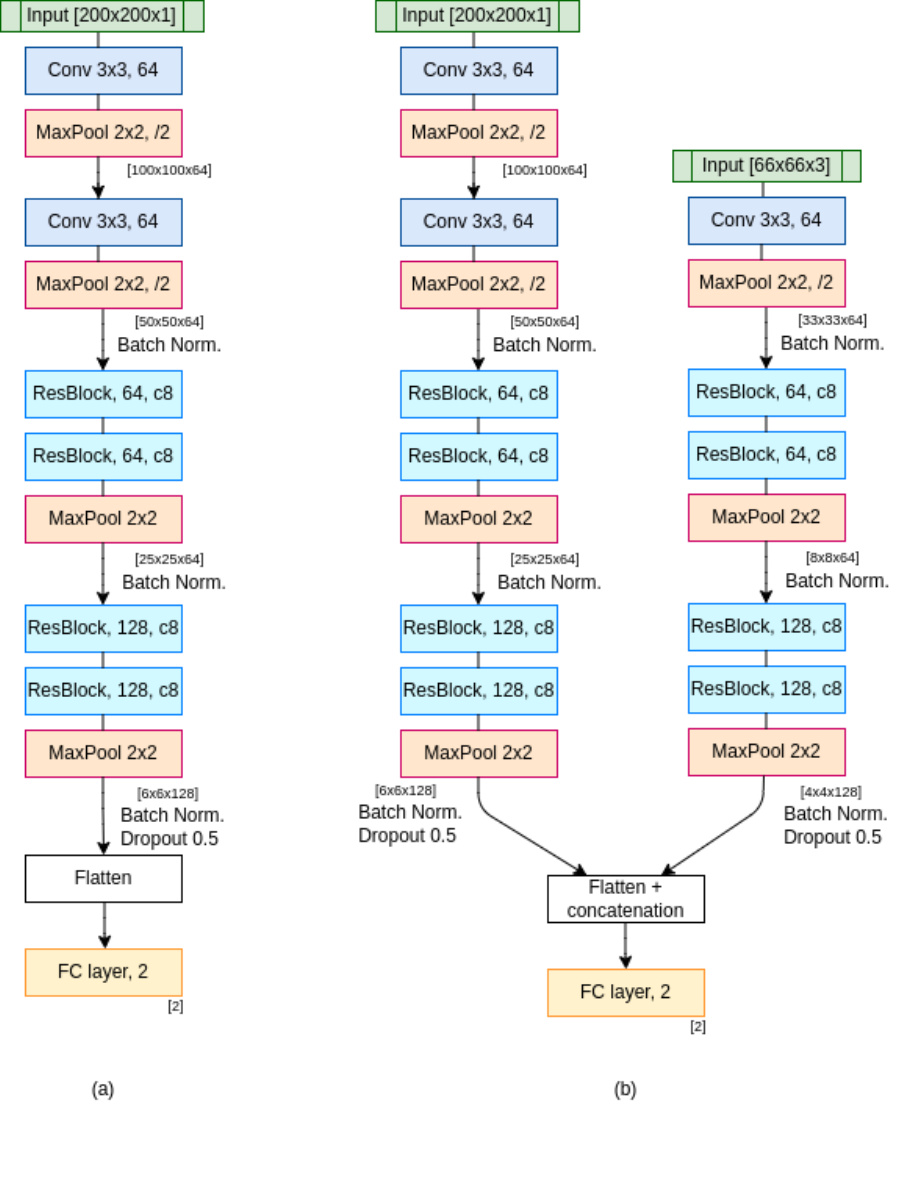}
    \caption{Residual network configurations tested on (a) VIS images and (b) multiband images. These diagrams use the same notation as in Fig. \ref{fig:my_vgg}. Every residual block (ResBlock) is built as described in subsection \ref{sec:resnet}: c8 refers to the cardinality of the block, which we set to be equal to eight.}
    \label{fig:my_resnext}
\end{figure*}

The three figures in this appendix show the architectures of the networks we implemented. In particular, Fig.\ref{fig:my_vgg} shows the VGG-like network, Fig. \ref{fig:my_incnet} shows the IncNet, and Fig.~\ref{fig:my_resnext} shows the ResNet.

\FloatBarrier

\FloatBarrier
\FloatBarrier

\section{Tables}\label{app:tables}
In this appendix, we summarize the main results of our tests. In Table~\ref{tab:summary_challenge} we show the results of training our models on VIS images, in Table~ \ref{tab:S3vsS6metrics} we compare the results of applying our models trained on S2 and on S4 to the test set S4, in Table~\ref{tab:s3_test} we show the results of two additional tests, S2/S3, and S2/S4, in Table~\ref{tab:summary_challenge_mw} we show the results of training our models on multiband images, and in Table~\ref{tab:test_trueprop} we present the results of a test with realistic proportions between lenses and nonlenses. 

\begin{table*}[h!]
\centering
\caption{Summary of the performance of the VGG-like network, the IncNet, and the ResNet in classifying the objects of the four selections of images in the $\IE$ band.}
\label{tab:summary_challenge}
\begin{tabular}{ccccccccc}
\hline
\hline\\ [-1.8ex]
\multicolumn{9}{c}{\textbf{VGG-like network}}                                                                         \\ \hline\\ [-2.0ex]
          & \multicolumn{2}{c}{S1}   & \multicolumn{2}{c}{S2}   & \multicolumn{2}{c}{S3}   & \multicolumn{2}{c}{S4}   \\ \hline\\ [-2.0ex]
         Class & 0           & 1          & 0           & 1          & 0           & 1          & 0           & 1          \\ \hline\\ [-2.0ex]
Precision & 0.95        & 0.98       & 0.94        & 0.97       & 0.92        & 0.94       & 0.79        & 0.89       \\ \hline\\ [-2.0ex]
Recall    & 0.98        & 0.94       & 0.98        & 0.94       & 0.94        & 0.92       & 0.90        & 0.77       \\ \hline\\ [-2.0ex]
F1-score  & 0.96        & 0.96       & 0.96        & 0.96       & 0.93        & 0.93       & 0.84        & 0.83       \\ \hline\\ [-2.0ex]
Accuracy  & \multicolumn{2}{c}{0.96} & \multicolumn{2}{c}{0.96} & \multicolumn{2}{c}{0.93} & \multicolumn{2}{c}{0.84} \\ \hline\\ [-2.0ex]
AUC       & \multicolumn{2}{c}{0.77} & \multicolumn{2}{c}{0.58} & \multicolumn{2}{c}{0.88} & \multicolumn{2}{c}{0.57} \\ \hline \hline\\ [-1.8ex]
\multicolumn{9}{c}{\textbf{Inception Network}}                                                                        \\ \hline\\ [-2.0ex]
          & \multicolumn{2}{c}{S1}   & \multicolumn{2}{c}{S2}   & \multicolumn{2}{c}{S3}   & \multicolumn{2}{c}{S4}   \\ \hline\\ [-2.0ex]
         Class & 0           & 1          & 0           & 1          & 0           & 1          & 0           & 1          \\ \hline\\ [-2.0ex]
Precision & 0.97        & 1.0        & 0.97        & 0.96       & 0.94        & 0.93       & 0.84        & 0.90       \\ \hline\\ [-2.0ex]
Recall    & 1.0         & 0.96       & 0.96        & 0.97       & 0.93        & 0.94       & 0.91        & 0.83       \\ \hline\\ [-2.0ex]
F1-score  & 0.98        & 0.98       & 0.96        & 0.96       & 0.93        & 0.94       & 0.87        & 0.86       \\ \hline\\ [-2.0ex]
Accuracy  & \multicolumn{2}{c}{0.98} & \multicolumn{2}{c}{0.96} & \multicolumn{2}{c}{0.94} & \multicolumn{2}{c}{0.87} \\ \hline\\ [-2.0ex]
AUC       & \multicolumn{2}{c}{0.92} & \multicolumn{2}{c}{0.88} & \multicolumn{2}{c}{0.90} & \multicolumn{2}{c}{0.81} \\ \hline\hline\\ [-1.8ex]
\multicolumn{9}{c}{\textbf{Residual Network}}                                                                         \\ \hline\\ [-2.0ex]
          & \multicolumn{2}{c}{S1}   & \multicolumn{2}{c}{S2}   & \multicolumn{2}{c}{S3}   & \multicolumn{2}{c}{S4}   \\ \hline\\ [-2.0ex]
         Class & 0           & 1          & 0           & 1          & 0           & 1          & 0           & 1          \\ \hline\\ [-2.0ex]
Precision & 0.93        & 0.97       & 0.90        & 0.92       & 0.86        & 0.89       & 0.71        & 0.84       \\ \hline\\ [-2.0ex]
Recall    & 0.97        & 0.92       & 0.92        & 0.89       & 0.89        & 0.85       & 0.87        & 0.66       \\ \hline\\ [-2.0ex]
F1-score  & 0.95        & 0.94       & 0.91        & 0.91       & 0.88        & 0.87       & 0.78        & 0.74       \\ \hline\\ [-2.0ex]
Accuracy  & \multicolumn{2}{c}{0.95} & \multicolumn{2}{c}{0.91} & \multicolumn{2}{c}{0.87} & \multicolumn{2}{c}{0.76} \\ \hline\\ [-2.0ex]
AUC       & \multicolumn{2}{c}{0.81} & \multicolumn{2}{c}{0.85} & \multicolumn{2}{c}{0.79} & \multicolumn{2}{c}{0.70} \\ \hline\\ [-2.0ex]
\end{tabular}
\tablefoot{The precision, recall, and F1-score are evaluated on the class of the nonlenses (0) and of the lenses (1) separately,  while accuracy and AUC are global quantities.}
\end{table*}

\begin{table*}
\centering
\caption{Comparison between the metrics of tests on the selection S2 with the models trained on S2 (top) and on S4 (bottom).}
\label{tab:S3vsS6metrics}
\begin{tabular}{ccccccc}
\hline\hline\\ [-2.0ex]
\multicolumn{7}{c}{\textbf{S2/S2}} \\ \hline\\ [-2.0ex]
 & \multicolumn{2}{c}{\textbf{VGG-like network}} & \multicolumn{2}{c}{\textbf{Inception Network}} & \multicolumn{2}{c}{\textbf{Residual Network}} \\ \hline\\ [-2.0ex]
Class & 0 & 1 & 0 & 1 & 0 & 1 \\ \hline\\ [-2.0ex]
Precision & 0.94 & 0.97 & 0.97 & 0.96 & 0.90 & 0.92 \\ \hline\\ [-2.0ex]
Recall & 0.98 & 0.94 & 0.96 & 0.97 & 0.92 & 0.89 \\ \hline\\ [-2.0ex]
F1-score & 0.96 & 0.96 & 0.96 & 0.96 & 0.91 & 0.91 \\ \hline\\ [-2.0ex]
Accuracy & \multicolumn{2}{c}{0.96} & \multicolumn{2}{c}{0.96} & \multicolumn{2}{c}{0.91} \\ \hline\\ [-2.0ex]
AUC & \multicolumn{2}{c}{0.58} & \multicolumn{2}{c}{0.88} & \multicolumn{2}{c}{0.85} \\ \hline\hline\\ [-2.0ex]
\multicolumn{7}{c}{\textbf{S4/S2}} \\ \hline\\ [-2.0ex]
 & \multicolumn{2}{c}{\textbf{VGG-like network}} & \multicolumn{2}{c}{\textbf{Inception Network}} & \multicolumn{2}{c}{\textbf{Residual Network}} \\ \hline\\ [-2.0ex]
 Class & 0 & 1 & 0 & 1 & 0 & 1 \\ \hline\\ [-2.0ex]
Precision & 0.96 & 0.74 & 0.99 & 0.77 & 0.95 & 0.67 \\ \hline\\ [-2.0ex]
Recall & 0.89 & 0.91 & 0.90 & 0.98 & 0.85 & 0.89 \\ \hline\\ [-2.0ex]
F1-score & 0.93 & 0.82 & 0.94 & 0.86 & 0.90 & 0.76 \\ \hline\\ [-2.0ex]
Accuracy & \multicolumn{2}{c}{0.89} & \multicolumn{2}{c}{0.92} & \multicolumn{2}{c}{0.86} \\ \hline\\ [-2.0ex]
AUC & \multicolumn{2}{c}{0.51} & \multicolumn{2}{c}{0.88} & \multicolumn{2}{c}{0.75} \\ \hline\\ [-2.0ex]
\end{tabular}
\tablefoot{Class 0 refers to the nonlenses, while class 1 refers to the lenses.}
\end{table*}

\begin{table*}[htpb]
\centering
\caption{Summary of the performance of the VGG-like network, the inception network, and the residual network, trained on the selection S2, in classifying the objects that are part of the selections S3 and S4.}
\label{tab:s3_test}
\begin{tabular}{ccccccccccccc}
\hline
\hline\\ [-2.0ex]
\textbf{} & \multicolumn{4}{c}{\textbf{VGG-like network}}                  & \multicolumn{4}{l}{\textbf{Inception Network}}                 & \multicolumn{4}{l}{\textbf{Residual Network}}                  \\ \hline\\ [-2.0ex]
          & \multicolumn{2}{c}{S2/S3} & \multicolumn{2}{c}{S2/S4} & \multicolumn{2}{c}{S2/S3} & \multicolumn{2}{c}{S2/S4} & \multicolumn{2}{c}{S2/S3} & \multicolumn{2}{c}{S2/S4} \\ \hline\\ [-2.0ex]
          Class & 0           & 1           & 0            & 1          & 0           & 1           & 0           & 1           & 0           & 1           & 0            & 1          \\ \hline\\ [-2.0ex]
Precision & 0.77        & 0.97        & 0.62         & 0.95       & 0.82        & 0.96        & 0.65        & 0.93        & 0.75        & 0.88        & 0.64         & 0.85       \\ \hline\\ [-2.0ex]
Recall    & 0.98        & 0.68        & 0.98         & 0.33       & 0.97        & 0.76        & 0.97        & 0.42        & 0.92        & 0.67        & 0.94         & 0.40       \\ \hline\\ [-2.0ex]
F1-score  & 0.86        & 0.80        & 0.76         & 0.48       & 0.89        & 0.85        & 0.78        & 0.58        & 0.83        & 0.76        & 0.76         & 0.55       \\ \hline\\ [-2.0ex]
Accuracy  & \multicolumn{2}{c}{0.83}  & \multicolumn{2}{c}{0.68}  & \multicolumn{2}{c}{0.87}  & \multicolumn{2}{c}{0.71}  & \multicolumn{2}{c}{0.80}  & \multicolumn{2}{c}{0.69}  \\ \hline\\ [-2.0ex]
AUC       & \multicolumn{2}{c}{0.57}  & \multicolumn{2}{c}{0.52}  & \multicolumn{2}{c}{0.81}  & \multicolumn{2}{c}{0.7}   & \multicolumn{2}{c}{0.78}  & \multicolumn{2}{c}{0.65}  \\ \hline\\ [-2.0ex]
\end{tabular}
\tablefoot{The precision, recall, and F1-score are evaluated on the class of the nonlenses (0) and of the lenses (1) separately.}
\end{table*}

\begin{table*}
\centering
\caption{Same as in Table \ref{tab:summary_challenge}, but using images in the VIS and NISP bands.}
\label{tab:summary_challenge_mw}
\begin{tabular}{ccccccccc}
\hline\hline\\ [-2.0ex]
\multicolumn{9}{c}{\textbf{VGG-like network}}                                                                         \\ \hline\\ [-2.0ex]
          & \multicolumn{2}{c}{S1}   & \multicolumn{2}{c}{S2}   & \multicolumn{2}{c}{S3}   & \multicolumn{2}{c}{S4}   \\ \hline\\ [-2.0ex]
          Class & 0           & 1          & 0           & 1          & 0           & 1          & 0           & 1          \\ \hline\\ [-2.0ex]
Precision & 0.99        & 0.97       & 0.98        & 0.97       & 0.91        & 0.96       & 0.81        & 0.91       \\ \hline\\ [-2.0ex]
Recall    & 0.97        & 0.99       & 0.97        & 0.98       & 0.96        & 0.91       & 0.92        & 0.79       \\ \hline\\ [-2.0ex]
F1-score  & 0.98        & 0.98       & 0.98        & 0.98       & 0.94        & 0.93       & 0.86        & 0.84       \\ \hline\\ [-2.0ex]
Accuracy  & \multicolumn{2}{c}{0.98} & \multicolumn{2}{c}{0.98} & \multicolumn{2}{c}{0.93} & \multicolumn{2}{c}{0.85} \\ \hline\\ [-2.0ex]
AUC       & \multicolumn{2}{c}{0.65} & \multicolumn{2}{c}{0.87} & \multicolumn{2}{c}{0.67} & \multicolumn{2}{c}{0.62} \\ \hline\hline\\ [-2.0ex]
\multicolumn{9}{c}{\textbf{Inception Network}}                                                                        \\ \hline\\ [-2.0ex]
          & \multicolumn{2}{c}{S1}   & \multicolumn{2}{c}{S2}   & \multicolumn{2}{c}{S3}   & \multicolumn{2}{c}{S4}   \\ \hline\\ [-2.0ex]
          Class & 0           & 1          & 0           & 1          & 0           & 1          & 0           & 1          \\ \hline\\ [-2.0ex]
Precision & 0.98        & 0.96       & 0.97        & 0.98       & 0.96        & 0.96       & 0.87        & 0.91       \\ \hline\\ [-2.0ex]
Recall    & 0.96        & 0.98       & 0.98        & 0.96       & 0.96        & 0.96       & 0.91        & 0.87       \\ \hline\\ [-2.0ex]
F1-score  & 0.97        & 0.97       & 0.97        & 0.97       & 0.96        & 0.96       & 0.89        & 0.89       \\ \hline\\ [-2.0ex]
Accuracy  & \multicolumn{2}{c}{0.97} & \multicolumn{2}{c}{0.97} & \multicolumn{2}{c}{0.96} & \multicolumn{2}{c}{0.89} \\ \hline\\ [-2.0ex]
AUC       & \multicolumn{2}{c}{0.77} & \multicolumn{2}{c}{0.9}  & \multicolumn{2}{c}{0.92} & \multicolumn{2}{c}{0.84} \\ \hline\hline\\ [-2.0ex]
\multicolumn{9}{c}{\textbf{Residual Network}}                                                                         \\ \hline\\ [-2.0ex]
          & \multicolumn{2}{c}{S1}   & \multicolumn{2}{c}{S2}   & \multicolumn{2}{c}{S3}   & \multicolumn{2}{c}{S4}   \\ \hline\\ [-2.0ex]
          Class & 0           & 1          & 0           & 1          & 0           & 1          & 0           & 1          \\ \hline\\ [-2.0ex]
Precision & 0.96        & 0.95       & 0.92        & 0.94       & 0.86        & 0.92       & 0.74        & 0.85       \\ \hline\\ [-2.0ex]
Recall    & 0.94        & 0.96       & 0.94        & 0.92       & 0.92        & 0.87       & 0.87        & 0.71       \\ \hline\\ [-2.0ex]
F1-score  & 0.95        & 0.95       & 0.93        & 0.93       & 0.90        & 0.89       & 0.80        & 0.77       \\ \hline\\ [-2.0ex]
Accuracy  & \multicolumn{2}{c}{0.95} & \multicolumn{2}{c}{0.93} & \multicolumn{2}{c}{0.90} & \multicolumn{2}{c}{0.78} \\ \hline\\ [-2.0ex]
AUC       & \multicolumn{2}{c}{0.81} & \multicolumn{2}{c}{0.88} & \multicolumn{2}{c}{0.81} & \multicolumn{2}{c}{0.72} \\ \hline\\ [-2.0ex]
\end{tabular}
\end{table*}

\begin{table*}
\centering
\caption{Results of testing our best-performing networks, trained on S1, on a test set with 200 lenses and $\num{80000}$ nonlenses.}
\label{tab:test_trueprop}
\begin{tabular}{ccccccccc}
\hline\hline\\ [-2.0ex] 
& \multicolumn{2}{c}{\textbf{VGG-like network}} & \multicolumn{2}{c}{\textbf{Inception Network}} & \multicolumn{2}{c}{\textbf{Residual Network}}  & \multicolumn{2}{c}{\textbf{Ensemble Network}}\\ \hline\\ [-2.0ex]
Class & 0 & 1 & 0 & 1 & 0 & 1 & 0 & 1 \\ \hline\\ [-2.0ex]
Precision & 1.0 & 0.15 & 1.0 & 0.45 & 1.0 & 0.13 & 1.0 & 0.46\\ \hline\\ [-2.0ex]
Recall & 0.98 & 0.94 & 0.99 & 0.96 & 0.98 & 0.92 & 1.0 & 0.97\\ \hline\\ [-2.0ex]
F1-score & 0.99 & 0.26 & 0.99 & 0.61 & 0.99 & 0.23 & 1.0 & 0.63\\ \hline\\ [-2.0ex]
Accuracy & \multicolumn{2}{c}{0.98} & \multicolumn{2}{c}{0.99} & \multicolumn{2}{c}{0.98}& \multicolumn{2}{c}{1.0} \\ \hline\\ [-2.0ex]
AUC & \multicolumn{2}{c}{0.76} & \multicolumn{2}{c}{0.83} & \multicolumn{2}{c}{0.81}& \multicolumn{2}{c}{0.99} \\ \hline\\ [-2.0ex]

\end{tabular}
\tablefoot{Class 0 refers to the nonlenses, while class 1 refers to the lenses. Ensemble network refers to the combination of the predictions of the three networks.}
\end{table*}

\end{appendix}

\end{document}